\documentclass[showpacs,amsmath,amssymb,onecolumn,pra,superscriptaddress]{revtex4-2}

\usepackage{graphicx}
\usepackage{dcolumn}
\usepackage{bm}
\usepackage[colorlinks = true]{hyperref}
\usepackage{mathtools}
\usepackage{braket}
\usepackage{tikz}
\usepackage{subfigure, epsfig}
\usepackage{amsthm}
\usepackage{comment}

\newtheorem{remark}{Remark}

\newtheorem{prop}{Proposition}
\newtheorem{lemma}{Lemma}

\newtheorem{fact}{Fact}

\newcommand{\mc}{\mathcal}

\newcommand{\mbb}{\mathbb}
\newcommand{\mr}{\mathrm}
\newcommand{\Hr}{\mathrm{Haar}}
\newcommand{\tr}{\mathrm{Tr}}

\newcommand{\id}{\mathbb{I}}
\newcommand{\zy}[1]{\textcolor{blue}{(zy: #1)}}

\newcommand{\comments}[1]{}

\definecolor{egg}{rgb}{.98,.97,.92}
\definecolor{dullblue}{rgb}{.29,.47,.77}

\begin{document}


\title{Quantum scrambling with classical shadows}

\author{Roy J. Garcia}
\thanks{These two authors contributed equally to this work.}
\affiliation{Department of Physics, Harvard University, Cambridge, Massachusetts 02138, USA}
\author{You Zhou}
\thanks{These two authors contributed equally to this work.}
\affiliation{Department of Physics, Harvard University, Cambridge, Massachusetts 02138, USA}
\affiliation{School of Physical and Mathematical Sciences, Nanyang Technological University, 637371, Singapore}
\author{Arthur Jaffe}
\affiliation{Department of Physics, Harvard University, Cambridge, Massachusetts 02138, USA}

\date{\today}

\begin{abstract}
Quantum dynamics is of fundamental interest and has implications in quantum information processing. The four-point out-of-time-ordered correlator (OTOC) is traditionally used to quantify quantum information scrambling under many-body dynamics. Due to the OTOC's unusual time ordering, its measurement is challenging. We propose higher-point OTOCs to reveal early-time scrambling behavior, and present protocols to measure any higher-point OTOC using the shadow estimation method. The protocols circumvent the need for time reversal evolution and ancillary control. They can be implemented in near-term quantum devices with single-qubit readout.
\end{abstract}

\maketitle

\section{Introduction}
Quantum scrambling describes the delocalization of quantum information in quantum chaotic systems \cite{Lewis-Swan2019,Shenker2014butterfly}. Much of the interest in scrambling derives from 
the study of black holes---the fastest scramblers in nature \cite{Hayden_2007,Sekino_2008,Shenker2014butterfly,Maldacena_2016}. The holographic duality permits the investigation of black hole scrambling via the probing of certain models in condensed matter physics, like the Sachdev-Ye-Kitaev model \cite{PhysRevLett.70.3339,Kitaev2015}. Scrambling can be studied by probing the four-point out-of-time-ordered correlator (OTOC)  \cite{PhysRevLett.115.131603,Swingle_2016,PhysRevB.95.060201,PhysRevD.96.065005}. 
These correlators can be used to quantify chaos in many-body systems ranging from a non-integrable Ising model \cite{PhysRevLett.106.050405} to the Dicke model \cite{PhysRev.93.99,Alavirad_2019,Lewis_Swan_2019}.
For fast scrambling systems \cite{PhysRevLett.70.3339,Kitaev2015,PhysRevLett.123.130601,Belyansky_2020}, this correlator decays exponentially within the scrambling time, according to a Lyapunov exponent. Measuring OTOCs in this regime can be used to investigate models with holographic duals. OTOCs can also be used to describe slow scrambling due to a small Lyapunov exponent, by probing many-body localized systems \cite{Huang_2016,Fan_2017,Chen2016Logarithmic,Chen_2016,Rong-Qiang2017localization,Swingle2017Slow}.

The time ordering of the OTOC makes its measurement tedious. Nevertheless, there are experimental protocols to measure the four-point OTOC. Protocols based on time reversal evolution have been demonstrated \cite{G_rttner_2017,Xuan2018Localization}. The OTOC has also been measured using a nuclear magnetic resonance quantum simulator \cite{PhysRevX.7.031011}. Recent investigations of scrambling have searched for measurement protocols that circumvent the need for time reversal. To distinguish between scrambling and decoherence, a teleportation protocol that measures the OTOC in the large-time limit has been developed in Ref.~\cite{Yoshida_2019} and demonstrated in Ref.~\cite{Landsman_2019}. A method based on statistical correlations computes the four-point OTOC in terms of experimentally-friendly correlators \cite{Vermersch_2019}; it has been demonstrated in Ref.~\cite{Trapped2020}. Although quantum chaos is often studied through the four-point OTOC, it is suspected that scrambling is sensitive to higher-point correlators \cite{Shenker_2014,Roberts_2017,Landsman_2019}. Thus, one desires a protocol to measure higher-point OTOCs, especially one without time reversal or ancillary control operations.

We present protocols to measure any higher-point out-of-time-ordered correlator using classical shadows \cite{aaronson2018shadow,Huang_2020}, which is an efficient scheme recently proposed to predict functions of quantum states. While nonlinear functions are often computed by preparing multiple copies of a state \cite{Ekert2002Direct,Daley2012Measuring,Abanin2012Measuring,Islam2015,Kaufmanen2016tanglement}, classical shadows allow us to bypass this preparation by measuring a single copy of the state. Our protocols avoid time reversal and can probe OTOCs at any time. While ancillary qubits are integrated, we do not require they exert an interaction on the system  \cite{yao2016interferometric}. We find that the eight-point OTOC reveals early-time information delocalization not present in the four-point OTOC, making it a promising candidate to probe scrambling dynamics. 

We provide a statistical error analysis to show that our protocols are more efficient than brute force tomography. Since our protocols express OTOCs in terms of nonlinear functions of a state, we give a refined variance analysis on nonlinear functions which rely on prior knowledge of 
the target state, establishing a tighter bound than previous works \cite{Huang_2020,elben2020mixedstate}\textendash a result which is of independent interest. We also numerically simulate our protocols in a non-integrable, mixed-field Ising model and show that they can be implemented by current experimental platforms containing a moderate number of qubits. 

\subsection{Higher-point correlators}
Scrambling is often quantified through operator spreading, described as follows. Consider a quantum many-body system consisting of $N$ qubits that evolves by a chaotic Hamiltonian, $H$. Let $W$ and $V$ be local, unitary operators on Hilbert space $\mathcal{H}_d$ of dimension $d=2^N$. `Local' refers to unitaries which act on different qubits. Assume $W$ and $V$ commute at time $t=0$. The Heisenberg operator 
\begin{equation}
    W(t)\equiv U_H^\dagger(t)WU_H(t)
\end{equation}
 evolves with $U_H(t)=e^{-iHt}$ ($\hbar=1$). The quantum butterfly effect \cite{Maldacena_2016,PhysRevLett.89.170405} states that $[W(t),V]$ grows as $W(t)$ delocalizes, i.e. spreads throughout the system. Intuitively, $W(t)$ eventually acts non-trivially at the location of $V$, at which point the commutator no longer vanishes. The size of the commutator can therefore be used to quantify the spreading of $W(t)$ and hence measure scrambling.

To measure the growth of the commutator, it is common to take its norm \cite{lieb1972}. The Hilbert–Schmidt norm is most often employed \cite{Swingle_2016,PhysRevB.95.060201,PhysRevD.96.065005},
\begin{equation}
||[W(t),V]||_{\mathrm{HS}}= [\tr\{|[W(t),V]|^2\}]^{1/2}.
\end{equation}
To measure this quantity, we interpret the trace as (up to a normalization) the expectation value of $|[W(t),V]|^2$ over the infinite temperature thermal state $\rho_\infty=\frac{1}{d}I_N$. By adopting the notation $\langle \bm{\cdot} \rangle \equiv \tr\{\rho_\infty \bm{\cdot}\}$, the expectation value is
\begin{equation}
\langle|[W(t),V]|^2\rangle=2(1-Re\{C_4(t)\}),
\end{equation}
where the four-point out-of-time-ordered correlator is defined as
\begin{equation}
C_4(t)=\langle W^\dagger(t)V^\dagger W(t)V\rangle.
\end{equation}
Scrambling causes this correlator to decay to near zero. Although $C_4(t)$ has been analyzed extensively in the literature, it does not describe the complete evolution of the commutator. Higher-point correlators are necessary to reveal new, early-time scrambling.

To extract higher-point correlators, we measure the commutator growth using the Schatten $2n$-norm for positive integer $n$ \cite{bhattacharya2018chaotic},
\begin{equation}
||[W(t),V]||_{2n}= [\tr\{|[W(t),V]|^{2n}\}]^{1/2n}.
\end{equation}
This is computed by measuring $\langle |[W(t),V]|^{2n}\rangle=\frac{1}{d}\tr\{|[W(t),V]|^{2n}\}$. By expanding out $|[W(t),V]|^{2n}$, the Schatten $2n$-norm can be expressed in terms of a linear combination of higher-point correlators
\begin{equation}
    ||[W(t),V]||_{2n}=\Bigg[d\sum_{k=0}^n b_{k}(n)Re\{C_{4k}(t)\}\Bigg]^{1/2n},
\end{equation}
for some coefficients $b_{k}(n)$. The $4k$-point OTOC, for a non-negative integer $k$, is defined as
\begin{equation}\label{C4k}
C_{4k}(t)\equiv\langle (W^\dagger(t) V^\dagger W(t)V)^k \rangle.
\end{equation}
We propose a physical interpretation of these correlators in Appendix~\ref{sec:interpret}. 

In this work, we develop measurement protocols to estimate $C_{4k}(t)$ by using randomized measurements \cite{van2012Measuring,Elben2019toolbox} via the classical shadow framework \cite{Huang_2020}. The applications of randomized measurements range from quantum many-body physics, such as the detection of topological order \cite{Elben2020topological,Cian2020Chern} and entanglement entropy \cite{Elben2018Random,Brydges2019Probing}, to quantum information and quantum foundations, such as the extraction of entanglement negativity \cite{elben2020mixedstate,Zhou_2020}, quantum benchmarking \cite{Elben2020Cross,zhang2020experimental}, and entanglement detection without reference frames \cite{Tran2015entanglement,Tran2016Correlations,Ketterer2019Multipartite,Ketterer2020entanglement,Knips2020}.

The paper is organized as follows. In Sec.~\ref{sec:global}, we adopt a global random ensemble to measure the eight-point correlator $C_8(t)$, which reveals scrambling earlier than $C_4(t)$. In Sec.~\ref{sec:shadow}, we propose three complementary protocols based on shadow tomography, only requiring random Pauli measurements on individual qubits. We give analytical variance upper bounds for a sample protocol in Sec.~\ref{sec:var} and show the protocol is more efficient than direct tomography. In Sec.~\ref{sec:numeric} we present numerical simulations for the predicted and estimated OTOC values. We give conclusions and provide an outlook in Sec.~\ref{sec:conclude}.

\section{Global protocol for eight-point OTOC}\label{sec:global}
We motivate our search for a novel protocol to measure higher-point correlators by first extending the four-point OTOC protocol developed in \cite{Vermersch_2019} to evaluate $C_8(t)$. The protocol implements global random unitaries and demonstrates new scrambling dynamics in $C_8(t)$. 

Assume $W$ and $V$ are unitary, traceless, and Hermitian operators on $\mathcal{H}_d$. Defining operators $A_1=W(t)$ and $A_2=V W(t)V$, the four-point and eight-point OTOCs are
\begin{equation}
    C_{4}(t)=\langle A_1 A_2\rangle, \quad C_{8}(t)=\langle A_1 A_2 A_1 A_2\rangle.
\end{equation}
$A_1$ and $A_2$ are also unitary, traceless, and Hermitian operators. Let $U$ be a unitary on $\mathcal{H}_d$ randomly sampled from the Haar measure on the unitary group. Define the notation for the integral over the Haar measure as $\overline{(\cdots)}=\int_{\Hr}dU(\cdots)$. Define the expectation value over the pure state $\rho_0\in\mathcal{H}_d$ as $\langle \bm{\cdot}\rangle_{\rho_0}=\tr\{\rho_0 \bm{\cdot}\}$.
One can prove that (see Appendix~\ref{sec:HaarProof})
\begin{equation}
    \langle A_1 A_2 \rangle=(d+1)\overline{\langle U^\dagger A_1 U\rangle_{\rho_0}\langle U^\dagger A_2 U\rangle_{\rho_0}},
\end{equation}
\begin{equation}
\langle A_1A_2A_1A_2\rangle=\frac{1}{2}(d+1)(d+2)(d+3)\overline{\langle U^\dagger A_1 U\rangle_{\rho_0}^2\langle U^\dagger A_2 U \rangle_{\rho_0}^2}
-d\langle A_1 A_2\rangle^2
-\frac{1}{2}(d+4),
\end{equation}
for any pure state $\rho_0$, even, for example, $\ket{0}^{\otimes N}$.
In terms of $W(t)$ and $V$, the OTOCs are
\begin{equation}
C_4(t)=(d+1)\overline{\langle U^\dagger W(t) U \rangle_{\rho_0}\langle U^\dagger V^\dagger W(t)V  U \rangle_{\rho_0}},
\end{equation}
\begin{equation}\label{eq:globalc8}
C_8(t)=\frac{1}{2}(d+1)(d+2)(d+3)\overline{\langle U^\dagger W(t) U\rangle_{\rho_0}^2\langle U^\dagger V^\dagger W(t)V U \rangle_{\rho_0}^2}
-dC^2_4(t)
-\frac{1}{2}(d+4).
\end{equation}
Eq.~\eqref{eq:globalc8} reveals $C_8(t)$ contains dynamics not captured by $C_4(t)$. Its third term is constant and does not affect the dynamics. The second term depends only on $C_4(t)$. The `hidden' dynamics arise due to the new term $\overline{\langle U^\dagger W(t) U\rangle_{\rho_0}^2\langle U^\dagger V^\dagger W(t)V U \rangle_{\rho_0}^2}$.

To evaluate the OTOCs, we measure $\langle U^\dagger W(t)U\rangle_{\rho_0}$ and $\langle U^\dagger V^\dagger W(t)V U\rangle_{\rho_0}$ with global random unitaries as follows.
\begin{enumerate}
\begin{samepage}
    \item Randomly sample a unitary $U$ from the Haar measure on the unitary group.  
    \item Prepare pure state $\rho_0$ and evolve with $U$. Then evolve with $U_H(t)$. Measure $W$.
    \item Repeat step 2 many times to compute the expectation value $\langle U^\dagger W(t)U\rangle_{\rho_0}$.
    \item Prepare $\rho_0$ and evolve with $U$. Apply $V$, then evolve with $U_H(t)$. Measure $W$. 
    \item Repeat step 4 many times to compute the expectation value $\langle U^\dagger V^\dagger W(t)V U\rangle_{\rho_0}$.
    \item Repeat steps 1-5 with many random unitaries.
\end{samepage}
\end{enumerate}
We compute $\overline{\langle U^\dagger W(t) U \rangle_{\rho_0}^2\langle U^\dagger V^\dagger W(t)V  U \rangle_{\rho_0}^2}$ by calculating $\langle U^\dagger W(t) U \rangle_{\rho_0}^2\langle U^\dagger V^\dagger W(t)V  U \rangle_{\rho_0}^2$ for each $U$, then averaging.

Although a sub-ensemble can be substituted in for the Haar measure on the unitary group via unitary t-design \cite{Gross2007review,Roberts_2017}, an ensemble forming a 4-design is needed to measure $C_8(t)$. That is, one must apply a more random (chaotic) unitary ensemble to access higher-point scrambling features. Note that the Clifford group forms a 3-design \cite{Webb2015design,Zhu_2016,Kueng2015proj}, but not a 4-design \cite{Zhu_2016}. One can generate an approximate t-design through a random local circuit \cite{Brando2016}, in particular, by inserting few $T$ gates into Clifford circuits \cite{Haferkamp2020homeopathy}. Since generating global random unitaries is experimentally challenging, Ref.~\cite{Vermersch_2019} adapts its global protocol to local unitaries, greatly simplifying the measurement procedure. However, higher-point correlators are difficult to evaluate using this local protocol, due to the lack of degrees of freedom needed to construct larger permutation operators. This kind of no-go result is also observed in the measurement of the higher moments of the density matrix \cite{Zhou_2020}. 

The global protocol serves as a proof of principle that higher-point OTOCs contain new scrambling dynamics not present in $C_4(t)$. This motivates the development of protocols based on classical shadows  \cite{Huang_2020} in the following section, which only utilize single-qubit, random Clifford unitaries. 

\section{Classical Shadow protocols}\label{sec:shadow}
We present three protocols to estimate $C_{4k}(t)$ using classical shadows generated through quantum shadow tomography \cite{aaronson2018shadow,Huang_2020}. We summarize the essentials of shadow tomography, but refer to \cite{Huang_2020} for further details. 

\subsection{Shadow Tomography}
The classical shadow protocols rely on the prediction of functions of an $N$-qubit state $\rho$. First, define a random unitary as a tensor product of local unitaries,
\begin{equation}
    U=\bigotimes_{i=1}^N u_i.
\end{equation}
Each single-qubit unitary $u_i$ is drawn randomly and independently from the Clifford group. Evolve $\rho$ to $U\rho U^\dagger$. Upon measurement in the computational basis, the state collapses to $\ket{\hat{b}}\bra{\hat{b}}$ by Born's rule, where $\ket{\hat{b}}\in\{0,1\}^{\otimes N}$ is an $N$-bit random variable. That is, the state is randomly measured in the Pauli basis, similar to traditional tomography. In shadow tomography, however, one is interested in estimating the properties of the state, not in reconstructing it. After measurement, the outcome is stored and processed classically. Then, we classically compute $U^\dagger\ket{\hat{b}}\bra{\hat{b}}U$ and apply the inverted channel $\mathcal{M}^{-1}=\bigotimes_{i=1}^N \mathcal{M}^{-1}_i$, where $\mathcal{M}^{-1}_i(\bm{\cdot})= 3(\bm{\cdot})-\tr(\cdot)I$ and $I$ is the identity on a single qubit. The result is a classical snapshot of $\rho$,

\begin{equation}
\hat{\rho}(U;\hat{b})=\mathcal{M}^{-1}(U^\dagger\ket{\hat{b}}\bra{\hat{b}}U).
\end{equation}

The classical snapshot satisfies $\mathbb{E}(\hat{\rho}(U;\hat{b}))=\rho$, where $\mathbb{E}$ is the average of the outcomes and of the unitaries over the local Clifford group. More formally, 
\begin{equation}
    \mathbb{E}(\hat{\rho}(U;\hat{b}))
    =\mathbb{E}_{U\sim\mathrm{Cl(2)}^{\otimes N}} \sum_{b\in\{0,1\}^{N}}\bra{b}U\rho U^\dagger\ket{b} \mathcal{M}^{-1}(U^\dagger\ket{b}\bra{b}U).
\end{equation}
As a result, an observable $O$ can be estimated using the snapshot through $\tr\{O\hat{\rho}\}$ with $\mathbb{E}(\tr\{O\hat{\rho}\})=\tr\{O\mathbb{E}(\hat{\rho})\}=\tr\{O\rho\}$. To reduce the statistical error of the estimation, one can repeat this process to generate $K$ independent classical snapshots. The \emph{classical shadow} of $\rho$ is defined as the set of these snapshots
\begin{equation}
    S(\rho,K)=\Big\{\hat{\rho}^{(i)}(U_i;\hat{b}_i)\Big\}_{i=1}^K.
\end{equation}
Each snapshot corresponds to a new measurement. $K$ is referred to as the size of the shadow. The shadow can also be used to estimate nonlinear functions of $\rho$, which are encountered in our measurement protocols. For example, a second-order function $f_2$ can be written as $f_2(\rho)=\tr\{O'\rho^{\otimes 2}\}$ for some observable $O'$ on $\mathcal{H}_d^{\otimes 2}$. This can be estimated using two distinct snapshots: $\tr\{O'\hat{\rho}\otimes\hat{\rho'}\}$.

\subsection{Multi-Bell state protocol}
We construct a protocol to measure $C_{4k}(t)$ by preparing multiple Bell states. Since $C_{4k}(t)$ is a function of the evolution unitary $U_H(t)$, we introduce Bell states to `store' $U_H(t)$ \cite{Hosur_2016,Jaffe_2018}. Consider a $2N$-qubit system. Let $\rho_{\mr{Bell}}=\ket{\Phi}\bra{\Phi}$ be the maximally entangled state, where
\begin{equation}
\ket{\Phi}=\frac{1}{d^{1/2}}\sum_{i=1}^{d}\ket{i} \otimes \ket{i}
\end{equation}
and $\ket{i}\in \mathcal{H}_d$. State $\ket{\Phi}$ consists of $N$ Bell states. To simplify computations, we introduce a graphical calculus \cite{Elben_2019} and refer to Appendix~\ref{sec:diagrams} for its complete description. With some stylistic adaptation from Ref.~\cite{Collins_2010}, $\rho_{\mr{Bell}}$ is expressed as 

\begin{equation}
\begin{tikzpicture}
	\node[] (v0) at (0,0) {$\rho_{\mr{Bell}}=\frac{1}{d}\cdot$};
	\draw [thick,color=dullblue] 
    (1,.5) -- (1.5,.5) -- (1.5,-.5) -- (1,-.5)
    (2.25,.5)--(1.75,.5)--(1.75,-.5)--(2.25,-.5);
\end{tikzpicture}.
\end{equation}
Evolving $\rho_{\mr{Bell}}$ with the unitary channel associated with $U_H\otimes I_N$, where $I_N$ is the identity on $N$ qubits, the resulting state and its diagram are, respectively,
\begin{equation}
\rho_H=(U_H\otimes I_N)\rho_{\mr{Bell}}(U^\dagger_H\otimes I_N),
\end{equation}
\begin{equation}\label{rhoH}
\begin{tikzpicture}
    \draw [thick,color=dullblue] (-1,.5)--(1,.5);
	\draw [thick,color=dullblue] (-1,-.5)--(1,-.5);
	\node[rectangle, minimum width=2em, minimum height =4.5em, fill=egg, rounded corners, draw] (v0) at (0,0) {$\rho_H$};

	\node[] (v1) at (1.75,0) {$=\frac{1}{d}\cdot$};
	\node[rectangle, minimum width=2em, minimum height =2em, fill=egg, rounded corners, draw] (v2) at (3.5,.5) {$U_H$};
	\node[rectangle, minimum width=2em, minimum height =2em, fill=egg, rounded corners, draw] (v3) at (5.75,.5) {$U^\dagger_H$};
	\draw [thick,color=dullblue] (2.5,.5)--(v2);
	\draw [thick,color=dullblue] 
    (v2) -- (4.5,.5) -- (4.5,-.5) -- (2.5,-.5)
    (6.75,.5)--(v3)--(4.75,.5)--(4.75,-.5)--(6.75,-.5);
\end{tikzpicture}.
\end{equation}
The time dependence of $\rho_H(t)$ is suppressed for conciseness. State $\rho_H$ is actually the channel-state duality for $U_H(t)$. 

Now write $C_{4k}(t)$ in terms of $\rho_H$. First, express the correlator diagrammatically,
\begin{equation}
C_{4k}(t)=\frac{1}{d}\tr\{(U_H^\dagger W^\dagger U_H V^\dagger U_H^\dagger W U_H V)^k\},
\end{equation}

\begin{equation}
\scalebox{.9}{
\begin{tikzpicture}

\node[] (c0) at (-4.2,-4) {\large $C_{4k}(t)=\frac{1}{d}\cdot$};

\node[rectangle, fill=egg, rounded corners, minimum width=2em, minimum height=2em, draw] (c1v1) at (0,0) {$U_H^\dagger$};
\node[rectangle, fill=egg, rounded corners, minimum width=2em, minimum height=2em, draw] (c1v2) at (0,-2){$U_H^\dagger$};

\draw [line width=0.4mm, dotted] (0,-3.75) -- (0,-4.3);

\node[rectangle, fill=egg, rounded corners, minimum width=2em, minimum height=2em, draw] (c1v3) at (0,-5) {$U_H^\dagger$};
\node[rectangle, fill=egg, rounded corners, minimum width=2em, minimum height=2em, draw] (c1v4) at (0,-7) {$U_H^\dagger$};

\node[rectangle, fill=egg, rounded corners, minimum width=2em, minimum height=2em, draw] (c2v1) at (1.5,0) {$W^\dagger$};
\node[rectangle, fill=egg, rounded corners, minimum width=2em, minimum height=2em, draw] (c2v2) at (1.5,-2){$W$};
\node[rectangle, fill=egg, rounded corners, minimum width=2em, minimum height=2em, draw] (c2v3) at (1.5,-5){$W^\dagger$};
\node[rectangle, fill=egg, rounded corners, minimum width=2em, minimum height=2em, draw] (c2v4) at (1.5,-7){$W$};

\node[rectangle, fill=egg, rounded corners, minimum width=2em, minimum height=2em, draw] (c3v1) at (3,0) {$U_H$};
\node[rectangle, fill=egg, rounded corners, minimum width=2em, minimum height=2em, draw] (c3v2) at (3,-2){$U_H$};
\node[rectangle, fill=egg, rounded corners, minimum width=2em, minimum height=2em, draw] (c3v3) at (3,-5){$U_H$};
\node[rectangle, fill=egg, rounded corners, minimum width=2em, minimum height=2em, draw] (c3v4) at (3,-7){$U_H$};

\node[rectangle, fill=egg, rounded corners, minimum width=2em, minimum height=2em, draw] (c4v1) at (4.5,0) {$V^\dagger$};
\node[rectangle, fill=egg, rounded corners, minimum width=2em, minimum height=2em, draw] (c4v2) at (4.5,-2){$V$};
\node[rectangle, fill=egg, rounded corners, minimum width=2em, minimum height=2em, draw] (c4v3) at (4.5,-5){$V^\dagger$};
\node[rectangle, fill=egg, rounded corners, minimum width=2em, minimum height=2em, draw] (c4v4) at (4.5,-7){$V$};

\draw [thick,color=dullblue] (-1,0)--(c1v1)--(c2v1)--(c3v1)--(c4v1)--(5.5,0) (6,0)--(6.5,0);
\draw [thick,color=dullblue] (-1,-1)--(5.5,-1) (6,-1)--(6.5,-1);
\draw [thick,color=dullblue] (-1,-2)--(c1v2)--(c2v2)--(c3v2)--(c4v2)--(5.5,-2) (6,-2)--(6.5,-2);
\draw [thick,color=dullblue] (-1,-3)--(5.5,-3) (6,-3)--(6.5,-3);
\draw [thick,color=dullblue](-1,-5)--(c1v3)--(c2v3)--(c3v3)--(c4v3)--(5.5,-5) (6,-5)--(6.5,-5);
\draw [thick,color=dullblue] (-1,-6)--(5.5,-6) (6,-6)--(6.5,-6);
\draw [thick,color=dullblue] (-1,-7)--(c1v4)--(c2v4)--(c3v4)--(c4v4)--(5.5,-7) (6,-7)--(6.5,-7);
\draw [thick,color=dullblue] (-1,-8)--(5.5,-8) (6,-8)--(6.5,-8);

\draw [thick,color=dullblue] (5.5,0)--(5.5,-1) (6,0)--(6,-1);
\draw [thick,color=dullblue] (5.5,-2)--(5.5,-3) (6,-2)--(6,-3);
\draw [thick,color=dullblue] (5.5,-5)--(5.5,-6) (6,-5)--(6,-6);
\draw [thick,color=dullblue] (5.5,-7)--(5.5,-8) (6,-7)--(6,-8);

\draw [thick,color=dullblue] (-1,0)--(-2,0)--(-3,0);
\draw [thick,color=dullblue] (-1,-1)--(-2,-3)--(-3,-3);
\draw [thick,color=dullblue] (-1,-2)--(-2,-2)--(-3,-2);
\draw [thick,color=dullblue] (-1,-3)--(-1.5,-4);
\draw [thick,color=dullblue] (-1,-5)--(-2,-5)--(-3,-5);
\draw [thick,color=dullblue] (-1.2,-4.5)--(-2,-6)--(-3,-6);
\draw [thick,color=dullblue] (-1,-6)--(-2,-8)--(-3,-8);
\draw [thick,color=dullblue] (-1,-7)--(-2,-7)--(-3,-7);
\draw [thick,color=dullblue] (-1,-8)--(-2,-1)--(-3,-1);

\end{tikzpicture}.
}
\end{equation}
`Slide' $U_H^\dagger$ leftwards and use the implied periodic boundary conditions. Introduce the following identity relating $A_1$ to its transpose,

\begin{equation}
\begin{tikzpicture}
	\node[rectangle, fill=egg, rounded corners, minimum width=1em, minimum size=2em, draw] (v0) at (.75,.5) {$A_1$};
	\draw [thick,color=dullblue] 
    (0,.5)--(v0) -- (1.5,.5) -- (1.5,-.5) -- (0,-.5);
    \node[] (v1) at (2,0) {$=$};
    \node[rectangle, fill=egg, rounded corners, minimum width=1em, minimum size=2em, draw] (v2) at (3.25,-.5) {$A_1^T$};
	\draw [thick,color=dullblue] 
    (2.5,.5) --(4,.5) -- (4,-.5) -- (v2)--(2.5,-.5);
\end{tikzpicture},
\end{equation}
for $A_1=V,V^\dagger$. The correlator is now drawn as

\begin{equation}\label{eq:C4kDiagram}
\scalebox{.9}{
\begin{tikzpicture}

\node[] (c0) at (-4.2,-4) {\large $C_{4k}(t)=\frac{1}{d}\cdot$};

\node[rectangle, fill=egg, rounded corners, minimum width=2em, minimum height=2em, draw] (c1v1) at (0,0) {$W^\dagger$};
\node[rectangle, fill=egg, rounded corners, minimum width=2em, minimum height=2em, draw] (c1v2) at (0,-1){$(V^\dagger)^T$};
\node[rectangle, fill=egg, rounded corners, minimum width=2em, minimum height=2em, draw] (c1v3) at (0,-2){$W$};
\node[rectangle, fill=egg, rounded corners, minimum width=2em, minimum height=2em, draw] (c1v4) at (0,-3){$V^T$};

\draw [line width=0.4mm, dotted] (0,-3.75) -- (0,-4.3);

\node[rectangle, fill=egg, rounded corners, minimum width=2em, minimum height=2em, draw] (c1v5) at (0,-5) {$W^\dagger$};
\node[rectangle, fill=egg, rounded corners, minimum width=2em, minimum height=2em, draw] (c1v6) at (0,-6){$(V^\dagger)^T$};
\node[rectangle, fill=egg, rounded corners, minimum width=2em, minimum height=2em, draw] (c1v7) at (0,-7) {$W$};
\node[rectangle, fill=egg, rounded corners, minimum width=2em, minimum height=2em, draw] (c1v8) at (0,-8){$V^T$};

\node[rectangle, fill=egg, rounded corners, minimum width=2em, minimum height=2em, draw] (c2v1) at (1.5,0) {$U_H$};
\node[rectangle, fill=egg, rounded corners, minimum width=2em, minimum height=2em, draw] (c2v2) at (1.5,-2){$U_H$};
\node[rectangle, fill=egg, rounded corners, minimum width=2em, minimum height=2em, draw] (c2v3) at (1.5,-5){$U_H$};
\node[rectangle, fill=egg, rounded corners, minimum width=2em, minimum height=2em, draw] (c2v4) at (1.5,-7){$U_H$};

\node[rectangle, fill=egg, rounded corners, minimum width=2em, minimum height=2em, draw] (c3v1) at (4,0) {$U_H^\dagger$};
\node[rectangle, fill=egg, rounded corners, minimum width=2em, minimum height=2em, draw] (c3v2) at (4,-2){$U_H^\dagger$};
\node[rectangle, fill=egg, rounded corners, minimum width=2em, minimum height=2em, draw] (c3v3) at (4,-5){$U_H^\dagger$};
\node[rectangle, fill=egg, rounded corners, minimum width=2em, minimum height=2em, draw] (c3v4) at (4,-7){$U_H^\dagger$};

\draw [thick,color=dullblue] (-1,0)--(c1v1)--(c2v1)--(2.5,0) (3,0)--(c3v1)--(5,0);
\draw [thick,color=dullblue] (-1,-1)--(c1v2)--(2.5,-1) (3,-1)--(5,-1);
\draw [thick,color=dullblue] (-1,-2)--(c1v3)--(c2v2)--(2.5,-2) (3,-2)--(c3v2)--(5,-2);
\draw [thick,color=dullblue] (-1,-3)--(c1v4)--(2.5,-3) (3,-3)--(5,-3);
\draw [thick,color=dullblue] (-1,-5)--(c1v5)--(c2v3)--(2.5,-5) (3,-5)--(c3v3)--(5,-5);
\draw [thick,color=dullblue] (-1,-6)--(c1v6)--(2.5,-6) (3,-6)--(5,-6);
\draw [thick,color=dullblue] (-1,-7)--(c1v7)--(c2v4)--(2.5,-7) (3,-7)--(c3v4)--(5,-7);
\draw [thick,color=dullblue] (-1,-8)--(c1v8)--(2.5,-8) (3,-8)--(5,-8);

\draw [thick,color=dullblue] (2.5,0)--(2.5,-1) (3,0)--(3,-1);
\draw [thick,color=dullblue] (2.5,-2)--(2.5,-3) (3,-2)--(3,-3);
\draw [thick,color=dullblue] (2.5,-5)--(2.5,-6) (3,-5)--(3,-6);
\draw [thick,color=dullblue] (2.5,-7)--(2.5,-8) (3,-7)--(3,-8);

\draw [thick,color=dullblue] (-1,0)--(-2,0)--(-3,0);
\draw [thick,color=dullblue] (-1,-1)--(-2,-3)--(-3,-3);
\draw [thick,color=dullblue] (-1,-2)--(-2,-2)--(-3,-2);
\draw [thick,color=dullblue] (-1,-3)--(-1.5,-4);
\draw [thick,color=dullblue] (-1,-5)--(-2,-5)--(-3,-5);
\draw [thick,color=dullblue] (-1.2,-4.5)--(-2,-6)--(-3,-6);
\draw [thick,color=dullblue] (-1,-6)--(-2,-8)--(-3,-8);
\draw [thick,color=dullblue] (-1,-7)--(-2,-7)--(-3,-7);
\draw [thick,color=dullblue] (-1,-8)--(-2,-1)--(-3,-1);

\draw [dashed,line width=0.25mm] (-2.3,.5)--(-.8,.5);
\draw [dashed,line width=0.25mm] (-2.3,.-8.5)--(-.8,-8.5);
\draw [dashed,line width=0.25mm] (-2.3,.5)--(-2.3,-8.5);
\draw [dashed,line width=0.25mm] (-.8,.5)--(-.8,-8.5);

\end{tikzpicture},
}
\end{equation}

\begin{equation}
\scalebox{.9}{
\begin{tikzpicture}
\node[] (c0) at (-6,-4) {\large $C_{4k}(t)=d^{2k-1}\cdot$};

\node[rectangle, fill=egg, rounded corners, minimum width=2em, minimum height=2em, draw] (c1v1) at (0,0) {$W^\dagger$};
\node[rectangle, fill=egg, rounded corners, minimum width=2em, minimum height=2em, draw] (c1v2) at (0,-1){$V^*$};
\node[rectangle, fill=egg, rounded corners, minimum width=2em, minimum height=2em, draw] (c1v3) at (0,-2){$W$};
\node[rectangle, fill=egg, rounded corners, minimum width=2em, minimum height=2em, draw] (c1v4) at (0,-3){$V^T$};

\draw [line width=0.4mm, dotted] (0,-3.75) -- (0,-4.3);

\node[rectangle, fill=egg, rounded corners, minimum width=2em, minimum height=2em, draw] (c1v5) at (0,-5) {$W^\dagger$};
\node[rectangle, fill=egg, rounded corners, minimum width=2em, minimum height=2em, draw] (c1v6) at (0,-6){$V^*$};
\node[rectangle, fill=egg, rounded corners, minimum width=2em, minimum height=2em, draw] (c1v7) at (0,-7) {$W$};
\node[rectangle, fill=egg, rounded corners, minimum width=2em, minimum height=2em, draw] (c1v8) at (0,-8){$V^T$};

\node[rectangle, fill=egg, rounded corners, minimum width=2em, minimum height=4.5em, draw] (c2v1) at (1.5,-.5) {$\rho_H$};
\node[rectangle, fill=egg, rounded corners, minimum width=2em, minimum height=4.5em, draw] (c2v2) at (1.5,-2.5){$\rho_H$};
\node[rectangle, fill=egg, rounded corners, minimum width=2em, minimum height=4.5em, draw] (c2v3) at (1.5,-5.5){$\rho_H$};
\node[rectangle, fill=egg, rounded corners, minimum width=2em, minimum height=4.5em, draw] (c2v4) at (1.5,-7.5){$\rho_H$};

\draw [thick,color=dullblue] (-4.5,0)--(-3.35,0) (-1.15,0)--(c1v1)--(1.11,0) (2-.11,0)--(3,0);
\draw [thick,color=dullblue] (-4.5,-1)--(-3.35,-1) (-1.15,-1)--(c1v2)--(1.11,-1) (2-.11,-1)--(3,-1);
\draw [thick,color=dullblue] (-4.5,-2)--(-3.35,-2) (-1.15,-2)--(c1v3)--(1.11,-2) (2-.11,-2)--(3,-2);
\draw [thick,color=dullblue] (-4.5,-3)--(-3.35,-3) (-1.15,-3)--(c1v4)--(1.11,-3) (2-.11,-3)--(3,-3);
\draw [thick,color=dullblue] (-4.5,-5)--(-3.35,-5) (-1.15,-5)--(c1v5)--(1.11,-5) (2-.11,-5)--(3,-5);
\draw [thick,color=dullblue] (-4.5,-6)--(-3.35,-6) (-1.15,-6)--(c1v6)--(1.11,-6) (2-.11,-6)--(3,-6);
\draw [thick,color=dullblue] (-4.5,-7)--(-3.35,-7) (-1.15,-7)--(c1v7)--(1.11,-7) (2-.11,-7)--(3,-7);
\draw [thick,color=dullblue] (-4.5,-8)--(-3.35,-8) (-1.15,-8)--(c1v8)--(1.11,-8) (2-.11,-8)--(3,-8);

\node[rectangle, fill=egg, rounded corners, minimum width=6.2em, minimum height=25em, draw] (vw) at (-2.25,-4) {$T_{(4k,\ldots,4,2)}$};

\end{tikzpicture}.
}
\end{equation}
The dashed rectangle in Eq.~\eqref{eq:C4kDiagram} is the cyclic permutation operator over even indexes, $T_{(4k,\ldots,4,2)}$. The correlator is now 
\begin{equation}\label{Eq:Bell4k}
C_{4k}(t)=d^{2k-1}\tr\{O_{4k}\rho_H^{\otimes 2k}\},
\end{equation}
where $O_{4k}=T_{(4k,\ldots,4,2)}(W^\dagger \otimes V^* \otimes W \otimes V^T)^{\otimes k}$. $O_{4k}$ has no time dependence; all time dependence is stored in $\rho_H$. $C_{4k}(t)$ can therefore be estimated by performing shadow tomography on $\rho_H$.

One can also define a general $4k$-point correlator as an expectation value over an arbitrary state $\rho$,
\begin{equation}
\langle (W^\dagger (t)V^\dagger W(t)V)^k\rangle_{\rho}\equiv \tr\{\rho (W^\dagger (t) V^\dagger W(t)V)^k\}.
\end{equation}
This can also be expressed in terms of $\rho_H$,
\begin{equation}
\langle (W^\dagger (t) V^\dagger W(t)V)^k\rangle_{\rho}=
 d^{2k}\tr\{T_{(4k,\ldots,4,2)}(I^{\otimes (4k-1)}\otimes\rho^T)(W^\dagger\otimes V^* \otimes W \otimes V^T)^{\otimes k} \rho_H^{\otimes 2k}\}.
\end{equation}
The appearance of $\rho^T$ may seem alarming, since the transpose is not a physical operation. However, since only $\rho_H$ is measured, $\rho^T$ is treated as an ordinary operator. This correlator enables an analysis of scrambling beyond the high temperature thermal state. Although our protocol can measure this general correlator, we focus on the maximally mixed state.

We use shadow tomography to construct an estimator for $C_{4k}(t)$. First, construct a classical shadow of size $K\geq 2k$ for state $\rho_H$,
\begin{equation}
S_H(\rho_H,K)=\Big\{\hat{\rho}_{H}^{(i)}(U_i;\hat{b}_i)\Big\}_{i=1}^{K}.    
\end{equation}
We suppress the arguments of each snapshot, $\hat{\rho}_{H}^{(i)}=\hat{\rho}_{H}^{(i)}(U_i;\hat{b}_i)$. Using the classical shadow, construct an unbiased estimator for $C_{4k}(t)$ through the U-statistic \cite{kotz1992breakthroughs,elben2020mixedstate}
\begin{equation}\label{eq:BellEst}
\hat{C}_{4k}(t)=\frac{d^{2k-1}}{(2k)!}
\binom{K}{2k}^{-1}
\sum_{i_1\neq i_2\neq\cdots \neq i_{2k}} \tr\Big\{O_{4k}\hat{\rho}_{H}^{(i_1)}\otimes \cdots  \otimes \hat{\rho}_{H}^{(i_{2k})}\Big\}.
\end{equation}
The sum is carried out over all size $2k$ permutations of classical snapshots in $S_H$. Each term in the sum is a function of $2k$ independent snapshots. Unitary $U_{i_j}$ and outcome $\hat{b}_{i_j}$ of snapshot $\hat{\rho}_{H}^{(i_j)}$ are independent of the unitaries and outcomes of any other snapshot $\hat{\rho}_{H}^{(i_{j'})}$. When averaging $\hat{C}_{4k}(t)$ over all Clifford unitaries and outcomes, each $\hat{\rho}_{H}^{(i_j)}$ is averaged individually
\begin{equation}
\mathbb{E}(\hat{C}_{4k}(t))=\frac{d^{2k-1}}{(2k)!}
\binom{K}{2k}^{-1}
\sum_{i_1\neq i_2\neq \cdots \neq i_{2k}} \tr\Big\{O_{4k}\mathbb{E}\Big(\hat{\rho}_{H}^{(i_1)}\Big)\otimes \cdots  \otimes \mathbb{E}\Big(\hat{\rho}_{H}^{(i_{2k})}\Big)\Big\}.
\end{equation}
Since all snapshots satisfy $\mathbb{E}\Big(\hat{\rho}_{H}^{(i_j)}\Big)=\rho_H$, then $\mathbb{E}(\hat{C}_{4k}(t))=C_{4k}(t)$. Thus, $\hat{C}_{4k}(t)$ is an unbiased estimator of $C_{4k}(t)$.

\begin{figure}[t]
  \centering
  \includegraphics[scale=.3]{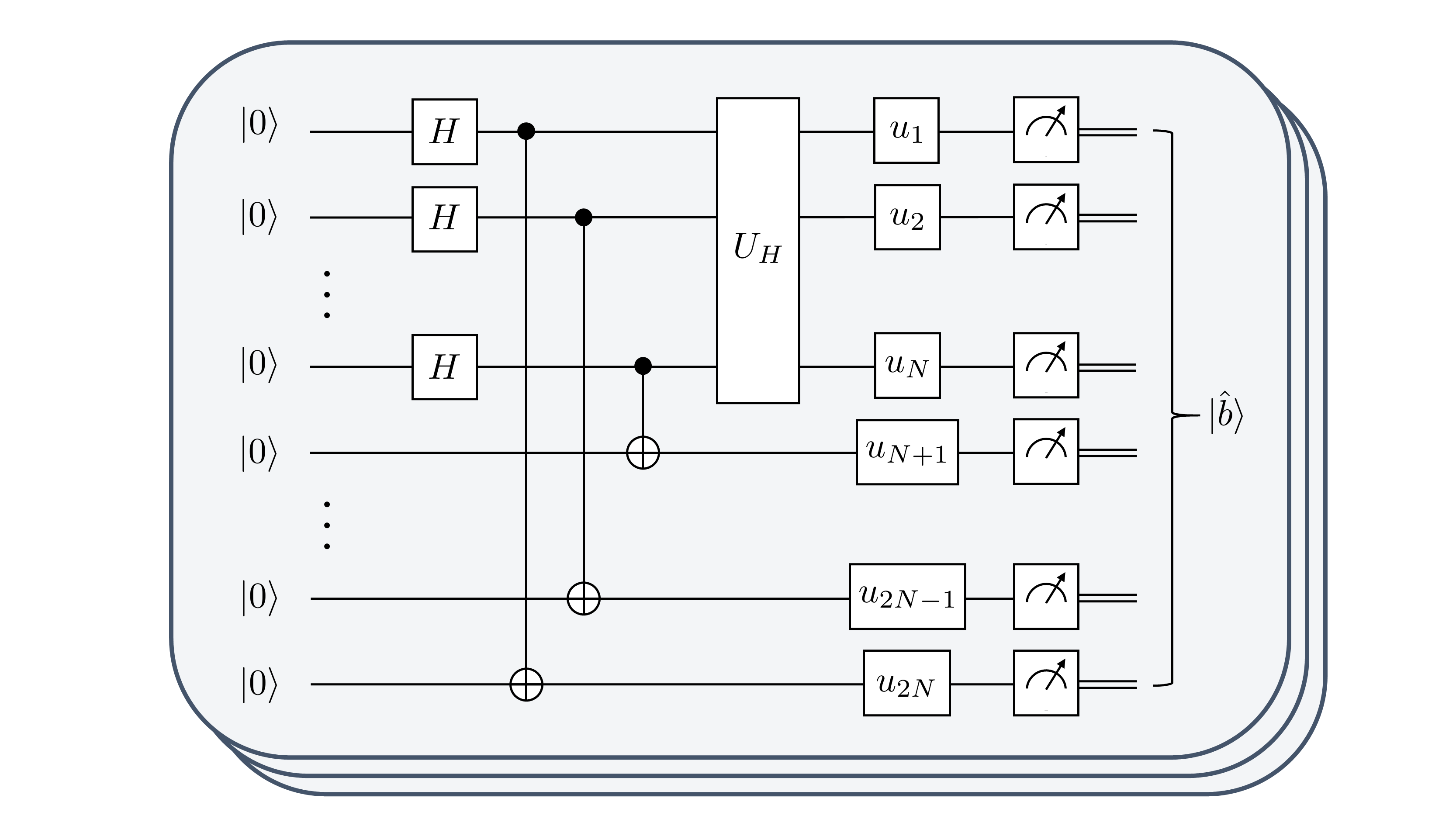}
  \vspace*{-3mm}
  \caption{Quantum circuit to generate a classical snapshot for the multi-Bell state protocol. $N$ Bell states are prepared. The subsystem composed of one qubit from each Bell state is evolved by $U_H(t)$. Each time this circuit runs, $u_i$ is randomly sampled anew from the Clifford group. A classical shadow is generated by running this circuit $K$ times.}
  \label{fig:Bell}
\end{figure}

We summarize the multi-Bell state protocol (see Fig.~\ref{fig:Bell}) to estimate $C_{4k}(t)$:
\begin{enumerate}
\begin{samepage}
    \item Prepare $N$ Bell states.
    \item Evolve the subsystem consisting of one qubit from each Bell state with $U_H(t)$.
    \item Create a classical shadow of size of $K\geq 2k$ for this state.
    \item Use the shadow post-processing in Eq.~\eqref{eq:BellEst} to compute $\hat{C}_{4k}(t)$.
\end{samepage}
\end{enumerate}
The initial state can be readily prepared. For instance, experiments with Rydberg-atom qubits and ultra cold-atoms have demonstrated high-fidelity control of many pairs of Bell states \cite{PhysRevLett.121.123603,Yang2020lattice}. The multi-Bell state protocol carries the advantage that no additional assumptions aside from unitarity and locality are made about $W$ or $V$. Furthermore, single-qubit Clifford unitaries can readily be implemented in experiments. The protocol can also be extended to measure a general correlator for an arbitrary state. Although the OTOC corresponds to an expectation value over an $N$-qubit state, this protocol requires a measurement of $2N$ qubits. For systems limited in size, a protocol requiring a measurement of only $N$ qubits without a preparation of Bell states is favorable. We develop such a protocol in the next section.

\subsection{Mixed state protocol}\label{Direct}
We introduce a protocol to estimate $C_{4k}(t)$ requiring a measurement on only $N$ qubits, by evolving the operator $V$ with some corresponding initial state. Using the cyclic property of the trace, the correlator can be written as
\begin{equation}\label{eq:cor4k}
C_{4k}(t)=\frac{1}{d}\tr\left\{\left(U_H(t)V U_H^\dagger(t) W U_H(t)V U_H^\dagger(t)W\right)^k\right\}.
\end{equation}
Set $W$ and $V$, for example, to
\begin{equation}
W =Z\otimes I^{\otimes N-1}, \quad
V=I^{\otimes N-1}\otimes Z.
\end{equation}
Writing $Z=2\ket{0}\bra{0}-I$ and introducing the initial state 
\begin{equation}
\rho_{in}=\frac{1}{2^{N-1}}I^{\otimes N-1}\otimes \ket{0}\bra{0},
\end{equation}
such that $V=d\rho_{in}-I^{\otimes N}$. Defining the time-evolved state
\begin{equation}
\rho_{V}=U_H(t)\rho_{in} U_H^\dagger(t)
\end{equation}
and using the expression for $V$, the correlator becomes
\begin{equation}\label{eq:MixedProtocolCorr}
C_{4k}(t)=\frac{1}{d}\tr\left\{\left(d^2\rho_{V} W \rho_{V}W-d\rho_V-dW\rho_VW+I^{\otimes N}\right)^k\right\}.
\end{equation}

Shadow tomography can be used to construct an estimator for $C_{4k}(t)$ for any $k$. For demonstration, we estimate the four-point and eight-point OTOCs. Evaluating Eq.~\eqref{eq:MixedProtocolCorr} at $k=1$, 
\begin{align}\label{DirectC4}
\begin{split}
C_{4}(t)&=d\ \tr\{\rho_V W\rho_V W\}-1\\
&=d\ \tr\{\rho_V^{\otimes 2}W^{\otimes 2}T_{(1,2)}\}-1.
\end{split}
\end{align}
By creating a shadow of size $K\geq 2$ for state $\rho_V$,
\begin{equation}
    S_V(\rho_V,K)=\Big\{\hat{\rho}_{V}^{(i)}(U_i;\hat{b}_i)\Big\}_{i=1}^{K},  
\end{equation}
an unbiased estimator of $C_4(t)$ can be constructed through the U-statistic
\begin{equation}\label{Eq:C4est}
\hat{C}_{4}(t)=\frac{1}{2}
\binom{K}{2}^{-1}
\sum_{ i_1\neq i_2}d\ \tr\Big\{\hat{\rho}_{V}^{(i_1)}\otimes \hat{\rho}_{V}^{(i_2)}W^{\otimes 2}T_{(1,2)}\Big\}-1.
\end{equation}
The sum is taken over all size 2 permutations of snapshots in $S_V$. 

Eq.~\eqref{eq:MixedProtocolCorr} evaluated at $k=2$ yields the eight-point OTOC,
\begin{equation}
C_{8}(t)=d^3 \tr\{\rho_VW\rho_VW\rho_VW\rho_VW\}-4d^2 \tr\{\rho_V W\rho_VW\rho_V\}+4d \ \tr\{\rho_VW\rho_VW\}+1.
\end{equation}
Noting $\rho_V^2=\frac{2}{d}\rho_V$, the correlator simplifies to
\begin{equation}\label{eq:8point}
    C_8(t)=L_8(t)-4C_4(t)-3,
\end{equation}
with the leading-order term
\begin{equation}\label{eq:Hidden}
    L_{8}(t)=d^3\tr\{\rho_VW\rho_VW\rho_VW\rho_VW\}
\end{equation}
which determines the scrambling dynamics not captured by the four-point correlator. Using shadow $S_V$ of size $K\geq 4$, the unbiased estimator for the leading-order term is
\begin{equation}\label{eq:HiddenEst}
    \hat{L}_{8}(t)=\frac{1}{4!}\binom{K}{4}^{-1}
    \sum_{i_1\neq i_2 \neq i_3 \neq i_4}
    d^3\tr\Big\{\hat{\rho}_{V}^{(i_1)}\otimes \hat{\rho}_{V}^{(i_2)}\otimes \hat{\rho}_{V}^{(i_3)}\otimes \hat{\rho}_{V}^{(i_4)}W^{\otimes 4} T_{(1,2,3,4)}\Big\}.
\end{equation}
The sum is over all size 4 permutations of snapshots in $S_V$. Thus, the unbiased estimator for the eight-point correlator is
\begin{equation}\label{DirC8est}
    \hat{C}_8(t)=\hat{L}_8(t)-4\hat{C}_4(t)-3,
\end{equation}
where $\hat{C}_4(t)$ is given by Eq.~\eqref{Eq:C4est}.

\begin{figure}[t]
  \centering
  \includegraphics[scale=.4]{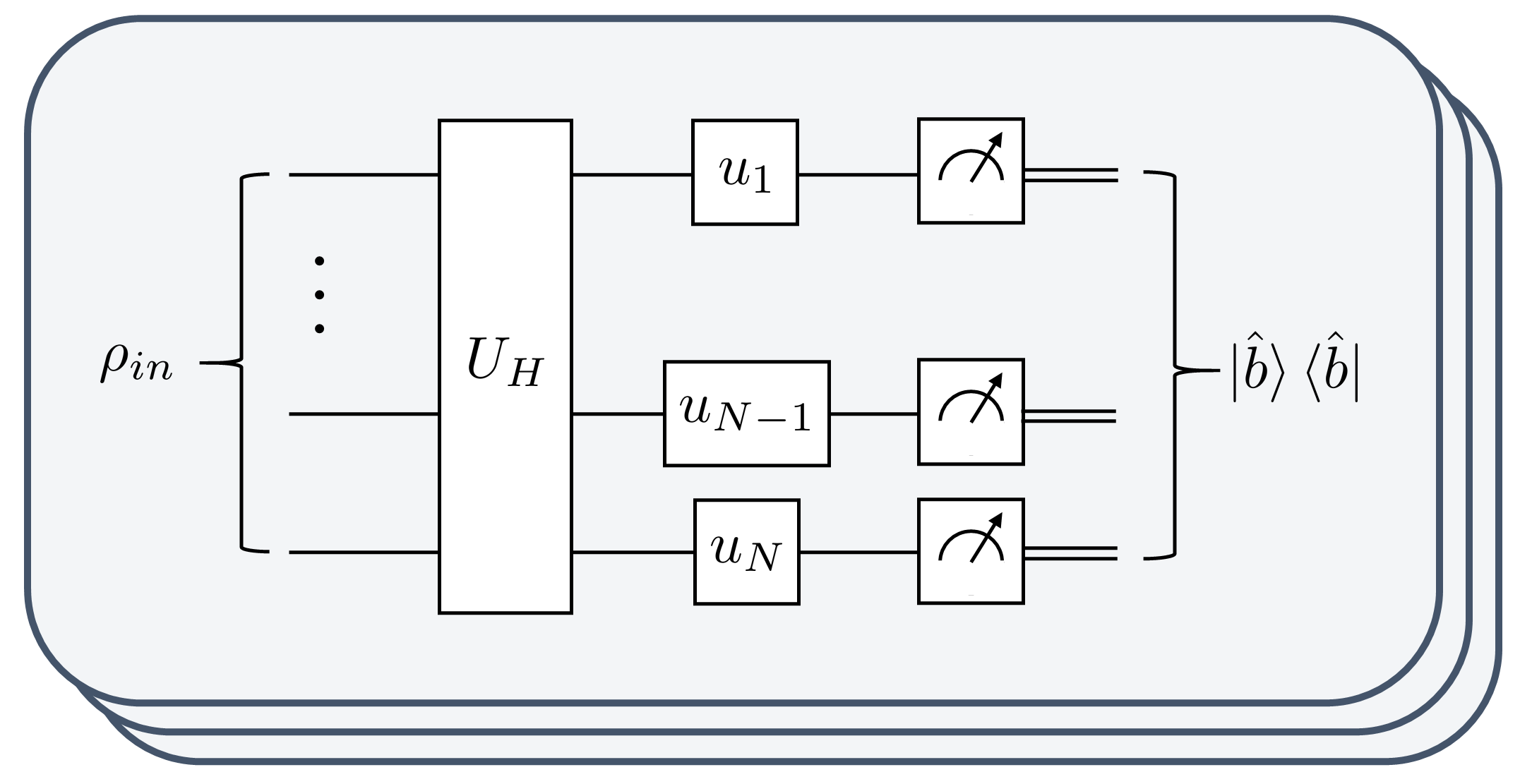}
  \vspace*{-3mm}
  \caption{Quantum circuit which generates a classical snapshot in the mixed state protocol. The initial state $\rho_{in}=\frac{1}{2^{N-1}}I^{\otimes N-1}\otimes \ket{0}\bra{0}$ evolves with $U_H(t)$, then local random Clifford unitaries $u_i$ act on each qubit, followed by computational basis measurement.}
  \label{fig:Pure}
\end{figure}

The mixed state protocol (see Fig.~\ref{fig:Pure}) to estimate $C_{4k}(t)$ is as follows:
\begin{enumerate}
\begin{samepage}
    \item Prepare state $\frac{1}{2^{N-1}}I^{\otimes N-1}\otimes\ket{0}\bra{0}$, where qubit $N$ is in $\ket{0}\bra{0}$ and the remaining qubits are in the maximally mixed state.
    \item Evolve with $U_H(t)$.
     \item Create a shadow of size $K\geq 2k$ for the state.
    \item Use the shadow to compute $\hat{C}_{4k}(t)$.
\end{samepage}
\end{enumerate} 
The advantage of the mixed state protocol over the multi-Bell state protocol is a measurement of only $N$ qubits is required and preparation of EPR pairs is avoided. Initial state $\rho_{in}$ can be prepared on a nuclear magnetic resonance quantum simulator \cite{PhysRevX.7.031011}.

\subsection{Single Bell state protocol}\label{sec:2qubit}
As a hybrid of the two protocols developed previously, we construct a protocol which introduces one Bell state by using just one ancillary qubit. The advantage of this protocol is that the operator $V$ is not predetermined by the initial input state as in Sec.~\ref{Direct}, but is selected in the final classical post-processing stage.

For simplicity, we first compute the estimator for $C_{4}(t)$, then generalize the result to $C_{4k}(t)$. For instance, by taking $W=Z_1$ and $V=X_N$, which both act on $\mathcal{H}_d$,
the four-point correlator in Eq.~\eqref{eq:cor4k} is
\begin{equation}
C_{4}(t)=\frac{1}{d}\tr\{U_H(t)X_N U_H^\dagger(t) Z_1 U_H(t)X_N U_H^\dagger(t)Z_1\}.    
\end{equation}
Introduce a diagram for the correlator where each tensor leg now represents an index for the single-qubit Hilbert space $\mathcal{H}_2$. A slash mark on a leg denotes all remaining qubits. The diagram with periodic boundary conditions shows
\begin{equation}
\begin{tikzpicture}
     \draw [thick,color=dullblue] 
    (-1,1)--(8,1)
    (-1,0)--(8,0)
    (-1,-1)--(8,-1);
    
    \draw[thick,color=dullblue]
    (-.7,-.1)--(-.6,.1)
    (-.7+1.65,-.1)--(-.6+1.65,.1)
    (-.7+3.65,-.1)--(-.6+3.65,.1)
    (-.7+5.65,-.1)--(-.6+5.65,.1)
    (-.7+7.65,-.1)--(-.6+7.65,.1);
    
    \node[] (vtext) at (-2,0) {$C_4(t)=\frac{1}{d}\cdot$};
	 \node[rectangle, fill=egg, rounded corners, minimum width=2em, minimum height =7.2em, draw] (v0) at (0,0) {$U_H$};
	 \node[rectangle, fill=egg, rounded corners, minimum width=2em, minimum size =2em, draw] (v1) at (1,-1) {$X$};      
    \node[rectangle, fill=egg, rounded corners, minimum width=2em, minimum height =7.2em, draw] (v2) at (2,0) {$U_H^\dagger$};
    \node[rectangle, fill=egg, rounded corners, minimum width=2em, minimum height =2em, draw] (v3) at (3,1) {$Z$};
    \node[rectangle, fill=egg, rounded corners, minimum width=2em, minimum height =7.2em, draw] (v4) at (4,0) {$U_H$};
	 \node[rectangle, fill=egg, rounded corners, minimum width=2em, minimum size =2em, draw] (v5) at (5,-1) {$X$};      
    \node[rectangle, fill=egg, rounded corners, minimum width=2em, minimum height =7.2em, draw] (v6) at (6,0) {$U_H^\dagger$};
    \node[rectangle, fill=egg, rounded corners, minimum width=2em, minimum height =2em, draw] (v7) at (7,1) {$Z$};
\end{tikzpicture},
\end{equation}

\begin{equation}\label{eq:SB1}
\begin{tikzpicture}
    \draw [thick,color=dullblue] 
    (-1,1)--(8,1)
    
    (-1,0)--(8,0)
    
    (-1,-1)--(0.75,-1) (1.25,-1)--(4.75,-1) (5.25,-1)--(8,-1)

    (-1,-2)--(0.75,-2) (1.25,-2)--(4.6,-2) (4.9,-2)--(5.1,-2) (5.4,-2)--(8,-2)
    
    (-1,-3)--(4.75,-3) (5.25,-3)--(8,-3);
    
    \draw [thick,color=dullblue] 
    (0.75,-1)--(0.75,-2)
    (1.25,-1)--(1.25,-2)
    (4.75,-1)--(4.75,-3)
    (5.25,-1)--(5.25,-3);

    \draw[thick,color=dullblue]
    (-.7,-.1)--(-.6,.1)
    (-.7+1.65,-.1)--(-.6+1.65,.1)
    (-.7+3.65,-.1)--(-.6+3.65,.1)
    (-.7+5.65,-.1)--(-.6+5.65,.1)
    (-.7+7.65,-.1)--(-.6+7.65,.1);
    
    \node[] (vtext) at (-2,-1) {$C_4(t)=\frac{1}{d}\cdot$};
	 \node[rectangle, fill=egg, rounded corners, minimum width=2em, minimum height =7.2em, draw] (v0) at (0,0) {$U_H$};
	 \node[rectangle, fill=egg, rounded corners, minimum width=2em, minimum size =2em, draw] (v1) at (3,-2) {$X^T$};      
    \node[rectangle, fill=egg, rounded corners, minimum width=2em, minimum height =7.2em, draw] (v2) at (2,0) {$U_H^\dagger$};
    \node[rectangle, fill=egg, rounded corners, minimum width=2em, minimum height =2em, draw] (v3) at (3,1) {$Z$};
    \node[rectangle, fill=egg, rounded corners, minimum width=2em, minimum height =7.2em, draw] (v4) at (4,0) {$U_H$};
	 \node[rectangle, fill=egg, rounded corners, minimum width=2em, minimum size =2em, draw] (v5) at (7,-3) {$X^T$};      
    \node[rectangle, fill=egg, rounded corners, minimum width=2em, minimum height =7.2em, draw] (v6) at (6,0) {$U_H^\dagger$};
    \node[rectangle, fill=egg, rounded corners, minimum width=2em, minimum height =2em, draw] (v7) at (7,1) {$Z$};
    
\end{tikzpicture}.
\end{equation}
To simplify notation, define $B\equiv Z_1 X^T_{a_1}$ with Pauli $Z$ on system qubit 1 and $X^T$ on an ancillary qubit denoted by $a_1$. The correlator is redrawn as

\begin{equation}\label{eq:SingleBellTensor}
\begin{tikzpicture}
    \draw [thick,color=dullblue] 
    
    (-1,0)--(4.2,0)
    
    (-1,-1)--(0.75,-1) (1.25,-1)--(4.2,-1)

    (-1,-2)--(0.75,-2) (1.25,-2)--(4.2,-2)
    
    (-1,-3)--(4.2,-3)
    
    (-1,-4)--(0.75,-4) (1.25,-4)--(4.2,-4)
    
    (-1,-5)--(0.75,-5) (1.25,-5)--(4.2,-5);
    
    \draw [thick,color=dullblue] 
    (0.75,-1)--(0.75,-2)
    (1.25,-1)--(1.25,-2)
    (0.75,-4)--(0.75,-5)
    (1.25,-4)--(1.25,-5);
    
    \draw [thick,color=dullblue] 
    (4.2,0)--(6.2,-3)--(7.2,-3)
    (4.2,-1)--(6.2,-4)--(7.2,-4)
    (4.2,-2)--(7.2,-2)
    (4.2,-3)--(6.2,0)--(7.2,0)
    (4.2,-4)--(6.2,-1)--(7.2,-1)
    (4.2,-5)--(7.2,-5);

    \draw[thick,color=dullblue]
    (-.7,-.1)--(-.6,.1)
    (-.7+1.65,-.1)--(-.6+1.65,.1)
    (-.7+4.55,-.1)--(-.6+4.55,.1)
    (-.7+7.3,-.1)--(-.6+7.3,.1)
    (-.7,-3.1)--(-.6,-2.9)
    (-.7+1.65,-3.1)--(-.6+1.65,-2.9)
    (-.7+4.55,-3.1)--(-.6+4.55,-2.9)
    (-.7+7.3,-3.1)--(-.6+7.3,-2.9);
    
    \node[] (vtext) at (-2,-2.5) {$C_4(t)=\frac{1}{d}\cdot$};
	 \node[rectangle, fill=egg, rounded corners, minimum width=2em, minimum height =4.5em, draw] (v0) at (0,-.5) {$U_H$};
	 \node[rectangle, fill=egg, rounded corners, minimum width=2em, minimum height =7.2em, draw] (v1) at (3,-1) {$B$};      
    \node[rectangle, fill=egg, rounded corners, minimum width=2em, minimum height =4.5em, draw] (v2) at (2,-.5) {$U_H^\dagger$};
	 \node[rectangle, fill=egg, rounded corners, minimum width=2em, minimum height =4.5em, draw] (v3) at (0,-3.5) {$U_H$};
	 \node[rectangle, fill=egg, rounded corners, minimum width=2em, minimum height =7.2em, draw] (v4) at (3,-4) {$B$};      
    \node[rectangle, fill=egg, rounded corners, minimum width=2em, minimum height =4.5em, draw] (v5) at (2,-3.5) {$U_H^\dagger$};

    \draw [dashed,line width=0.25mm]
    (.55,0.35)--(1.45,0.35)--(1.45,-2.35)--(.55,-2.35)--(.55,0.35);
    
    \draw [dashed,line width=0.25mm]
    (3.6,0.35)--(7,0.35)--(7,-5.35)--(3.6,-5.35)--(3.6,0.35);
    
\end{tikzpicture}.
\end{equation}
To clarify, the slash marks in Eq.~\eqref{eq:SB1} and \eqref{eq:SingleBellTensor} represent $N-2$ and $N-1$ system qubits, respectively. Define $\rho_{N,a_1}$ as the initial state where $N$-th system qubit forms a Bell state with the ancillary qubit. The remaining $N-1$ system qubits are in the maximally mixed state. The small dotted rectangle in Eq.~\eqref{eq:SingleBellTensor}  represents $\rho_{N,a_1}$ up to a normalization factor $d$. The large dotted rectangle represents the permutation operator $\prod_{l=1}^{N}T_{(l,N+1+l)}$, which is a product of swap operators. Define the time-evolved state
\begin{equation}
    \rho_{H,N,a_1}=(U_H\otimes I_{a_1})\rho_{N,a_1}(U_H^\dagger\otimes I_{a_1}),
\end{equation}
where $I_{a_1}$ is the identity on the ancillary qubit. The correlator is
\begin{equation}
    C_4(t)=d\cdot \tr\Bigg\{(\rho_{H,N,a_1}\otimes \rho_{H,N,a_1})B^{\otimes 2}\prod_{l=1}^{N}T_{(l,N+1+l)}\Bigg\}.
\end{equation}

To construct an estimator for $C_4(t)$, create a shadow of size $K\geq 2$ for state $\rho_{H,N,a_1}$:
\begin{equation}
    S_{SB}(\rho_{H,N,a_1},K)=\Big\{\hat{\rho}_{H,N,a_1}^{(i)}(U_i;\hat{b}_i)\Big\}_{i=1}^{K}.
\end{equation}
Use this shadow to compute 
\begin{equation}
    \hat{C}_4(t)=\frac{d}{2}\binom{K}{2}^{-1}\sum_{i\neq j} \tr\Bigg\{\left(\hat{\rho}_{H,N,a_1}^{(i)}\otimes \hat{\rho}_{H,N,a_1}^{(j)}\right)B^{\otimes 2}\prod_{l=1}^{N}T_{(l,N+1+l)}\Bigg\}.
\end{equation}
This procedure can be generalized to construct an estimator for $C_{4k}(t)$ for $k\geq 2$,
\begin{equation}
\hat{C}_{4k}(t)=\frac{d^{2k-1}}{(2k)!}\binom{K}{2k}^{-1}\sum_{i_1\neq\cdots \neq i_{2k}} \tr\Bigg\{\left(\otimes_{j=1}^{2k}\hat{\rho}_{H,N,a_1}^{(i_j)}\right)B^{\otimes 2k}\prod_{l=1}^{N}T_{(l,N+1+l,\ldots,(2k-1)(N+1)+l)}\Bigg\}.
\end{equation}
The sum is over all size $2k$ permutations of snapshots in $S_{SB}$.

\begin{figure}[t]
  \centering
  \includegraphics[scale=.4]{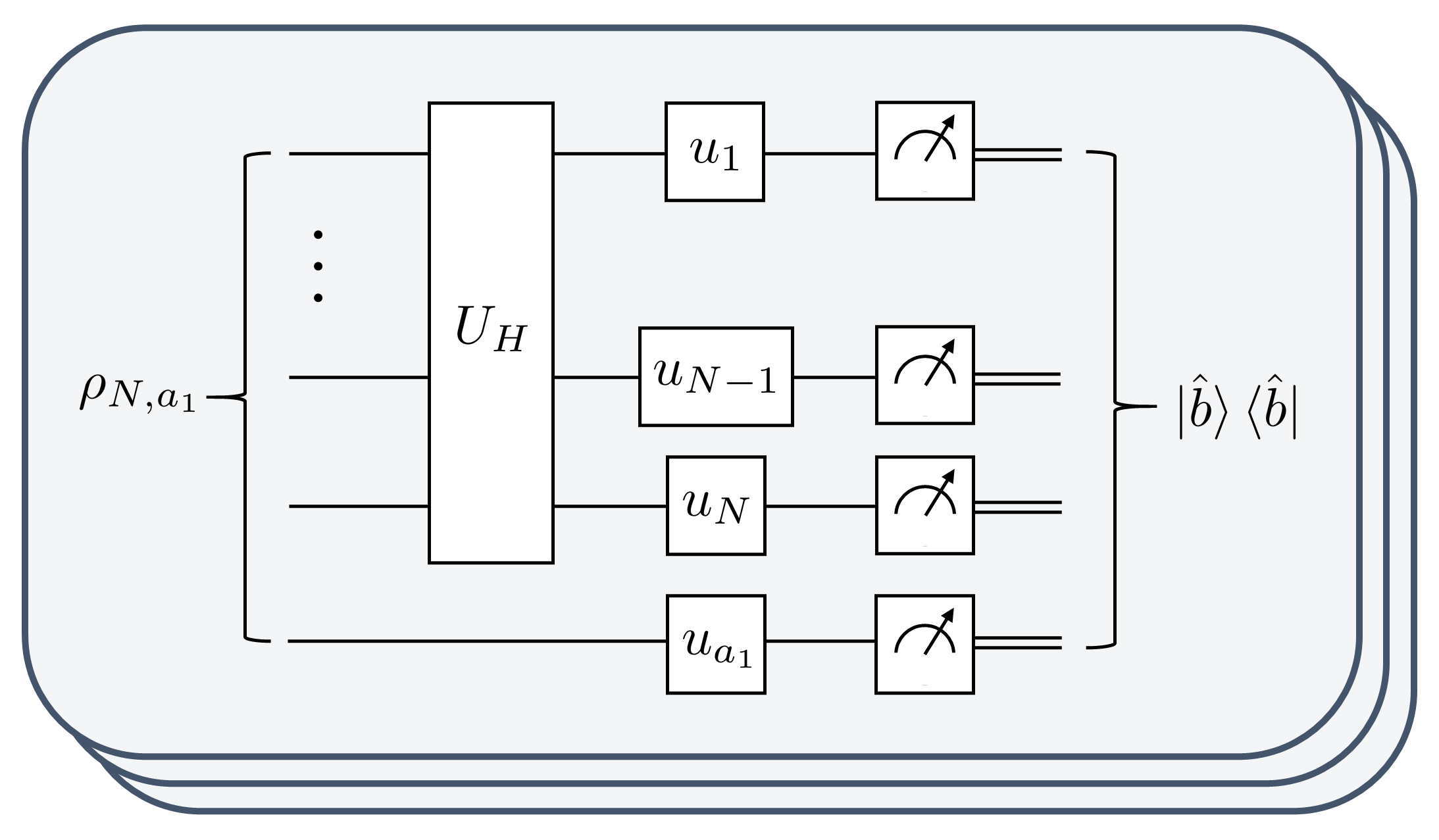}\label{fig:SingleBellState}
  \vspace*{-3mm}
  \caption{Quantum circuit which generates a classical snapshot for the single Bell state protocol. The initial state is $\rho_{N,a_1}$, where system qubit $N$ forms a Bell state with an ancillary qubit, while the remaining qubits are in the maximally mixed state. Unitary $U_H(t)$ evolves the $N$ system qubits. Then local random Clifford unitaries $u_i$ act on each qubit, followed by a measurement.}
  \label{fig:SingleBell}
\end{figure}

The single Bell state protocol (see Fig.~\ref{fig:SingleBell}) to estimate $C_{4k}(t)$ is:
\begin{enumerate}
\begin{samepage}
    \item Prepare $N$ system qubits and $1$ ancillary qubit. Create a Bell state between system qubit $N$ and the ancillary qubit. Prepare the remaining system qubits in the maximally mixed state. 
    \item Evolve the system qubits with $U_H(t)$.
    \item Construct a shadow of size $K\geq 2k$ for the state.
    \item Use the shadow to compute $\hat{C}_{4k}$.
\end{samepage}
\end{enumerate}
The single Bell state protocol carries the advantage that the OTOCs are given by a single trace, rather than a sum of traces as in the mixed state protocol. This feature makes the single Bell state protocol more appropriate for the commutator type correlators discussed in Appendix~\ref{sec:NonComEst}. This protocol is ideal for systems with a limited number of qubits, since it only introduces one ancillary qubit. Similar to the multi-Bell state protocol, a manipulation of diagrams can easily yield an estimator for a general $4k$-point correlator over an arbitrary state. The same classical shadow can be used to compute OTOCs for different choices of $W$ and $V$, allowing for the calculation of multiple OTOCs with the same batch of measurements.

\section{Statistical error analysis}\label{sec:var}
We analyze the statistical error for estimators due to the finite shadow size $K$. We focus on the mixed state protocol from Sec.~\ref{Direct} for its experimental practicality. We state the following results on the variance in shadow tomography:
\begin{fact}\label{Fact:Var}
(Proposition 3 in \cite{Huang_2020}). For a state $\rho\in \mathcal{H}_d$  and a linear function $\tr(O\rho)$, the single-shot variance
of the function obeys
\begin{equation}
\mathrm{Var}[\tr\{O\hat{\rho}\}]\leq d\tr(O^2),
\end{equation}
where $\hat{\rho}$ is a snapshot of $\rho$.
\end{fact}
\begin{lemma}\label{Lemma:2copy}
For a state $\rho^{\otimes 2}\in \mathcal{H}_d^{\otimes 2}$ and a nonlinear function $\tr\left\{ O_2 \rho^{\otimes2}\right\}$ with observable $O_2=T_{(1,2)}W^{\otimes 2}$, the single-shot variance
of the function obeys
\begin{equation}
\mathrm{Var}\left[\tr\left\{ O_2 \hat{\rho}\otimes\hat{\rho}'\right\}\right]\leq d^3,
\end{equation}
where $W$ is any Pauli operator and $\hat{\rho}, \hat{\rho}'$ are two distinct snapshots of $\rho$.
\end{lemma}
The proof of Lemma \ref{Lemma:2copy} is left in Appendix~\ref{ApLem2copy}. The lemma shows a significant enhancement when compared with using Fact \ref{Fact:Var} on the doubled Hilbert space, which yields $d^2\tr\{O_2^2\}=d^2\tr\{T_{(1,2)}^2\}=d^4$. Lemma \ref{Lemma:2copy} is also suitable for $O_2=T_{(1,2)}$, which can tighten the variance for the purity measurement performed in \cite{Huang_2020,elben2020mixedstate}. The improvement relies on the prior knowledge that the target state here is in the tensor product form $\rho\otimes \rho$. We believe a modification of Lemma \ref{Lemma:2copy} with an appropriate $O$ on a few-copy state can enhance shadow tomography in other scenarios. 

\subsection{Variance of four-point OTOC}\label{Sec:Var4}
We compute the variance of $\hat{C}_4(t)$ in Eq.~\eqref{Eq:C4est}. The summation index constraint $i_1\neq i_2$ in $\hat{C}_4(t)$ can be changed to $i_1<i_2$ by symmetrizing the observable as follows
\begin{equation}\label{}
\begin{aligned}
\hat{C}_{4}(t)&=\frac{d}{2}
\binom{K}{2}^{-1}
\sum_{ i_1\neq i_2}\ \tr\Big\{\hat{\rho}_{V}^{(i_1)}\otimes \hat{\rho}_{V}^{(i_2)}W^{\otimes 2}T_{(1,2)}\Big\}-1\\
&=
d\binom{K}{2}^{-1}
\sum_{ i_1< i_2}\ \tr\left\{  \frac{W^{\otimes 2}T_{(1,2)}+T_{(1,2)}W^{\otimes 2}}{2}\hat{\rho}_{V}^{(i_1)}\otimes \hat{\rho}_{V}^{(i_2)}\right\}-1\\
&=d\binom{K}{2}^{-1}
\sum_{ i_1< i_2}\ \tr\left\{T_{(1,2)}W^{\otimes 2}\hat{\rho}_{V}^{(i_1)}\otimes \hat{\rho}_{V}^{(i_2)}\right\}-1\\
&=d\binom{K}{2}^{-1}
 \sum_{ i_1< i_2}\hat{D}_{4}(i_1,i_2) -1.
\end{aligned}
\end{equation}
Define  $\hat{D}_{4}(i_1,i_2)=\tr\left\{T_{(1,2)}W^{\otimes 2}\hat{\rho}_{V}^{(i_1)}\otimes \hat{\rho}_{V}^{(i_2)}\right\}$. The third line is due to $T_{(1,2)}$ commuting with $W^{\otimes 2}$.
To simplify notation, we suppress all time dependence in this section. The variance of $\hat{C}_4$ satisfies 
\begin{equation}
\mathrm{Var}(\hat{C}_{4})=\left[
d\binom{K}{2}^{-1}\right]^2\mathrm{Var}\left(\sum_{ i_1< i_2}\hat{D}_{4}(i_1,i_2)\right),    
\end{equation}
with 
\begin{equation}\label{VarD4}
\begin{aligned}
\mathrm{Var}\left(\sum_{ i_1< i_2}\hat{D}_{4}(i_1,i_2)\right)&=\mathbb{E}\left[\left(\sum_{ i_1< i_2}\hat{D}_{4}(i_1,i_2)\right)^2\right]-\mathbb{E}\left[\sum_{ i_1< i_2}\hat{D}_{4}(i_1,i_2)\right]^2\\
&=\sum_{\substack{i_1< i_2 \\ j_1< j_2}}\left\{\mathbb{E}\left[\hat{D}_{4}(i_1,i_2)\hat{D}_{4}(j_1,j_2)\right]-D_4^2 \right\}\\
&=\sum_{\substack{i_1< i_2 \\ j_1< j_2}}\left[V_4(i_1,i_2,j_1,j_2)-D_4^2\right].
\end{aligned}
\end{equation}
Define $D_4=\mathbb{E}[\hat{D}_{4}(i_1,i_2)]$ for any $i_1,i_2$ and $V_4(i_1,i_2,j_1,j_2)=\mathbb{E}\left[\hat{D}_{4}(i_1,i_2)\hat{D}_{4}(j_1,j_2)\right]$. $V_4$ depends on the coincidences between indices $i_1,i_2$ and $j_1, j_2$, respectively. A coincidence indicates that the two $\hat{D}_{4}$ in the second line of Eq.~\eqref{VarD4} share the same snapshot, and thus are not independent. There are three possible coincidence cases of the indices discussed as follows:
\begin{itemize}
    \item No coincidence: $V_4=\mathbb{E}\left[\hat{D}_{4}(i_1,i_2)\right]\mathbb{E}\left[\hat{D}_{4}(j_1,j_2)\right]=D_4^2$ since all snapshots are independent.
    \item One coincidence: take for example $i_1\neq j_1$ and $i_2=j_2$. There are a total of $2\binom{K}{1}\binom{K-1}{2}$ such terms. We simplify
\begin{equation}\label{Var1coin}
\begin{aligned}
V_4&=\mathbb{E}\left[\tr\left\{T_{(1,2)}W^{\otimes 2} \hat{\rho}^{(i_1)}_{V}\otimes\hat{\rho}^{(i_2)}_{V}\right\}\tr\left\{T_{(1,2)}W^{\otimes 2} \hat{\rho}^{(j_1)}_{V}\otimes\hat{\rho}^{(i_2)}_{V}\right\}\right]\\
&=\mathbb{E}\left[\tr\left\{T_{(1,2)}W^{\otimes 2} \rho_{V}\otimes\hat{\rho}_{V}\right\}^2\right]\\
&=\mathbb{E}\left[\tr\left\{ W\rho_{V}W \hat{\rho}_{V}\right\}^2\right].
\end{aligned}
\end{equation}
In the second line, we take the expectation of $\hat{\rho}_V^{(i_1)}$ and $\hat{\rho}_V^{(j_1)}$, then denote $\hat{\rho}_V^{(i_2)}$ as $\hat{\rho}_V$ without ambiguity.
Set $O_1=W\rho_{V}W$ so that $\mathbb{E}\left[\tr\left\{ O_1 \hat{\rho}_{V}\right\}\right]=D_4$. We can analyze $V_4-D_4^2=\mathrm{Var}[\tr\left\{ O_1 \hat{\rho}_{V}\right\}]$ using Fact \ref{Fact:Var}.

\item Two coincidences: $i_1=j_1$ and $i_2=j_2$.
There are $\binom{K}{2}$ such terms in total. We can write
\begin{equation}
\begin{aligned}
V_4&=\mathbb{E}\left[\tr\left\{T_{(1,2)}W^{\otimes 2} \hat{\rho}_{V}\otimes\hat{\rho}'_{V}\right\}^2\right],
\end{aligned}
\end{equation}
where $\hat{\rho}_V$ and $\hat{\rho}_V'$ are independent snapshots.
Set $O_2=T_{(1,2)}W^{\otimes 2}$ for state $\rho_{V}^{\otimes 2}\in\mc{H}_d^{\otimes 2}$ so that Lemme \ref{Lemma:2copy} bounds $V_4-D_4^2=\mathrm{Var}[\tr\left\{ O_2 \hat{\rho}_{V}\otimes\hat{\rho}'_{V}\right\}]$.

\end{itemize}

Inserting the above three cases into Eq.~\eqref{VarD4}, 
\begin{equation}
\begin{aligned}
\mathrm{Var}(\hat{C}_{4})&=d^2\binom{K}{2}^{-2}\mathrm{Var}\left(\sum_{ i_1< i_2}\hat{D}_{4}(i_1,i_2)\right)\\
&=d^2\binom{K}{2}^{-1}\left\{2(K-2)\mathrm{Var}[\tr\left\{ O_1 \hat{\rho}_{V}\right\}]+\mathrm{Var}[\tr\left\{ O_2 \hat{\rho}_{V}\otimes\hat{\rho}'_{V}\right\}]\right\}\\
&\leq d^2\left[\frac{4(K-2)}{K(K-1)}\tr\{O_1^2\} d+\frac{2d^3}{K(K-1)}\right]\\
&=d^2\left[\frac{8(K-2)}{K(K-1)} +\frac{2d^3}{K(K-1)}\right]\\
&\leq \frac{8d^2}{K}+\frac{3d^5}{K^2}.
\end{aligned}
\end{equation}
In the third line, we apply Fact \ref{Fact:Var} and Lemma \ref{Lemma:2copy} to the variances respectively. The fourth line is due to $\tr\{O_1^2\}=\tr\{\rho_{V}^2\}=\frac{2}{d}$.
By using Chebyshev’s inequality, one arrives at
\begin{prop}\label{prop:K:C4}
To estimate $C_4=d\tr\left\{T_{(1,2)}W^{\otimes 2}\rho_V^{\otimes 2}\right\}-1$ under confidence level $\delta$ and error $\epsilon$, the shadow size $K$ satisfing
\begin{align}\label{Eq:K:C4}
K\geq2\max\left\{\frac{8d^2}{\epsilon^2\delta}, \frac{\sqrt{3}d^{2.5}}{\epsilon\sqrt{\delta}}\right\}
\end{align}
is sufficient to let $\mathrm{Prob}(|C_4-\hat{C}_4|\leq\epsilon)\geq1-\delta$.
\end{prop}
\begin{remark}\label{remark:tomo}
One may apply full quantum state tomography \cite{Haah2016,Yu2020Pauli} on $\rho_V$ to calculate $C_4=d\tr\{T_{(1,2)}W^{\otimes 2}\rho_V^{\otimes 2}\}-1$ directly. To make the error of $C_4$ less than $\epsilon$, the error on the state $\rho_V$ should be around $\epsilon'=\epsilon/d$. As a result, the necessary number of measurements on $\rho_V$ with quantum tomography is $K= \Omega(d^2/\epsilon'^2)=\Omega(d^4/\epsilon^2)$ .
Even with the optimistic estimation by taking the trace distance comparable to the infidelity,  $K=\Omega(d^2/\epsilon')=\Omega(d^3/\epsilon)$, which is still worse than Eq.~\eqref{Eq:K:C4}. Furthermore, if one is restricted to independent measurements on a single copy of $\rho_V$ (like in the classical shadow protocols), the scaling worsens: $K=\Omega(d^3/\epsilon'^2)=\Omega(d^5/\epsilon^2)$.
\end{remark}
\subsection{Variance of eight-point OTOC}
We compute the variance of the unbiased estimator in Eq.~\eqref{DirC8est}. By equivalently symmetrizing the observable, the estimator is
\begin{equation}\label{C8D8est}
\begin{aligned}
\hat{C}_{8}=&
d^3\binom{K}{4}^{-1}
\sum_{i_1<i_2<i_3<i_4}\tr\left\{\Theta_4(T_{(1,2,3,4)} W^{\otimes 4}) \hat{\rho}^{(i_1)}_{V}\otimes\hat{\rho}^{(i_2)}_{V}\otimes\hat{\rho}^{(i_3)}_{V}\otimes\hat{\rho}^{(i_4)}_{V}\right\}\\
&-4d\binom{K}{2}^{-1}
\sum_{j_1<j_2} \tr\{\Theta_2(T_{(1,2)}W^{\otimes 2})\hat{\rho}^{(j_1)}_{V}\otimes\hat{\rho}^{(j_2)}_{V}\}+1\\
=&d^3\binom{K}{4}^{-1}
\sum_{i_1<i_2<i_3<i_4}\hat{D}_{8}(i_1,i_2,i_3,i_4)-4d\binom{K}{2}^{-1}
 \sum_{j_1<j_2}\hat{D}_{4}(j_1,j_2)+1\\
\end{aligned}
\end{equation}
where $\hat{D}_{8}(i_1,i_2,i_3,i_4)$ and $\hat{D}_{4}(j_1,j_2)$ denote the random variables under the summations. 
\begin{equation}
\begin{aligned}
\Theta_t(\cdot)=\frac1{t!}\sum_{\pi\in S_t} T_{\pi}(\cdot)T_{\pi^{-1}}
\end{aligned}
\end{equation}
is the twirling channel of the symmetry group on $\mc{H}_d^{\otimes t}$ and $T_{\pi}$ is the permutation operator for permutation $\pi$. 
$\Theta_t(T_{(1,2,\cdots,t)}W^{\otimes t})=\Theta_t(T_{(1,2,\cdots,t)})W^{\otimes t}$, as $W^{\otimes t}$ commutes with all $T_{\pi}$. Since permutation $(1,2,\cdots,t)$ is an element of $S_t$, the twirling $\Phi_t(T_{(1,2,\cdots,t)})$ returns the average on the elements in the same conjugate class of $T_{(1,2,\cdots,t)}$. For the $t=2$ case, $\Phi_t(T_{(1,2)})=T_{(1,2)}$. For $t=4$, there are 6 elements in this class, so one need only consider six permutations of the snapshots: $\{i_1i_2i_3i_4, i_1i_2i_4i_3, i_1i_3i_2i_4,i_1i_3i_4i_2,i_1i_4i_2i_3,i_1i_4i_3i_2\}$. This improves the classical computation time when calculating $\hat{C}_8(t)$. 

Similar to the analysis of $\hat{C}_{4}$ in Sec.~\ref{Sec:Var4}, the variance $\mathrm{Var}(\hat{C}_{8})=\mathbb{E}[\hat{C}_{8}^2]-\mathbb{E}[\hat{C}_{8}]^2$ depends on the coincidences of the indices. For simplicity, we consider the leading contribution to $\mathrm{Var}(\hat{C}_{8})$, namely, the variance of the first term $\hat{L}_{8}$:
\begin{equation}\label{VarD8}
\begin{aligned}
\mathrm{Var}(\hat{L}_{8})=&\left[
d^3\binom{K}{4}^{-1}\right]^2\mathrm{Var}\left(\sum_{\vec{i}  }\hat{D}_{8}\left(\vec{i}\right)\right)  \\ 
=&
d^6\binom{K}{4}^{-2}\sum_{\vec{i},\vec{j}}\left\{\mathbb{E}\left[\hat{D}_{8}\left(\vec{i}\right)\hat{D}_{8}\left(\vec{j}\right)\right]-D_8^2 \right\}\\
=&
d^6\binom{K}{4}^{-2}\sum_{\vec{i},\vec{j}}\left[V_8\left(\vec{i},\vec{j}\right)-D_{8}^2\right].
\end{aligned}
\end{equation}
The summation over $\vec{i}$ labels the summation over $i_1<i_2<i_3<i_4$, and similarly for $\vec{j}$. Define $V_8\left(\vec{i},\vec{j}\right)=\mathbb{E}\left[\hat{D}_{8}\left(\vec{i}\right)\hat{D}_{8}\left(\vec{j}\right)\right]$. There are sub-leading terms like $\mathbb{E}[\hat{D}_{8}\hat{D}_{4}]-\mathbb{E}[\hat{D}_{8}]\mathbb{E}[\hat{D}_{4}]$ and $\mathrm{Var}(\hat{D}_{4})$ contributing to $\mathrm{Var}(\hat{C}_{8})$, and they can be calculated similarly. As in the four-point case, we consider the coincidences of $\vec{i}, \vec{j}$. We present the result, but leave the derivation in Appendix~\ref{Ap:D8Var}.
\begin{prop}\label{prop:D8}
The variance of the estimator of the leading term  $\hat{L}_8(t)$ for the eight-point OTOC in Eq.~\eqref{DirC8est} can be upper bounded by 
\begin{align*}
\mathrm{Var}(\hat{L}_{8})\leq &\frac{64d^5D_8}{K}+ \frac{16(4dD_4+d^2D_4^2+8D_2^2+2)}{K^2}+\frac{32[d^{10}(1+D_2^2)+3d^8]}{K^3}+\frac{4(d^{14}+5d^6)}{K^4},
\end{align*}
where $D_{4k}=\tr\{(W\rho_V)^{2k}\}$.
In the early-time limit, the bound becomes
\begin{align*}
\frac{512d^2}{K}+\frac{352}{K^2}+\frac{32(2d^{10}+3d^8)}{K^3}+\frac{4(d^{14}+5d^6)}{K^4}.
\end{align*}
\end{prop}
The variance behaves like $O(d^{14}/K^4)$ and still outperforms full quantum state tomography (see Appendix~\ref{Ap:D8Var}). We apply Fact \ref{Fact:Var} in the proof of Proposition \ref{prop:D8}, but expect an improvement for large coincidence cases by using similar techniques as Lemma \ref{Lemma:2copy}.

\section{Numerical simulations}\label{sec:numeric}
We present simulations of predicted and estimated OTOCs using an $N$-qubit mixed-field Ising spin chain with Hamiltonian
\begin{equation}
    H=-\frac{1}{E_0}\Big(J\sum_{i=1}^{N-1} Z_i Z_{i+1}+h_x\sum_{i=1}^N X_i +h_{z}\sum_{i=1}^N Z_i \Big).
\end{equation}
The parameters are $J=1$, $h_x=1.05$, $h_z=0.5$, and $E_0=\sqrt{4J^2+2h_x^2+2h_z^2}$. $X_i$ and $Z_i$ are Pauli operators on the $i$-th qubit. This non-integrable model 
has been shown to exhibit chaotic behavior \cite{PhysRevLett.106.050405}.

\begin{figure}[t]
    \begin{center}
    \subfigure[\centering \ N=4]{{\includegraphics[scale=.32]{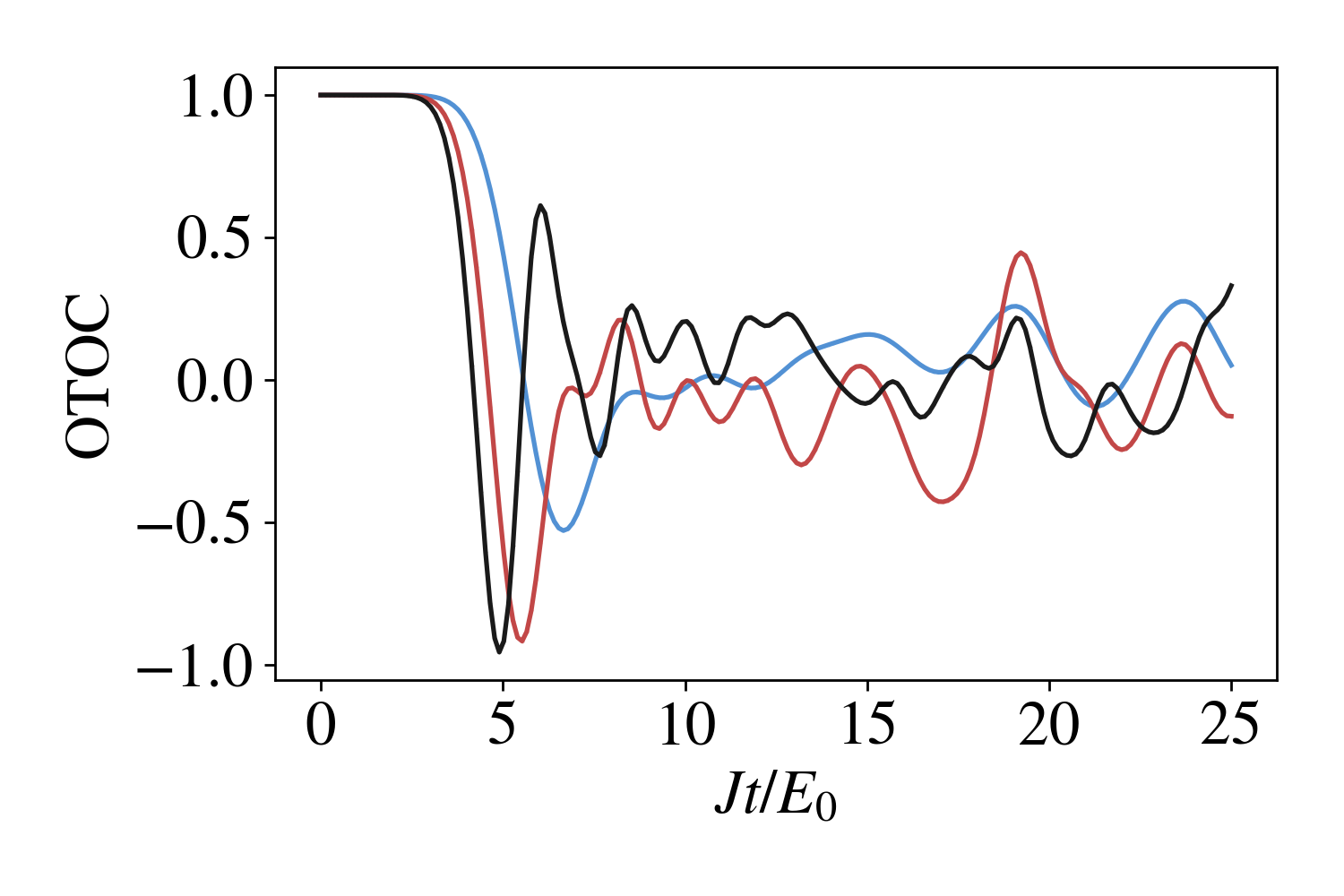} }}%
    \subfigure[\centering \ N=8]{{\includegraphics[scale=.32]{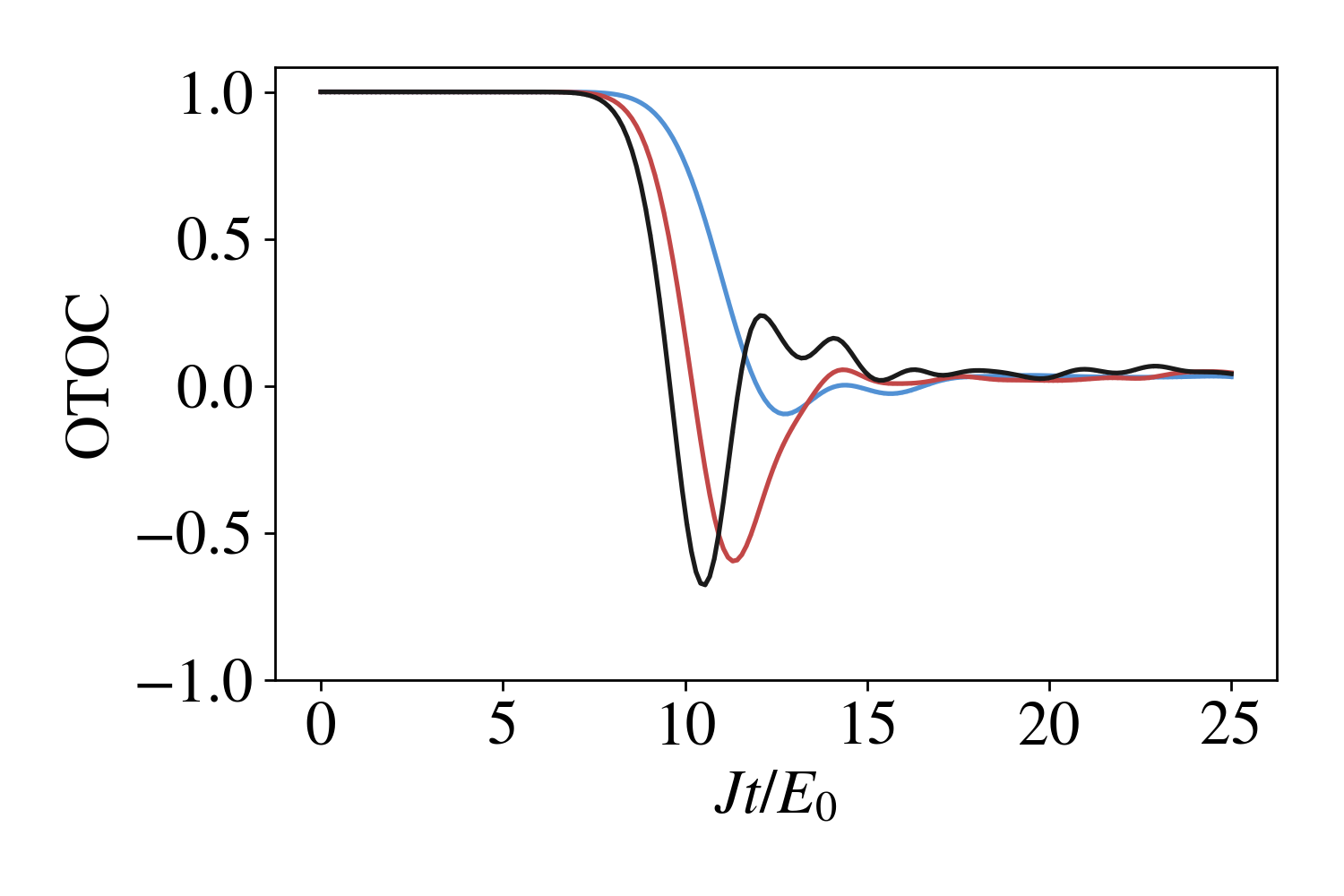} }}%
    \subfigure[\centering \  N=10]{{\includegraphics[scale=.32]{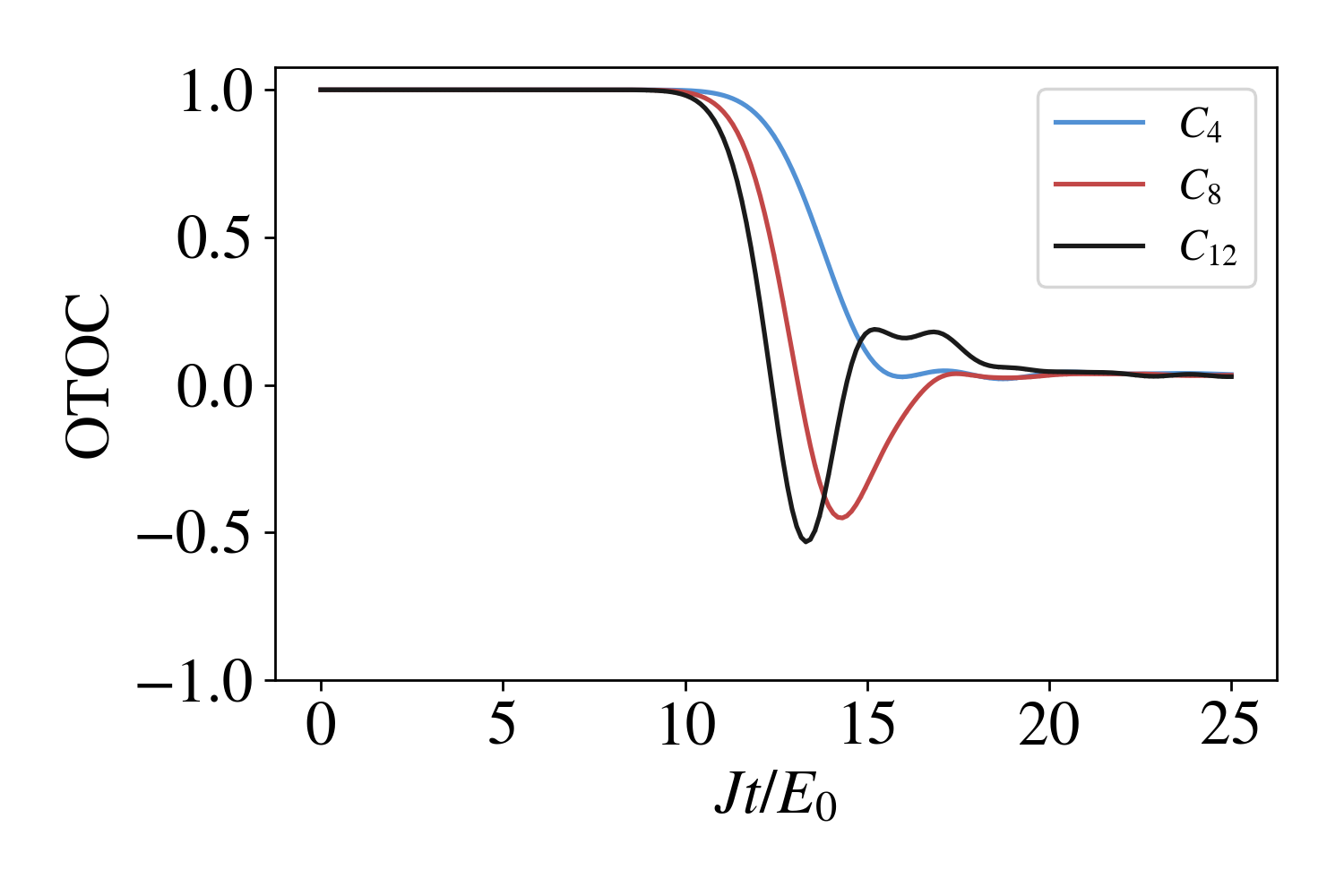} }}%
    \vspace*{-3mm}
    \caption{Predicted values of first three OTOCs for different qubit numbers $N$ for a mixed-field Ising spin chain. In all cases, $W=Z_1$ and $V=Z_N$.}%
    \label{fig:Exact10}%
    \end{center}
\end{figure}

We simulate the evolution of $C_{4k}(t)=\langle (W^\dagger(t)V^\dagger W(t) V)^k \rangle$, where $W=Z_1$ and $V=Z_N$, for the first three OTOCs ($k=1,2,3$) and various $N$ in Fig.~\ref{fig:Exact10}. In the early-time limit, all OTOCs experience an initial steep decay. As the qubit number increases, the time taken for any OTOC to exhibit this decay increases. Since the separation between $W$ and $V$ is larger, $W$ takes longer to spread to the support of $V$, delaying the growth of $[W(t),V]$. Intuitively, since all OTOCs are related to a Schatten $2n$-norm of this commutator, the decay of every OTOC is delayed. Referring to the $N=8$ and $N=10$ cases of Fig.~\ref{fig:Exact10}, all OTOCs decay to the same value in the large-time limit, reflecting the system's degree of scrambling. This point is discussed further in Appendix~\ref{sec:type}. These three characteristics: an initial steep decay, the delay in decay for larger qubit systems, and the large-time decay to a floor value (for sufficiently large systems) are well-known features of the famous four-point OTOC for chaotic systems. Higher-point OTOCs also display these characteristics, making them reliable stand-alone quantities to study scrambling.

\begin{figure}[t]
    \begin{center}
    \subfigure{{\includegraphics[scale=.32]{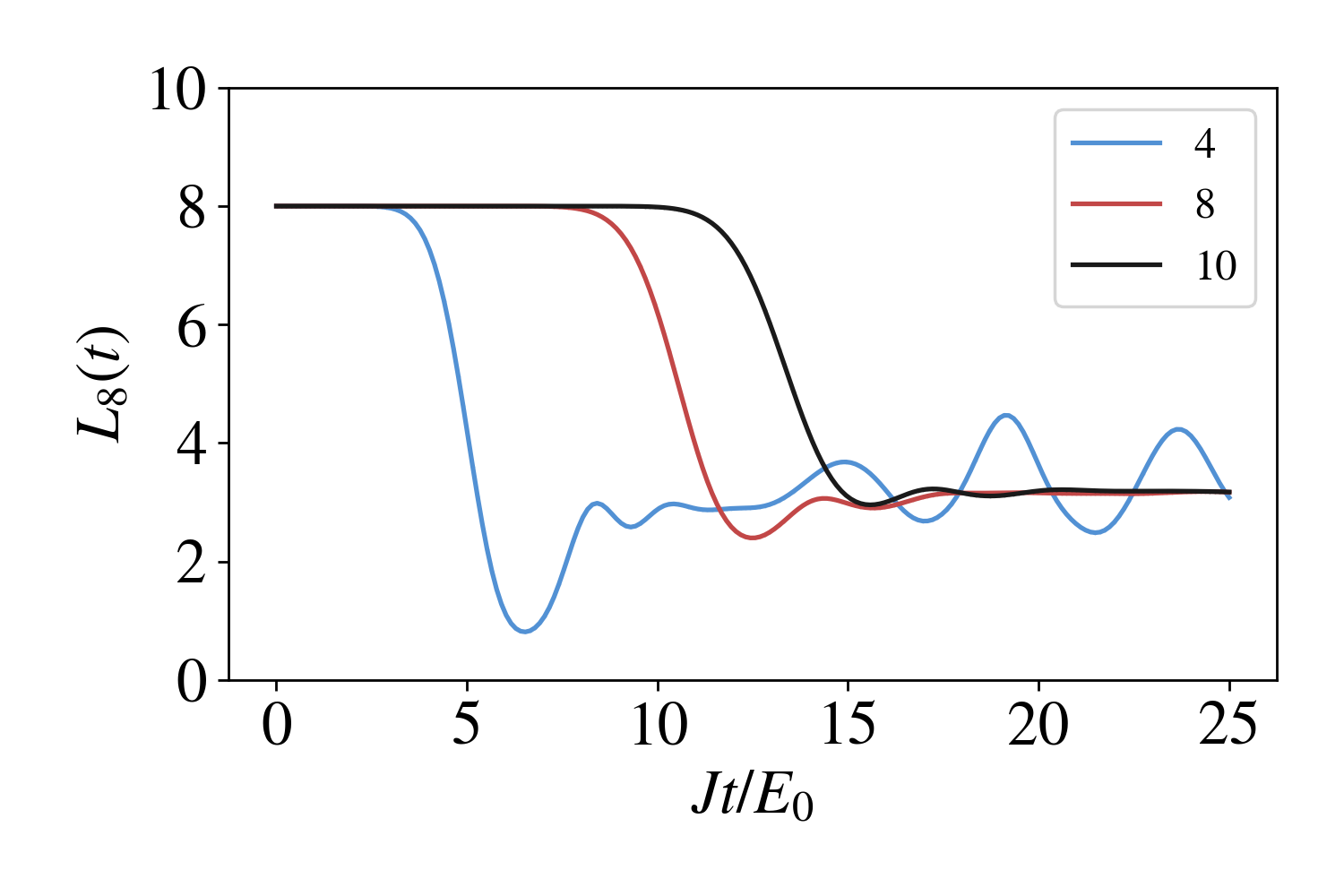} }} %
    \subfigure{{\includegraphics[scale=.32]{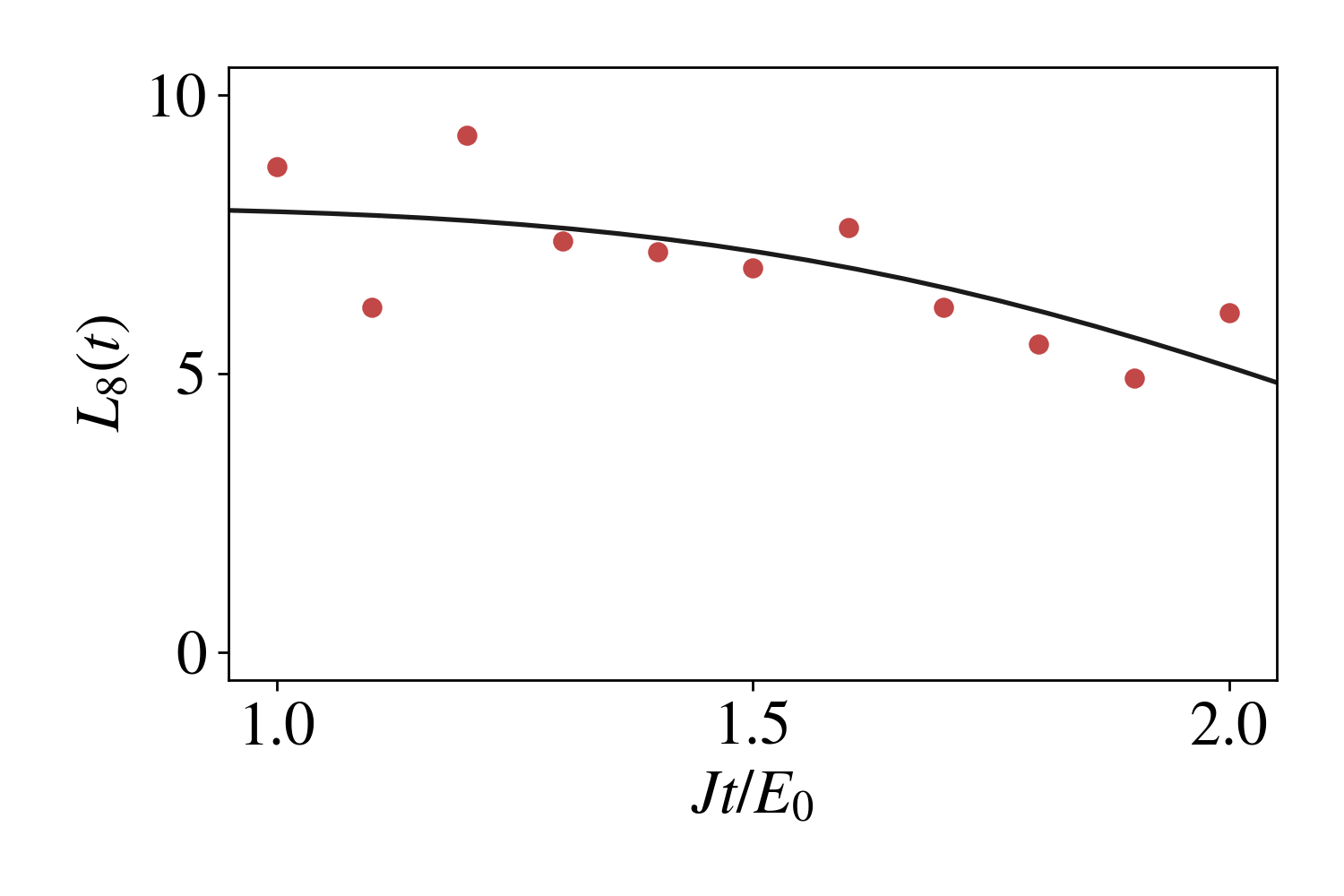} }}
    \vspace*{-3mm}
    \caption{(Left) Dynamics of the predicted leading-order term $L_8(t)$ for $N=4,8,10$ (Right) Early-time behavior of the predicted leading-order term $L_8(t)$ (black line) plotted against its estimator $\hat{L}_8(t)$ (red dots) using the mixed state protocol for a 2-qubit chain. Each point represents an average over 25 estimators, each computed with a different shadow of size 150.}
    \label{fig:Hidden}%
    \end{center}
\end{figure}

\begin{figure}[t]
    \begin{center}
    \subfigure[\centering \ K=5,000]{{\includegraphics[scale=.32]{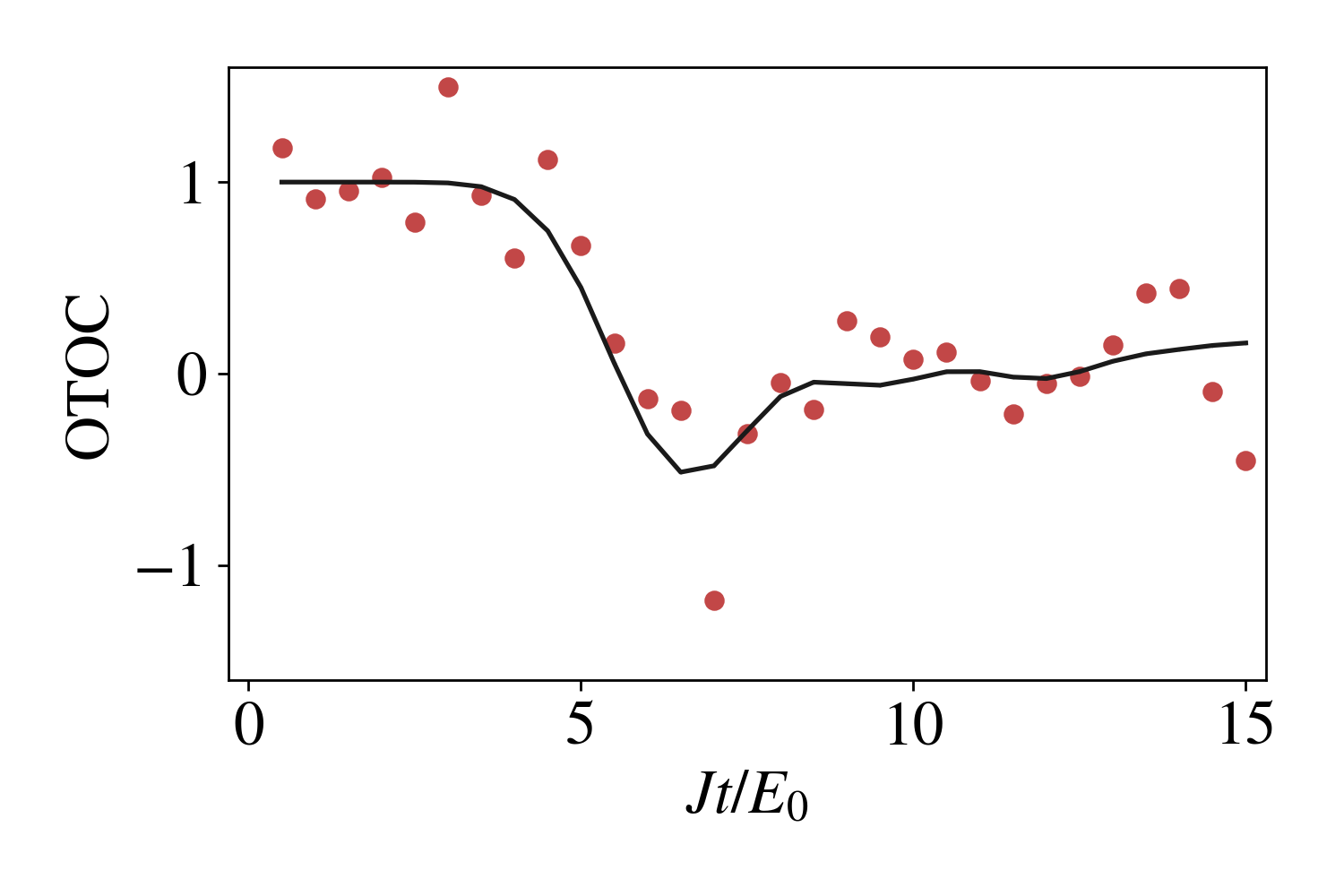} }}%
    \subfigure[\centering \ K=10,000]{{\includegraphics[scale=.32]{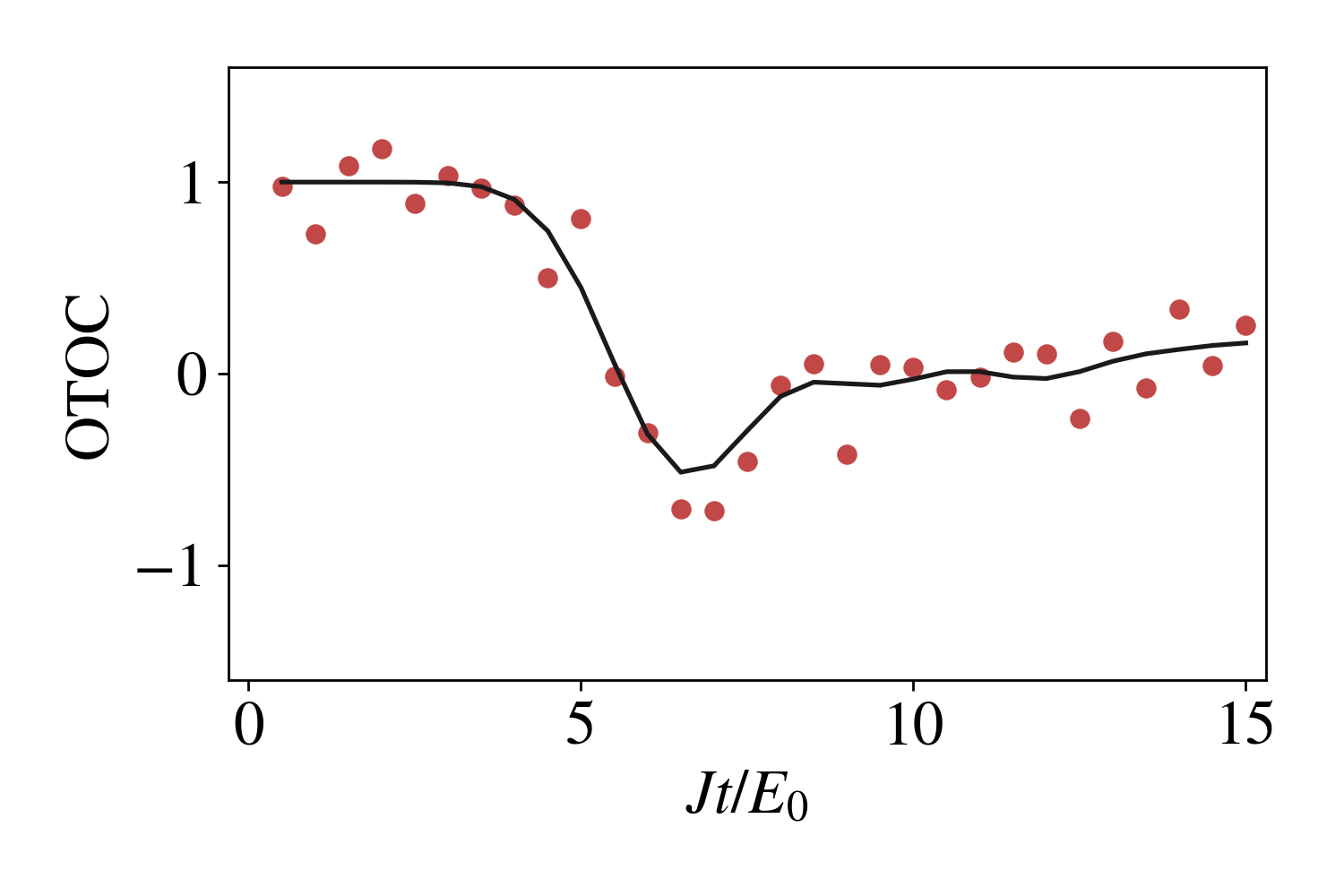} }}%
    \subfigure[\centering \ K=15,000]{{\includegraphics[scale=.32]{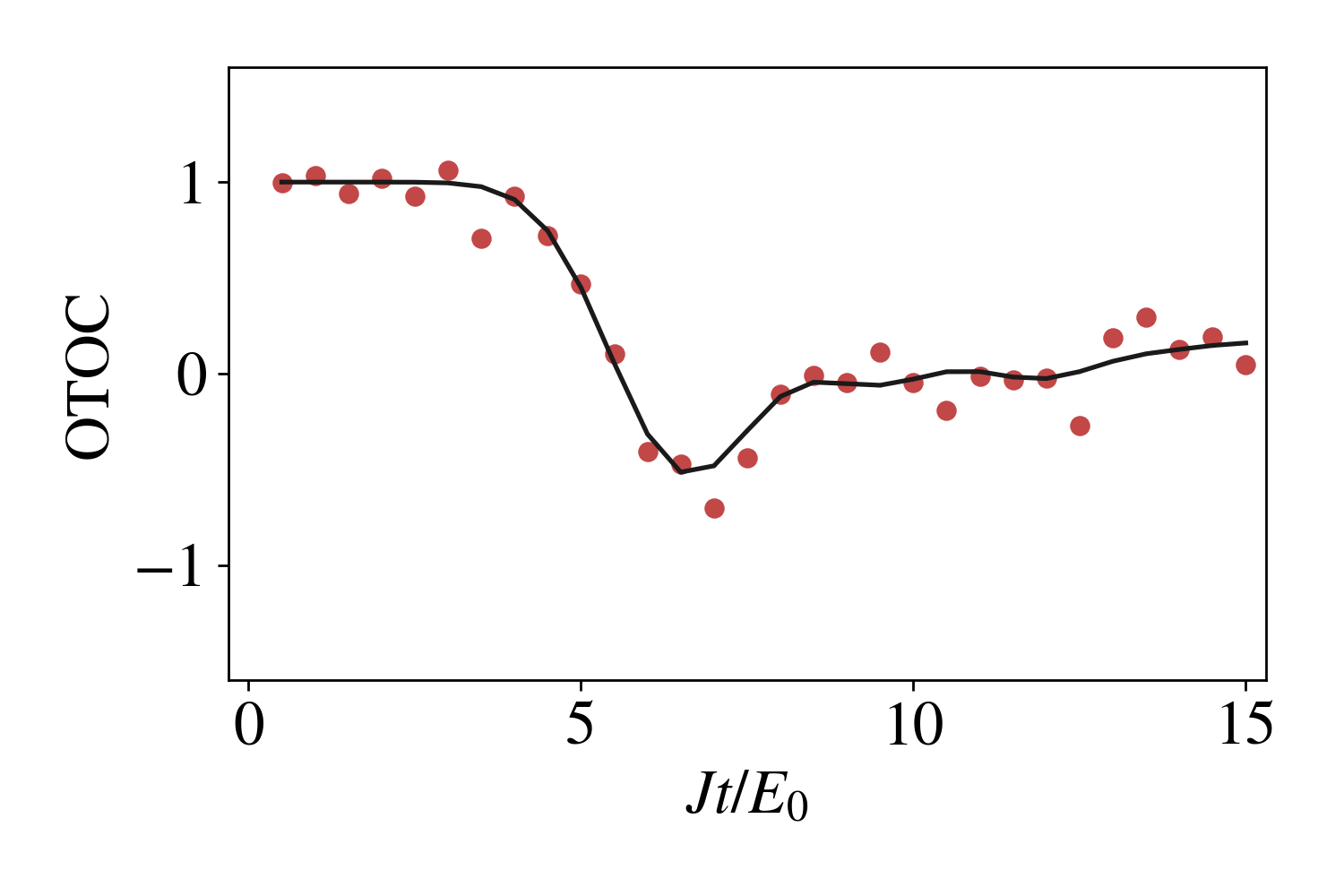} }}%
    \vspace*{-3mm}
    \caption{Predicted four-point OTOC $C_{4}(t)$ (black line) plotted against its estimator $\hat{C}_{4}(t)$ (red dots) constructed from the mixed state protocol for different shadow sizes $K$. A mixed-field Ising spin chain of $N=4$ qubits is used.}
    \label{fig:Shadow}
    \end{center}
\end{figure}

Studying the early-time behavior of OTOCs sheds light on the initial rate of information delocalization \cite{Rozenbaum_2017,Lin_2018,Xu_2019}. In Fig.~\ref{fig:Hidden}, we numerically simulate the eight-point correlator's leading-order term $L_8(t)$ from Eq.~\eqref{eq:Hidden} for various Ising spin-chain lengths. The early-time dynamics of this term contribute to the faster initial decay of the eight-point correlator relative to the four-point OTOC in Fig.~\ref{fig:Exact10}. This is consistent with the discussion in Appendix~\ref{sec:interpret} in which the multiple perturbations found in higher-point OTOCs are expected to result in faster information delocalization. In Fig.~\ref{fig:Hidden}, we also simulate the measurement of $\hat{L}_8(t)$ from Eq.~\eqref{eq:HiddenEst} with the mixed state protocol. To reduce the post-processing computations, we run the protocol with a fixed shadow size $K$ several times and average the results. In Fig.~\ref{fig:Shadow}, the predicted four-point OTOC is plotted against its estimator from Eq.~\eqref{Eq:C4est} using the mixed state protocol for various shadow sizes. As the shadow size increases, agreement between the two improves.

\section{Conclusion}\label{sec:conclude}
We present a definition of higher-point out-of-time-ordered correlators to describe the dynamics of quantum scrambling in chaotic systems. In the early-time limit, higher-point correlators exhibit faster decay and reveal information delocalization earlier than the four-point OTOC. We present protocols using classical shadows to estimate these correlators and show they can outperform full quantum state tomography. For sufficiently many measurements, good agreement between the predicted and estimated values can be achieved. The protocols avoid time reversal and can probe the dynamics of OTOCs at any time. They can be implemented using single-qubit Pauli measurements, making them ideal for experiments with single-qubit control. The protocols here extract nonlinear functions by measuring only a single-copy of a target state, thus avoiding the preparation of multiple identical copies and the joint control and readout on them. In addition, the same classical shadow can be used to compute multiple correlators by classical post-processing. Furthermore, the protocols can be used to construct estimators for general correlators over an arbitrary initial state $\langle (W(t)VW(t)V)^k\rangle_\rho$, allowing for the study of OTOCs beyond the thermal state background. 



There are a few interesting points which merit further investigation. First, our protocols can be directly extended to the noisy evolution scenario, where the system dynamics are described by general quantum channels. Adopting noise mitigation methods \cite{Chen_2020,Koh_2020} may be an intriguing approach to distinguish quantum scrambling from classical decoherence effects \cite{PhysRevA.97.062113,Zhang_2019,Kudler_Flam_2020}. Second, the enhancement of the variance analysis based on prior knowledge of the state's tensor structure can be generalized to other scenarios, which can further improve the practicality of shadow tomography for nonlinear functions.
Third, it is possible to extend the protocols to measure other quantities in quantum scrambling, such as the operator weight distribution \cite{Roberts2018growth,Qi_2019}. Finally, it is important to explore the operational and physical implications of higher-point OTOCs, and demonstrate our protocols on near-term quantum platforms.


\begin{acknowledgments}
We would like to thank Beni Yoshida for helpful discussions on non-commutator type correlators, Xun Gao for discussions on diagrammatic representations of quantum objects, and Pei Zeng for discussions on shadow tomography. This work was supported in part by ARO Grants W911NF-19-1-0302 and W911NF-20-1-0082. Y.~Zhou is also supported by the National Research Foundation (NRF), Singapore, under its NRFF Fellow program (Award No. NRF-NRFF2016-02), the Quantum Engineering Program Grant QEP-SF3, the Singapore Ministry of Education Tier 1 Grants No. MOE2017-T1-002-043, and No. FQXi-RFP-1809 from the Foundational Questions Institute and Fetzer Franklin Fund (a donor-advised fund of Silicon Valley Community Foundation). 
\end{acknowledgments}

\bibliography{BibOTOC}

\begin{thebibliography}{89}%
\makeatletter
\providecommand \@ifxundefined [1]{%
 \@ifx{#1\undefined}
}%
\providecommand \@ifnum [1]{%
 \ifnum #1\expandafter \@firstoftwo
 \else \expandafter \@secondoftwo
 \fi
}%
\providecommand \@ifx [1]{%
 \ifx #1\expandafter \@firstoftwo
 \else \expandafter \@secondoftwo
 \fi
}%
\providecommand \natexlab [1]{#1}%
\providecommand \enquote  [1]{``#1''}%
\providecommand \bibnamefont  [1]{#1}%
\providecommand \bibfnamefont [1]{#1}%
\providecommand \citenamefont [1]{#1}%
\providecommand \href@noop [0]{\@secondoftwo}%
\providecommand \href [0]{\begingroup \@sanitize@url \@href}%
\providecommand \@href[1]{\@@startlink{#1}\@@href}%
\providecommand \@@href[1]{\endgroup#1\@@endlink}%
\providecommand \@sanitize@url [0]{\catcode `\\12\catcode `\$12\catcode
  `\&12\catcode `\#12\catcode `\^12\catcode `\_12\catcode `\%12\relax}%
\providecommand \@@startlink[1]{}%
\providecommand \@@endlink[0]{}%
\providecommand \url  [0]{\begingroup\@sanitize@url \@url }%
\providecommand \@url [1]{\endgroup\@href {#1}{\urlprefix }}%
\providecommand \urlprefix  [0]{URL }%
\providecommand \Eprint [0]{\href }%
\providecommand \doibase [0]{https://doi.org/}%
\providecommand \selectlanguage [0]{\@gobble}%
\providecommand \bibinfo  [0]{\@secondoftwo}%
\providecommand \bibfield  [0]{\@secondoftwo}%
\providecommand \translation [1]{[#1]}%
\providecommand \BibitemOpen [0]{}%
\providecommand \bibitemStop [0]{}%
\providecommand \bibitemNoStop [0]{.\EOS\space}%
\providecommand \EOS [0]{\spacefactor3000\relax}%
\providecommand \BibitemShut  [1]{\csname bibitem#1\endcsname}%
\let\auto@bib@innerbib\@empty
\bibitem [{\citenamefont {Lewis-Swan}\ \emph
  {et~al.}(2019{\natexlab{a}})\citenamefont {Lewis-Swan}, \citenamefont
  {Safavi-Naini}, \citenamefont {Kaufman},\ and\ \citenamefont
  {Rey}}]{Lewis-Swan2019}%
  \BibitemOpen
  \bibfield  {author} {\bibinfo {author} {\bibfnamefont {R.~J.}\ \bibnamefont
  {Lewis-Swan}}, \bibinfo {author} {\bibfnamefont {A.}~\bibnamefont
  {Safavi-Naini}}, \bibinfo {author} {\bibfnamefont {A.~M.}\ \bibnamefont
  {Kaufman}},\ and\ \bibinfo {author} {\bibfnamefont {A.~M.}\ \bibnamefont
  {Rey}},\ }\href {https://doi.org/10.1038/s42254-019-0090-y} {\bibfield
  {journal} {\bibinfo  {journal} {Nature Reviews Physics}\ }\textbf {\bibinfo
  {volume} {1}},\ \bibinfo {pages} {627} (\bibinfo {year}
  {2019}{\natexlab{a}})}\BibitemShut {NoStop}%
\bibitem [{\citenamefont {Shenker}\ and\ \citenamefont
  {Stanford}(2014{\natexlab{a}})}]{Shenker2014butterfly}%
  \BibitemOpen
  \bibfield  {author} {\bibinfo {author} {\bibfnamefont {S.~H.}\ \bibnamefont
  {Shenker}}\ and\ \bibinfo {author} {\bibfnamefont {D.}~\bibnamefont
  {Stanford}},\ }\href {https://doi.org/10.1007/JHEP03(2014)067} {\bibfield
  {journal} {\bibinfo  {journal} {Journal of High Energy Physics}\ }\textbf
  {\bibinfo {volume} {2014}},\ \bibinfo {pages} {67} (\bibinfo {year}
  {2014}{\natexlab{a}})}\BibitemShut {NoStop}%
\bibitem [{\citenamefont {Hayden}\ and\ \citenamefont
  {Preskill}(2007)}]{Hayden_2007}%
  \BibitemOpen
  \bibfield  {author} {\bibinfo {author} {\bibfnamefont {P.}~\bibnamefont
  {Hayden}}\ and\ \bibinfo {author} {\bibfnamefont {J.}~\bibnamefont
  {Preskill}},\ }\href {https://doi.org/10.1088/1126-6708/2007/09/120}
  {\bibfield  {journal} {\bibinfo  {journal} {Journal of High Energy Physics}\
  }\textbf {\bibinfo {volume} {2007}},\ \bibinfo {pages} {120} (\bibinfo {year}
  {2007})}\BibitemShut {NoStop}%
\bibitem [{\citenamefont {Sekino}\ and\ \citenamefont
  {Susskind}(2008)}]{Sekino_2008}%
  \BibitemOpen
  \bibfield  {author} {\bibinfo {author} {\bibfnamefont {Y.}~\bibnamefont
  {Sekino}}\ and\ \bibinfo {author} {\bibfnamefont {L.}~\bibnamefont
  {Susskind}},\ }\href {https://doi.org/10.1088/1126-6708/2008/10/065}
  {\bibfield  {journal} {\bibinfo  {journal} {Journal of High Energy Physics}\
  }\textbf {\bibinfo {volume} {2008}},\ \bibinfo {pages} {065–065} (\bibinfo
  {year} {2008})}\BibitemShut {NoStop}%
\bibitem [{\citenamefont {Maldacena}\ \emph {et~al.}(2016)\citenamefont
  {Maldacena}, \citenamefont {Shenker},\ and\ \citenamefont
  {Stanford}}]{Maldacena_2016}%
  \BibitemOpen
  \bibfield  {author} {\bibinfo {author} {\bibfnamefont {J.}~\bibnamefont
  {Maldacena}}, \bibinfo {author} {\bibfnamefont {S.~H.}\ \bibnamefont
  {Shenker}},\ and\ \bibinfo {author} {\bibfnamefont {D.}~\bibnamefont
  {Stanford}},\ }\bibfield  {journal} {\bibinfo  {journal} {Journal of High
  Energy Physics}\ }\textbf {\bibinfo {volume} {2016}},\ \href
  {https://doi.org/10.1007/jhep08(2016)106} {10.1007/jhep08(2016)106} (\bibinfo
  {year} {2016})\BibitemShut {NoStop}%
\bibitem [{\citenamefont {Sachdev}\ and\ \citenamefont
  {Ye}(1993)}]{PhysRevLett.70.3339}%
  \BibitemOpen
  \bibfield  {author} {\bibinfo {author} {\bibfnamefont {S.}~\bibnamefont
  {Sachdev}}\ and\ \bibinfo {author} {\bibfnamefont {J.}~\bibnamefont {Ye}},\
  }\href {https://doi.org/10.1103/PhysRevLett.70.3339} {\bibfield  {journal}
  {\bibinfo  {journal} {Phys. Rev. Lett.}\ }\textbf {\bibinfo {volume} {70}},\
  \bibinfo {pages} {3339} (\bibinfo {year} {1993})}\BibitemShut {NoStop}%
\bibitem [{\citenamefont {Kitaev}(2015)}]{Kitaev2015}%
  \BibitemOpen
  \bibfield  {author} {\bibinfo {author} {\bibfnamefont {A.}~\bibnamefont
  {Kitaev}},\ }\href {https://online.kitp.ucsb.edu/online/entangled15/}
  {\bibinfo {title} {A simple model of quantum holography}} (\bibinfo {year}
  {2015})\BibitemShut {NoStop}%
\bibitem [{\citenamefont {Roberts}\ and\ \citenamefont
  {Stanford}(2015)}]{PhysRevLett.115.131603}%
  \BibitemOpen
  \bibfield  {author} {\bibinfo {author} {\bibfnamefont {D.~A.}\ \bibnamefont
  {Roberts}}\ and\ \bibinfo {author} {\bibfnamefont {D.}~\bibnamefont
  {Stanford}},\ }\href {https://doi.org/10.1103/PhysRevLett.115.131603}
  {\bibfield  {journal} {\bibinfo  {journal} {Phys. Rev. Lett.}\ }\textbf
  {\bibinfo {volume} {115}},\ \bibinfo {pages} {131603} (\bibinfo {year}
  {2015})}\BibitemShut {NoStop}%
\bibitem [{\citenamefont {Swingle}\ \emph {et~al.}(2016)\citenamefont
  {Swingle}, \citenamefont {Bentsen}, \citenamefont {Schleier-Smith},\ and\
  \citenamefont {Hayden}}]{Swingle_2016}%
  \BibitemOpen
  \bibfield  {author} {\bibinfo {author} {\bibfnamefont {B.}~\bibnamefont
  {Swingle}}, \bibinfo {author} {\bibfnamefont {G.}~\bibnamefont {Bentsen}},
  \bibinfo {author} {\bibfnamefont {M.}~\bibnamefont {Schleier-Smith}},\ and\
  \bibinfo {author} {\bibfnamefont {P.}~\bibnamefont {Hayden}},\ }\bibfield
  {journal} {\bibinfo  {journal} {Physical Review A}\ }\textbf {\bibinfo
  {volume} {94}},\ \href {https://doi.org/10.1103/physreva.94.040302}
  {10.1103/physreva.94.040302} (\bibinfo {year} {2016})\BibitemShut {NoStop}%
\bibitem [{\citenamefont {Swingle}\ and\ \citenamefont
  {Chowdhury}(2017{\natexlab{a}})}]{PhysRevB.95.060201}%
  \BibitemOpen
  \bibfield  {author} {\bibinfo {author} {\bibfnamefont {B.}~\bibnamefont
  {Swingle}}\ and\ \bibinfo {author} {\bibfnamefont {D.}~\bibnamefont
  {Chowdhury}},\ }\href {https://doi.org/10.1103/PhysRevB.95.060201} {\bibfield
   {journal} {\bibinfo  {journal} {Phys. Rev. B}\ }\textbf {\bibinfo {volume}
  {95}},\ \bibinfo {pages} {060201} (\bibinfo {year}
  {2017}{\natexlab{a}})}\BibitemShut {NoStop}%
\bibitem [{\citenamefont {Chowdhury}\ and\ \citenamefont
  {Swingle}(2017)}]{PhysRevD.96.065005}%
  \BibitemOpen
  \bibfield  {author} {\bibinfo {author} {\bibfnamefont {D.}~\bibnamefont
  {Chowdhury}}\ and\ \bibinfo {author} {\bibfnamefont {B.}~\bibnamefont
  {Swingle}},\ }\href {https://doi.org/10.1103/PhysRevD.96.065005} {\bibfield
  {journal} {\bibinfo  {journal} {Phys. Rev. D}\ }\textbf {\bibinfo {volume}
  {96}},\ \bibinfo {pages} {065005} (\bibinfo {year} {2017})}\BibitemShut
  {NoStop}%
\bibitem [{\citenamefont {Ba\~nuls}\ \emph {et~al.}(2011)\citenamefont
  {Ba\~nuls}, \citenamefont {Cirac},\ and\ \citenamefont
  {Hastings}}]{PhysRevLett.106.050405}%
  \BibitemOpen
  \bibfield  {author} {\bibinfo {author} {\bibfnamefont {M.~C.}\ \bibnamefont
  {Ba\~nuls}}, \bibinfo {author} {\bibfnamefont {J.~I.}\ \bibnamefont
  {Cirac}},\ and\ \bibinfo {author} {\bibfnamefont {M.~B.}\ \bibnamefont
  {Hastings}},\ }\href {https://doi.org/10.1103/PhysRevLett.106.050405}
  {\bibfield  {journal} {\bibinfo  {journal} {Phys. Rev. Lett.}\ }\textbf
  {\bibinfo {volume} {106}},\ \bibinfo {pages} {050405} (\bibinfo {year}
  {2011})}\BibitemShut {NoStop}%
\bibitem [{\citenamefont {Dicke}(1954)}]{PhysRev.93.99}%
  \BibitemOpen
  \bibfield  {author} {\bibinfo {author} {\bibfnamefont {R.~H.}\ \bibnamefont
  {Dicke}},\ }\href {https://doi.org/10.1103/PhysRev.93.99} {\bibfield
  {journal} {\bibinfo  {journal} {Phys. Rev.}\ }\textbf {\bibinfo {volume}
  {93}},\ \bibinfo {pages} {99} (\bibinfo {year} {1954})}\BibitemShut {NoStop}%
\bibitem [{\citenamefont {Alavirad}\ and\ \citenamefont
  {Lavasani}(2019)}]{Alavirad_2019}%
  \BibitemOpen
  \bibfield  {author} {\bibinfo {author} {\bibfnamefont {Y.}~\bibnamefont
  {Alavirad}}\ and\ \bibinfo {author} {\bibfnamefont {A.}~\bibnamefont
  {Lavasani}},\ }\bibfield  {journal} {\bibinfo  {journal} {Physical Review A}\
  }\textbf {\bibinfo {volume} {99}},\ \href
  {https://doi.org/10.1103/physreva.99.043602} {10.1103/physreva.99.043602}
  (\bibinfo {year} {2019})\BibitemShut {NoStop}%
\bibitem [{\citenamefont {Lewis-Swan}\ \emph
  {et~al.}(2019{\natexlab{b}})\citenamefont {Lewis-Swan}, \citenamefont
  {Safavi-Naini}, \citenamefont {Bollinger},\ and\ \citenamefont
  {Rey}}]{Lewis_Swan_2019}%
  \BibitemOpen
  \bibfield  {author} {\bibinfo {author} {\bibfnamefont {R.~J.}\ \bibnamefont
  {Lewis-Swan}}, \bibinfo {author} {\bibfnamefont {A.}~\bibnamefont
  {Safavi-Naini}}, \bibinfo {author} {\bibfnamefont {J.~J.}\ \bibnamefont
  {Bollinger}},\ and\ \bibinfo {author} {\bibfnamefont {A.~M.}\ \bibnamefont
  {Rey}},\ }\bibfield  {journal} {\bibinfo  {journal} {Nature Communications}\
  }\textbf {\bibinfo {volume} {10}},\ \href
  {https://doi.org/10.1038/s41467-019-09436-y} {10.1038/s41467-019-09436-y}
  (\bibinfo {year} {2019}{\natexlab{b}})\BibitemShut {NoStop}%
\bibitem [{\citenamefont {Bentsen}\ \emph {et~al.}(2019)\citenamefont
  {Bentsen}, \citenamefont {Hashizume}, \citenamefont {Buyskikh}, \citenamefont
  {Davis}, \citenamefont {Daley}, \citenamefont {Gubser},\ and\ \citenamefont
  {Schleier-Smith}}]{PhysRevLett.123.130601}%
  \BibitemOpen
  \bibfield  {author} {\bibinfo {author} {\bibfnamefont {G.}~\bibnamefont
  {Bentsen}}, \bibinfo {author} {\bibfnamefont {T.}~\bibnamefont {Hashizume}},
  \bibinfo {author} {\bibfnamefont {A.~S.}\ \bibnamefont {Buyskikh}}, \bibinfo
  {author} {\bibfnamefont {E.~J.}\ \bibnamefont {Davis}}, \bibinfo {author}
  {\bibfnamefont {A.~J.}\ \bibnamefont {Daley}}, \bibinfo {author}
  {\bibfnamefont {S.~S.}\ \bibnamefont {Gubser}},\ and\ \bibinfo {author}
  {\bibfnamefont {M.}~\bibnamefont {Schleier-Smith}},\ }\href
  {https://doi.org/10.1103/PhysRevLett.123.130601} {\bibfield  {journal}
  {\bibinfo  {journal} {Phys. Rev. Lett.}\ }\textbf {\bibinfo {volume} {123}},\
  \bibinfo {pages} {130601} (\bibinfo {year} {2019})}\BibitemShut {NoStop}%
\bibitem [{\citenamefont {Belyansky}\ \emph {et~al.}(2020)\citenamefont
  {Belyansky}, \citenamefont {Bienias}, \citenamefont {Kharkov}, \citenamefont
  {Gorshkov},\ and\ \citenamefont {Swingle}}]{Belyansky_2020}%
  \BibitemOpen
  \bibfield  {author} {\bibinfo {author} {\bibfnamefont {R.}~\bibnamefont
  {Belyansky}}, \bibinfo {author} {\bibfnamefont {P.}~\bibnamefont {Bienias}},
  \bibinfo {author} {\bibfnamefont {Y.~A.}\ \bibnamefont {Kharkov}}, \bibinfo
  {author} {\bibfnamefont {A.~V.}\ \bibnamefont {Gorshkov}},\ and\ \bibinfo
  {author} {\bibfnamefont {B.}~\bibnamefont {Swingle}},\ }\bibfield  {journal}
  {\bibinfo  {journal} {Physical Review Letters}\ }\textbf {\bibinfo {volume}
  {125}},\ \href {https://doi.org/10.1103/physrevlett.125.130601}
  {10.1103/physrevlett.125.130601} (\bibinfo {year} {2020})\BibitemShut
  {NoStop}%
\bibitem [{\citenamefont {Huang}\ \emph {et~al.}(2016)\citenamefont {Huang},
  \citenamefont {Zhang},\ and\ \citenamefont {Chen}}]{Huang_2016}%
  \BibitemOpen
  \bibfield  {author} {\bibinfo {author} {\bibfnamefont {Y.}~\bibnamefont
  {Huang}}, \bibinfo {author} {\bibfnamefont {Y.-L.}\ \bibnamefont {Zhang}},\
  and\ \bibinfo {author} {\bibfnamefont {X.}~\bibnamefont {Chen}},\ }\href
  {https://doi.org/10.1002/andp.201600318} {\bibfield  {journal} {\bibinfo
  {journal} {Annalen der Physik}\ }\textbf {\bibinfo {volume} {529}},\ \bibinfo
  {pages} {1600318} (\bibinfo {year} {2016})}\BibitemShut {NoStop}%
\bibitem [{\citenamefont {Fan}\ \emph {et~al.}(2017)\citenamefont {Fan},
  \citenamefont {Zhang}, \citenamefont {Shen},\ and\ \citenamefont
  {Zhai}}]{Fan_2017}%
  \BibitemOpen
  \bibfield  {author} {\bibinfo {author} {\bibfnamefont {R.}~\bibnamefont
  {Fan}}, \bibinfo {author} {\bibfnamefont {P.}~\bibnamefont {Zhang}}, \bibinfo
  {author} {\bibfnamefont {H.}~\bibnamefont {Shen}},\ and\ \bibinfo {author}
  {\bibfnamefont {H.}~\bibnamefont {Zhai}},\ }\href
  {https://doi.org/10.1016/j.scib.2017.04.011} {\bibfield  {journal} {\bibinfo
  {journal} {Science Bulletin}\ }\textbf {\bibinfo {volume} {62}},\ \bibinfo
  {pages} {707–711} (\bibinfo {year} {2017})}\BibitemShut {NoStop}%
\bibitem [{\citenamefont {Chen}(2016)}]{Chen2016Logarithmic}%
  \BibitemOpen
  \bibfield  {author} {\bibinfo {author} {\bibfnamefont {Y.}~\bibnamefont
  {Chen}},\ }\href@noop {} {\bibinfo {title} {Universal logarithmic scrambling
  in many body localization}} (\bibinfo {year} {2016}),\ \Eprint
  {https://arxiv.org/abs/1608.02765} {arXiv:1608.02765 [quant-ph]} \BibitemShut
  {NoStop}%
\bibitem [{\citenamefont {Chen}\ \emph {et~al.}(2016)\citenamefont {Chen},
  \citenamefont {Zhou}, \citenamefont {Huse},\ and\ \citenamefont
  {Fradkin}}]{Chen_2016}%
  \BibitemOpen
  \bibfield  {author} {\bibinfo {author} {\bibfnamefont {X.}~\bibnamefont
  {Chen}}, \bibinfo {author} {\bibfnamefont {T.}~\bibnamefont {Zhou}}, \bibinfo
  {author} {\bibfnamefont {D.~A.}\ \bibnamefont {Huse}},\ and\ \bibinfo
  {author} {\bibfnamefont {E.}~\bibnamefont {Fradkin}},\ }\href
  {https://doi.org/10.1002/andp.201600332} {\bibfield  {journal} {\bibinfo
  {journal} {Annalen der Physik}\ }\textbf {\bibinfo {volume} {529}},\ \bibinfo
  {pages} {1600332} (\bibinfo {year} {2016})}\BibitemShut {NoStop}%
\bibitem [{\citenamefont {He}\ and\ \citenamefont
  {Lu}(2017)}]{Rong-Qiang2017localization}%
  \BibitemOpen
  \bibfield  {author} {\bibinfo {author} {\bibfnamefont {R.-Q.}\ \bibnamefont
  {He}}\ and\ \bibinfo {author} {\bibfnamefont {Z.-Y.}\ \bibnamefont {Lu}},\
  }\href {https://doi.org/10.1103/PhysRevB.95.054201} {\bibfield  {journal}
  {\bibinfo  {journal} {Phys. Rev. B}\ }\textbf {\bibinfo {volume} {95}},\
  \bibinfo {pages} {054201} (\bibinfo {year} {2017})}\BibitemShut {NoStop}%
\bibitem [{\citenamefont {Swingle}\ and\ \citenamefont
  {Chowdhury}(2017{\natexlab{b}})}]{Swingle2017Slow}%
  \BibitemOpen
  \bibfield  {author} {\bibinfo {author} {\bibfnamefont {B.}~\bibnamefont
  {Swingle}}\ and\ \bibinfo {author} {\bibfnamefont {D.}~\bibnamefont
  {Chowdhury}},\ }\href {https://doi.org/10.1103/PhysRevB.95.060201} {\bibfield
   {journal} {\bibinfo  {journal} {Phys. Rev. B}\ }\textbf {\bibinfo {volume}
  {95}},\ \bibinfo {pages} {060201} (\bibinfo {year}
  {2017}{\natexlab{b}})}\BibitemShut {NoStop}%
\bibitem [{\citenamefont {Gärttner}\ \emph {et~al.}(2017)\citenamefont
  {Gärttner}, \citenamefont {Bohnet}, \citenamefont {Safavi-Naini},
  \citenamefont {Wall}, \citenamefont {Bollinger},\ and\ \citenamefont
  {Rey}}]{G_rttner_2017}%
  \BibitemOpen
  \bibfield  {author} {\bibinfo {author} {\bibfnamefont {M.}~\bibnamefont
  {Gärttner}}, \bibinfo {author} {\bibfnamefont {J.~G.}\ \bibnamefont
  {Bohnet}}, \bibinfo {author} {\bibfnamefont {A.}~\bibnamefont
  {Safavi-Naini}}, \bibinfo {author} {\bibfnamefont {M.~L.}\ \bibnamefont
  {Wall}}, \bibinfo {author} {\bibfnamefont {J.~J.}\ \bibnamefont
  {Bollinger}},\ and\ \bibinfo {author} {\bibfnamefont {A.~M.}\ \bibnamefont
  {Rey}},\ }\href {https://doi.org/10.1038/nphys4119} {\bibfield  {journal}
  {\bibinfo  {journal} {Nature Physics}\ }\textbf {\bibinfo {volume} {13}},\
  \bibinfo {pages} {781–786} (\bibinfo {year} {2017})}\BibitemShut {NoStop}%
\bibitem [{\citenamefont {Wei}\ \emph {et~al.}(2018)\citenamefont {Wei},
  \citenamefont {Ramanathan},\ and\ \citenamefont
  {Cappellaro}}]{Xuan2018Localization}%
  \BibitemOpen
  \bibfield  {author} {\bibinfo {author} {\bibfnamefont {K.~X.}\ \bibnamefont
  {Wei}}, \bibinfo {author} {\bibfnamefont {C.}~\bibnamefont {Ramanathan}},\
  and\ \bibinfo {author} {\bibfnamefont {P.}~\bibnamefont {Cappellaro}},\
  }\href {https://doi.org/10.1103/PhysRevLett.120.070501} {\bibfield  {journal}
  {\bibinfo  {journal} {Phys. Rev. Lett.}\ }\textbf {\bibinfo {volume} {120}},\
  \bibinfo {pages} {070501} (\bibinfo {year} {2018})}\BibitemShut {NoStop}%
\bibitem [{\citenamefont {Li}\ \emph {et~al.}(2017)\citenamefont {Li},
  \citenamefont {Fan}, \citenamefont {Wang}, \citenamefont {Ye}, \citenamefont
  {Zeng}, \citenamefont {Zhai}, \citenamefont {Peng},\ and\ \citenamefont
  {Du}}]{PhysRevX.7.031011}%
  \BibitemOpen
  \bibfield  {author} {\bibinfo {author} {\bibfnamefont {J.}~\bibnamefont
  {Li}}, \bibinfo {author} {\bibfnamefont {R.}~\bibnamefont {Fan}}, \bibinfo
  {author} {\bibfnamefont {H.}~\bibnamefont {Wang}}, \bibinfo {author}
  {\bibfnamefont {B.}~\bibnamefont {Ye}}, \bibinfo {author} {\bibfnamefont
  {B.}~\bibnamefont {Zeng}}, \bibinfo {author} {\bibfnamefont {H.}~\bibnamefont
  {Zhai}}, \bibinfo {author} {\bibfnamefont {X.}~\bibnamefont {Peng}},\ and\
  \bibinfo {author} {\bibfnamefont {J.}~\bibnamefont {Du}},\ }\href
  {https://doi.org/10.1103/PhysRevX.7.031011} {\bibfield  {journal} {\bibinfo
  {journal} {Phys. Rev. X}\ }\textbf {\bibinfo {volume} {7}},\ \bibinfo {pages}
  {031011} (\bibinfo {year} {2017})}\BibitemShut {NoStop}%
\bibitem [{\citenamefont {Yoshida}\ and\ \citenamefont
  {Yao}(2019)}]{Yoshida_2019}%
  \BibitemOpen
  \bibfield  {author} {\bibinfo {author} {\bibfnamefont {B.}~\bibnamefont
  {Yoshida}}\ and\ \bibinfo {author} {\bibfnamefont {N.~Y.}\ \bibnamefont
  {Yao}},\ }\bibfield  {journal} {\bibinfo  {journal} {Physical Review X}\
  }\textbf {\bibinfo {volume} {9}},\ \href
  {https://doi.org/10.1103/physrevx.9.011006} {10.1103/physrevx.9.011006}
  (\bibinfo {year} {2019})\BibitemShut {NoStop}%
\bibitem [{\citenamefont {Landsman}\ \emph {et~al.}(2019)\citenamefont
  {Landsman}, \citenamefont {Figgatt}, \citenamefont {Schuster}, \citenamefont
  {Linke}, \citenamefont {Yoshida}, \citenamefont {Yao},\ and\ \citenamefont
  {Monroe}}]{Landsman_2019}%
  \BibitemOpen
  \bibfield  {author} {\bibinfo {author} {\bibfnamefont {K.~A.}\ \bibnamefont
  {Landsman}}, \bibinfo {author} {\bibfnamefont {C.}~\bibnamefont {Figgatt}},
  \bibinfo {author} {\bibfnamefont {T.}~\bibnamefont {Schuster}}, \bibinfo
  {author} {\bibfnamefont {N.~M.}\ \bibnamefont {Linke}}, \bibinfo {author}
  {\bibfnamefont {B.}~\bibnamefont {Yoshida}}, \bibinfo {author} {\bibfnamefont
  {N.~Y.}\ \bibnamefont {Yao}},\ and\ \bibinfo {author} {\bibfnamefont
  {C.}~\bibnamefont {Monroe}},\ }\href
  {https://doi.org/10.1038/s41586-019-0952-6} {\bibfield  {journal} {\bibinfo
  {journal} {Nature}\ }\textbf {\bibinfo {volume} {567}},\ \bibinfo {pages}
  {61–65} (\bibinfo {year} {2019})}\BibitemShut {NoStop}%
\bibitem [{\citenamefont {Vermersch}\ \emph {et~al.}(2019)\citenamefont
  {Vermersch}, \citenamefont {Elben}, \citenamefont {Sieberer}, \citenamefont
  {Yao},\ and\ \citenamefont {Zoller}}]{Vermersch_2019}%
  \BibitemOpen
  \bibfield  {author} {\bibinfo {author} {\bibfnamefont {B.}~\bibnamefont
  {Vermersch}}, \bibinfo {author} {\bibfnamefont {A.}~\bibnamefont {Elben}},
  \bibinfo {author} {\bibfnamefont {L.}~\bibnamefont {Sieberer}}, \bibinfo
  {author} {\bibfnamefont {N.}~\bibnamefont {Yao}},\ and\ \bibinfo {author}
  {\bibfnamefont {P.}~\bibnamefont {Zoller}},\ }\bibfield  {journal} {\bibinfo
  {journal} {Physical Review X}\ }\textbf {\bibinfo {volume} {9}},\ \href
  {https://doi.org/10.1103/physrevx.9.021061} {10.1103/physrevx.9.021061}
  (\bibinfo {year} {2019})\BibitemShut {NoStop}%
\bibitem [{\citenamefont {Joshi}\ \emph {et~al.}(2020)\citenamefont {Joshi},
  \citenamefont {Elben}, \citenamefont {Vermersch}, \citenamefont {Brydges},
  \citenamefont {Maier}, \citenamefont {Zoller}, \citenamefont {Blatt},\ and\
  \citenamefont {Roos}}]{Trapped2020}%
  \BibitemOpen
  \bibfield  {author} {\bibinfo {author} {\bibfnamefont {M.~K.}\ \bibnamefont
  {Joshi}}, \bibinfo {author} {\bibfnamefont {A.}~\bibnamefont {Elben}},
  \bibinfo {author} {\bibfnamefont {B.}~\bibnamefont {Vermersch}}, \bibinfo
  {author} {\bibfnamefont {T.}~\bibnamefont {Brydges}}, \bibinfo {author}
  {\bibfnamefont {C.}~\bibnamefont {Maier}}, \bibinfo {author} {\bibfnamefont
  {P.}~\bibnamefont {Zoller}}, \bibinfo {author} {\bibfnamefont
  {R.}~\bibnamefont {Blatt}},\ and\ \bibinfo {author} {\bibfnamefont {C.~F.}\
  \bibnamefont {Roos}},\ }\href
  {https://doi.org/10.1103/PhysRevLett.124.240505} {\bibfield  {journal}
  {\bibinfo  {journal} {Phys. Rev. Lett.}\ }\textbf {\bibinfo {volume} {124}},\
  \bibinfo {pages} {240505} (\bibinfo {year} {2020})}\BibitemShut {NoStop}%
\bibitem [{\citenamefont {Shenker}\ and\ \citenamefont
  {Stanford}(2014{\natexlab{b}})}]{Shenker_2014}%
  \BibitemOpen
  \bibfield  {author} {\bibinfo {author} {\bibfnamefont {S.~H.}\ \bibnamefont
  {Shenker}}\ and\ \bibinfo {author} {\bibfnamefont {D.}~\bibnamefont
  {Stanford}},\ }\bibfield  {journal} {\bibinfo  {journal} {Journal of High
  Energy Physics}\ }\textbf {\bibinfo {volume} {2014}},\ \href
  {https://doi.org/10.1007/jhep12(2014)046} {10.1007/jhep12(2014)046} (\bibinfo
  {year} {2014}{\natexlab{b}})\BibitemShut {NoStop}%
\bibitem [{\citenamefont {Roberts}\ and\ \citenamefont
  {Yoshida}(2017)}]{Roberts_2017}%
  \BibitemOpen
  \bibfield  {author} {\bibinfo {author} {\bibfnamefont {D.~A.}\ \bibnamefont
  {Roberts}}\ and\ \bibinfo {author} {\bibfnamefont {B.}~\bibnamefont
  {Yoshida}},\ }\bibfield  {journal} {\bibinfo  {journal} {Journal of High
  Energy Physics}\ }\textbf {\bibinfo {volume} {2017}},\ \href
  {https://doi.org/10.1007/jhep04(2017)121} {10.1007/jhep04(2017)121} (\bibinfo
  {year} {2017})\BibitemShut {NoStop}%
\bibitem [{\citenamefont {Aaronson}(2018)}]{aaronson2018shadow}%
  \BibitemOpen
  \bibfield  {author} {\bibinfo {author} {\bibfnamefont {S.}~\bibnamefont
  {Aaronson}},\ }\href@noop {} {\bibinfo {title} {Shadow tomography of quantum
  states}} (\bibinfo {year} {2018}),\ \Eprint
  {https://arxiv.org/abs/1711.01053} {arXiv:1711.01053 [quant-ph]} \BibitemShut
  {NoStop}%
\bibitem [{\citenamefont {Huang}\ \emph {et~al.}(2020)\citenamefont {Huang},
  \citenamefont {Kueng},\ and\ \citenamefont {Preskill}}]{Huang_2020}%
  \BibitemOpen
  \bibfield  {author} {\bibinfo {author} {\bibfnamefont {H.-Y.}\ \bibnamefont
  {Huang}}, \bibinfo {author} {\bibfnamefont {R.}~\bibnamefont {Kueng}},\ and\
  \bibinfo {author} {\bibfnamefont {J.}~\bibnamefont {Preskill}},\ }\bibfield
  {journal} {\bibinfo  {journal} {Nature Physics}\ }\href
  {https://doi.org/10.1038/s41567-020-0932-7} {10.1038/s41567-020-0932-7}
  (\bibinfo {year} {2020})\BibitemShut {NoStop}%
\bibitem [{\citenamefont {Ekert}\ \emph {et~al.}(2002)\citenamefont {Ekert},
  \citenamefont {Alves}, \citenamefont {Oi}, \citenamefont {Horodecki},
  \citenamefont {Horodecki},\ and\ \citenamefont {Kwek}}]{Ekert2002Direct}%
  \BibitemOpen
  \bibfield  {author} {\bibinfo {author} {\bibfnamefont {A.~K.}\ \bibnamefont
  {Ekert}}, \bibinfo {author} {\bibfnamefont {C.~M.}\ \bibnamefont {Alves}},
  \bibinfo {author} {\bibfnamefont {D.~K.~L.}\ \bibnamefont {Oi}}, \bibinfo
  {author} {\bibfnamefont {M.}~\bibnamefont {Horodecki}}, \bibinfo {author}
  {\bibfnamefont {P.}~\bibnamefont {Horodecki}},\ and\ \bibinfo {author}
  {\bibfnamefont {L.~C.}\ \bibnamefont {Kwek}},\ }\href
  {https://doi.org/10.1103/PhysRevLett.88.217901} {\bibfield  {journal}
  {\bibinfo  {journal} {Phys. Rev. Lett.}\ }\textbf {\bibinfo {volume} {88}},\
  \bibinfo {pages} {217901} (\bibinfo {year} {2002})}\BibitemShut {NoStop}%
\bibitem [{\citenamefont {Daley}\ \emph {et~al.}(2012)\citenamefont {Daley},
  \citenamefont {Pichler}, \citenamefont {Schachenmayer},\ and\ \citenamefont
  {Zoller}}]{Daley2012Measuring}%
  \BibitemOpen
  \bibfield  {author} {\bibinfo {author} {\bibfnamefont {A.~J.}\ \bibnamefont
  {Daley}}, \bibinfo {author} {\bibfnamefont {H.}~\bibnamefont {Pichler}},
  \bibinfo {author} {\bibfnamefont {J.}~\bibnamefont {Schachenmayer}},\ and\
  \bibinfo {author} {\bibfnamefont {P.}~\bibnamefont {Zoller}},\ }\href
  {https://doi.org/10.1103/PhysRevLett.109.020505} {\bibfield  {journal}
  {\bibinfo  {journal} {Phys. Rev. Lett.}\ }\textbf {\bibinfo {volume} {109}},\
  \bibinfo {pages} {020505} (\bibinfo {year} {2012})}\BibitemShut {NoStop}%
\bibitem [{\citenamefont {Abanin}\ and\ \citenamefont
  {Demler}(2012)}]{Abanin2012Measuring}%
  \BibitemOpen
  \bibfield  {author} {\bibinfo {author} {\bibfnamefont {D.~A.}\ \bibnamefont
  {Abanin}}\ and\ \bibinfo {author} {\bibfnamefont {E.}~\bibnamefont
  {Demler}},\ }\href {https://doi.org/10.1103/PhysRevLett.109.020504}
  {\bibfield  {journal} {\bibinfo  {journal} {Phys. Rev. Lett.}\ }\textbf
  {\bibinfo {volume} {109}},\ \bibinfo {pages} {020504} (\bibinfo {year}
  {2012})}\BibitemShut {NoStop}%
\bibitem [{\citenamefont {Islam}\ \emph {et~al.}(2015)\citenamefont {Islam},
  \citenamefont {Ma}, \citenamefont {Preiss}, \citenamefont {Eric~Tai},
  \citenamefont {Lukin}, \citenamefont {Rispoli},\ and\ \citenamefont
  {Greiner}}]{Islam2015}%
  \BibitemOpen
  \bibfield  {author} {\bibinfo {author} {\bibfnamefont {R.}~\bibnamefont
  {Islam}}, \bibinfo {author} {\bibfnamefont {R.}~\bibnamefont {Ma}}, \bibinfo
  {author} {\bibfnamefont {P.~M.}\ \bibnamefont {Preiss}}, \bibinfo {author}
  {\bibfnamefont {M.}~\bibnamefont {Eric~Tai}}, \bibinfo {author}
  {\bibfnamefont {A.}~\bibnamefont {Lukin}}, \bibinfo {author} {\bibfnamefont
  {M.}~\bibnamefont {Rispoli}},\ and\ \bibinfo {author} {\bibfnamefont
  {M.}~\bibnamefont {Greiner}},\ }\href {https://doi.org/10.1038/nature15750}
  {\bibfield  {journal} {\bibinfo  {journal} {Nature}\ }\textbf {\bibinfo
  {volume} {528}},\ \bibinfo {pages} {77} (\bibinfo {year} {2015})}\BibitemShut
  {NoStop}%
\bibitem [{\citenamefont {Kaufman}\ \emph {et~al.}(2016)\citenamefont
  {Kaufman}, \citenamefont {Tai}, \citenamefont {Lukin}, \citenamefont
  {Rispoli}, \citenamefont {Schittko}, \citenamefont {Preiss},\ and\
  \citenamefont {Greiner}}]{Kaufmanen2016tanglement}%
  \BibitemOpen
  \bibfield  {author} {\bibinfo {author} {\bibfnamefont {A.~M.}\ \bibnamefont
  {Kaufman}}, \bibinfo {author} {\bibfnamefont {M.~E.}\ \bibnamefont {Tai}},
  \bibinfo {author} {\bibfnamefont {A.}~\bibnamefont {Lukin}}, \bibinfo
  {author} {\bibfnamefont {M.}~\bibnamefont {Rispoli}}, \bibinfo {author}
  {\bibfnamefont {R.}~\bibnamefont {Schittko}}, \bibinfo {author}
  {\bibfnamefont {P.~M.}\ \bibnamefont {Preiss}},\ and\ \bibinfo {author}
  {\bibfnamefont {M.}~\bibnamefont {Greiner}},\ }\href
  {https://doi.org/10.1126/science.aaf6725} {\bibfield  {journal} {\bibinfo
  {journal} {Science}\ }\textbf {\bibinfo {volume} {353}},\ \bibinfo {pages}
  {794} (\bibinfo {year} {2016})},\ \Eprint
  {https://arxiv.org/abs/https://science.sciencemag.org/content/353/6301/794.full.pdf}
  {https://science.sciencemag.org/content/353/6301/794.full.pdf} \BibitemShut
  {NoStop}%
\bibitem [{\citenamefont {Yao}\ \emph {et~al.}(2016)\citenamefont {Yao},
  \citenamefont {Grusdt}, \citenamefont {Swingle}, \citenamefont {Lukin},
  \citenamefont {Stamper-Kurn}, \citenamefont {Moore},\ and\ \citenamefont
  {Demler}}]{yao2016interferometric}%
  \BibitemOpen
  \bibfield  {author} {\bibinfo {author} {\bibfnamefont {N.~Y.}\ \bibnamefont
  {Yao}}, \bibinfo {author} {\bibfnamefont {F.}~\bibnamefont {Grusdt}},
  \bibinfo {author} {\bibfnamefont {B.}~\bibnamefont {Swingle}}, \bibinfo
  {author} {\bibfnamefont {M.~D.}\ \bibnamefont {Lukin}}, \bibinfo {author}
  {\bibfnamefont {D.~M.}\ \bibnamefont {Stamper-Kurn}}, \bibinfo {author}
  {\bibfnamefont {J.~E.}\ \bibnamefont {Moore}},\ and\ \bibinfo {author}
  {\bibfnamefont {E.~A.}\ \bibnamefont {Demler}},\ }\href@noop {} {\bibinfo
  {title} {Interferometric approach to probing fast scrambling}} (\bibinfo
  {year} {2016}),\ \Eprint {https://arxiv.org/abs/1607.01801} {arXiv:1607.01801
  [quant-ph]} \BibitemShut {NoStop}%
\bibitem [{\citenamefont {Elben}\ \emph
  {et~al.}(2020{\natexlab{a}})\citenamefont {Elben}, \citenamefont {Kueng},
  \citenamefont {Huang}, \citenamefont {van Bijnen}, \citenamefont {Kokail},
  \citenamefont {Dalmonte}, \citenamefont {Calabrese}, \citenamefont {Kraus},
  \citenamefont {Preskill}, \citenamefont {Zoller},\ and\ \citenamefont
  {Vermersch}}]{elben2020mixedstate}%
  \BibitemOpen
  \bibfield  {author} {\bibinfo {author} {\bibfnamefont {A.}~\bibnamefont
  {Elben}}, \bibinfo {author} {\bibfnamefont {R.}~\bibnamefont {Kueng}},
  \bibinfo {author} {\bibfnamefont {H.-Y.~R.}\ \bibnamefont {Huang}}, \bibinfo
  {author} {\bibfnamefont {R.}~\bibnamefont {van Bijnen}}, \bibinfo {author}
  {\bibfnamefont {C.}~\bibnamefont {Kokail}}, \bibinfo {author} {\bibfnamefont
  {M.}~\bibnamefont {Dalmonte}}, \bibinfo {author} {\bibfnamefont
  {P.}~\bibnamefont {Calabrese}}, \bibinfo {author} {\bibfnamefont
  {B.}~\bibnamefont {Kraus}}, \bibinfo {author} {\bibfnamefont
  {J.}~\bibnamefont {Preskill}}, \bibinfo {author} {\bibfnamefont
  {P.}~\bibnamefont {Zoller}},\ and\ \bibinfo {author} {\bibfnamefont
  {B.}~\bibnamefont {Vermersch}},\ }\href
  {https://doi.org/10.1103/PhysRevLett.125.200501} {\bibfield  {journal}
  {\bibinfo  {journal} {Phys. Rev. Lett.}\ }\textbf {\bibinfo {volume} {125}},\
  \bibinfo {pages} {200501} (\bibinfo {year} {2020}{\natexlab{a}})}\BibitemShut
  {NoStop}%
\bibitem [{\citenamefont {Karkuszewski}\ \emph {et~al.}(2002)\citenamefont
  {Karkuszewski}, \citenamefont {Jarzynski},\ and\ \citenamefont
  {Zurek}}]{PhysRevLett.89.170405}%
  \BibitemOpen
  \bibfield  {author} {\bibinfo {author} {\bibfnamefont {Z.~P.}\ \bibnamefont
  {Karkuszewski}}, \bibinfo {author} {\bibfnamefont {C.}~\bibnamefont
  {Jarzynski}},\ and\ \bibinfo {author} {\bibfnamefont {W.~H.}\ \bibnamefont
  {Zurek}},\ }\href {https://doi.org/10.1103/PhysRevLett.89.170405} {\bibfield
  {journal} {\bibinfo  {journal} {Phys. Rev. Lett.}\ }\textbf {\bibinfo
  {volume} {89}},\ \bibinfo {pages} {170405} (\bibinfo {year}
  {2002})}\BibitemShut {NoStop}%
\bibitem [{\citenamefont {Lieb}\ and\ \citenamefont
  {Robinson}(1972)}]{lieb1972}%
  \BibitemOpen
  \bibfield  {author} {\bibinfo {author} {\bibfnamefont {E.~H.}\ \bibnamefont
  {Lieb}}\ and\ \bibinfo {author} {\bibfnamefont {D.~W.}\ \bibnamefont
  {Robinson}},\ }\href {https://projecteuclid.org:443/euclid.cmp/1103858407}
  {\bibfield  {journal} {\bibinfo  {journal} {Comm. Math. Phys.}\ }\textbf
  {\bibinfo {volume} {28}},\ \bibinfo {pages} {251} (\bibinfo {year}
  {1972})}\BibitemShut {NoStop}%
\bibitem [{\citenamefont {Bhattacharya}\ \emph {et~al.}(2018)\citenamefont
  {Bhattacharya}, \citenamefont {Jatkar},\ and\ \citenamefont
  {Kundu}}]{bhattacharya2018chaotic}%
  \BibitemOpen
  \bibfield  {author} {\bibinfo {author} {\bibfnamefont {R.}~\bibnamefont
  {Bhattacharya}}, \bibinfo {author} {\bibfnamefont {D.~P.}\ \bibnamefont
  {Jatkar}},\ and\ \bibinfo {author} {\bibfnamefont {A.}~\bibnamefont
  {Kundu}},\ }\href@noop {} {\bibinfo {title} {Chaotic correlation functions
  with complex fermions}} (\bibinfo {year} {2018}),\ \Eprint
  {https://arxiv.org/abs/1810.13217} {arXiv:1810.13217 [hep-th]} \BibitemShut
  {NoStop}%
\bibitem [{\citenamefont {van Enk}\ and\ \citenamefont
  {Beenakker}(2012)}]{van2012Measuring}%
  \BibitemOpen
  \bibfield  {author} {\bibinfo {author} {\bibfnamefont {S.~J.}\ \bibnamefont
  {van Enk}}\ and\ \bibinfo {author} {\bibfnamefont {C.~W.~J.}\ \bibnamefont
  {Beenakker}},\ }\href {https://doi.org/10.1103/PhysRevLett.108.110503}
  {\bibfield  {journal} {\bibinfo  {journal} {Phys. Rev. Lett.}\ }\textbf
  {\bibinfo {volume} {108}},\ \bibinfo {pages} {110503} (\bibinfo {year}
  {2012})}\BibitemShut {NoStop}%
\bibitem [{\citenamefont {Elben}\ \emph
  {et~al.}(2019{\natexlab{a}})\citenamefont {Elben}, \citenamefont {Vermersch},
  \citenamefont {Roos},\ and\ \citenamefont {Zoller}}]{Elben2019toolbox}%
  \BibitemOpen
  \bibfield  {author} {\bibinfo {author} {\bibfnamefont {A.}~\bibnamefont
  {Elben}}, \bibinfo {author} {\bibfnamefont {B.}~\bibnamefont {Vermersch}},
  \bibinfo {author} {\bibfnamefont {C.~F.}\ \bibnamefont {Roos}},\ and\
  \bibinfo {author} {\bibfnamefont {P.}~\bibnamefont {Zoller}},\ }\href
  {https://doi.org/10.1103/PhysRevA.99.052323} {\bibfield  {journal} {\bibinfo
  {journal} {Phys. Rev. A}\ }\textbf {\bibinfo {volume} {99}},\ \bibinfo
  {pages} {052323} (\bibinfo {year} {2019}{\natexlab{a}})}\BibitemShut
  {NoStop}%
\bibitem [{\citenamefont {Elben}\ \emph
  {et~al.}(2020{\natexlab{b}})\citenamefont {Elben}, \citenamefont {Yu},
  \citenamefont {Zhu}, \citenamefont {Hafezi}, \citenamefont {Pollmann},
  \citenamefont {Zoller},\ and\ \citenamefont
  {Vermersch}}]{Elben2020topological}%
  \BibitemOpen
  \bibfield  {author} {\bibinfo {author} {\bibfnamefont {A.}~\bibnamefont
  {Elben}}, \bibinfo {author} {\bibfnamefont {J.}~\bibnamefont {Yu}}, \bibinfo
  {author} {\bibfnamefont {G.}~\bibnamefont {Zhu}}, \bibinfo {author}
  {\bibfnamefont {M.}~\bibnamefont {Hafezi}}, \bibinfo {author} {\bibfnamefont
  {F.}~\bibnamefont {Pollmann}}, \bibinfo {author} {\bibfnamefont
  {P.}~\bibnamefont {Zoller}},\ and\ \bibinfo {author} {\bibfnamefont
  {B.}~\bibnamefont {Vermersch}},\ }\bibfield  {journal} {\bibinfo  {journal}
  {Science Advances}\ }\textbf {\bibinfo {volume} {6}},\ \href
  {https://doi.org/10.1126/sciadv.aaz3666} {10.1126/sciadv.aaz3666} (\bibinfo
  {year} {2020}{\natexlab{b}}),\ \Eprint
  {https://arxiv.org/abs/https://advances.sciencemag.org/content/6/15/eaaz3666.full.pdf}
  {https://advances.sciencemag.org/content/6/15/eaaz3666.full.pdf} \BibitemShut
  {NoStop}%
\bibitem [{\citenamefont {Cian}\ \emph {et~al.}(2020)\citenamefont {Cian},
  \citenamefont {Dehghani}, \citenamefont {Andreas~Elben}, \citenamefont {Zhu},
  \citenamefont {Barkeshli}, \citenamefont {Zoller},\ and\ \citenamefont
  {Hafezi}}]{Cian2020Chern}%
  \BibitemOpen
  \bibfield  {author} {\bibinfo {author} {\bibfnamefont {Z.-P.}\ \bibnamefont
  {Cian}}, \bibinfo {author} {\bibfnamefont {H.}~\bibnamefont {Dehghani}},
  \bibinfo {author} {\bibfnamefont {B.~V.}\ \bibnamefont {Andreas~Elben}},
  \bibinfo {author} {\bibfnamefont {G.}~\bibnamefont {Zhu}}, \bibinfo {author}
  {\bibfnamefont {M.}~\bibnamefont {Barkeshli}}, \bibinfo {author}
  {\bibfnamefont {P.}~\bibnamefont {Zoller}},\ and\ \bibinfo {author}
  {\bibfnamefont {M.}~\bibnamefont {Hafezi}},\ }\href@noop {} {\bibinfo {title}
  {Many-body chern number from statistical correlations of randomized
  measurements}} (\bibinfo {year} {2020}),\ \Eprint
  {https://arxiv.org/abs/2005.13543} {arXiv:2005.13543 [quant-ph]} \BibitemShut
  {NoStop}%
\bibitem [{\citenamefont {Elben}\ \emph {et~al.}(2018)\citenamefont {Elben},
  \citenamefont {Vermersch}, \citenamefont {Dalmonte}, \citenamefont {Cirac},\
  and\ \citenamefont {Zoller}}]{Elben2018Random}%
  \BibitemOpen
  \bibfield  {author} {\bibinfo {author} {\bibfnamefont {A.}~\bibnamefont
  {Elben}}, \bibinfo {author} {\bibfnamefont {B.}~\bibnamefont {Vermersch}},
  \bibinfo {author} {\bibfnamefont {M.}~\bibnamefont {Dalmonte}}, \bibinfo
  {author} {\bibfnamefont {J.~I.}\ \bibnamefont {Cirac}},\ and\ \bibinfo
  {author} {\bibfnamefont {P.}~\bibnamefont {Zoller}},\ }\href
  {https://doi.org/10.1103/PhysRevLett.120.050406} {\bibfield  {journal}
  {\bibinfo  {journal} {Phys. Rev. Lett.}\ }\textbf {\bibinfo {volume} {120}},\
  \bibinfo {pages} {050406} (\bibinfo {year} {2018})}\BibitemShut {NoStop}%
\bibitem [{\citenamefont {Brydges}\ \emph {et~al.}(2019)\citenamefont
  {Brydges}, \citenamefont {Elben}, \citenamefont {Jurcevic}, \citenamefont
  {Vermersch}, \citenamefont {Maier}, \citenamefont {Lanyon}, \citenamefont
  {Zoller}, \citenamefont {Blatt},\ and\ \citenamefont
  {Roos}}]{Brydges2019Probing}%
  \BibitemOpen
  \bibfield  {author} {\bibinfo {author} {\bibfnamefont {T.}~\bibnamefont
  {Brydges}}, \bibinfo {author} {\bibfnamefont {A.}~\bibnamefont {Elben}},
  \bibinfo {author} {\bibfnamefont {P.}~\bibnamefont {Jurcevic}}, \bibinfo
  {author} {\bibfnamefont {B.}~\bibnamefont {Vermersch}}, \bibinfo {author}
  {\bibfnamefont {C.}~\bibnamefont {Maier}}, \bibinfo {author} {\bibfnamefont
  {B.~P.}\ \bibnamefont {Lanyon}}, \bibinfo {author} {\bibfnamefont
  {P.}~\bibnamefont {Zoller}}, \bibinfo {author} {\bibfnamefont
  {R.}~\bibnamefont {Blatt}},\ and\ \bibinfo {author} {\bibfnamefont {C.~F.}\
  \bibnamefont {Roos}},\ }\href {https://doi.org/10.1126/science.aau4963}
  {\bibfield  {journal} {\bibinfo  {journal} {Science}\ }\textbf {\bibinfo
  {volume} {364}},\ \bibinfo {pages} {260} (\bibinfo {year} {2019})},\ \Eprint
  {https://arxiv.org/abs/https://science.sciencemag.org/content/364/6437/260.full.pdf}
  {https://science.sciencemag.org/content/364/6437/260.full.pdf} \BibitemShut
  {NoStop}%
\bibitem [{\citenamefont {Zhou}\ \emph {et~al.}(2020)\citenamefont {Zhou},
  \citenamefont {Zeng},\ and\ \citenamefont {Liu}}]{Zhou_2020}%
  \BibitemOpen
  \bibfield  {author} {\bibinfo {author} {\bibfnamefont {Y.}~\bibnamefont
  {Zhou}}, \bibinfo {author} {\bibfnamefont {P.}~\bibnamefont {Zeng}},\ and\
  \bibinfo {author} {\bibfnamefont {Z.}~\bibnamefont {Liu}},\ }\href
  {https://doi.org/10.1103/PhysRevLett.125.200502} {\bibfield  {journal}
  {\bibinfo  {journal} {Phys. Rev. Lett.}\ }\textbf {\bibinfo {volume} {125}},\
  \bibinfo {pages} {200502} (\bibinfo {year} {2020})}\BibitemShut {NoStop}%
\bibitem [{\citenamefont {Elben}\ \emph
  {et~al.}(2020{\natexlab{c}})\citenamefont {Elben}, \citenamefont {Vermersch},
  \citenamefont {van Bijnen}, \citenamefont {Kokail}, \citenamefont {Brydges},
  \citenamefont {Maier}, \citenamefont {Joshi}, \citenamefont {Blatt},
  \citenamefont {Roos},\ and\ \citenamefont {Zoller}}]{Elben2020Cross}%
  \BibitemOpen
  \bibfield  {author} {\bibinfo {author} {\bibfnamefont {A.}~\bibnamefont
  {Elben}}, \bibinfo {author} {\bibfnamefont {B.}~\bibnamefont {Vermersch}},
  \bibinfo {author} {\bibfnamefont {R.}~\bibnamefont {van Bijnen}}, \bibinfo
  {author} {\bibfnamefont {C.}~\bibnamefont {Kokail}}, \bibinfo {author}
  {\bibfnamefont {T.}~\bibnamefont {Brydges}}, \bibinfo {author} {\bibfnamefont
  {C.}~\bibnamefont {Maier}}, \bibinfo {author} {\bibfnamefont {M.~K.}\
  \bibnamefont {Joshi}}, \bibinfo {author} {\bibfnamefont {R.}~\bibnamefont
  {Blatt}}, \bibinfo {author} {\bibfnamefont {C.~F.}\ \bibnamefont {Roos}},\
  and\ \bibinfo {author} {\bibfnamefont {P.}~\bibnamefont {Zoller}},\ }\href
  {https://doi.org/10.1103/PhysRevLett.124.010504} {\bibfield  {journal}
  {\bibinfo  {journal} {Phys. Rev. Lett.}\ }\textbf {\bibinfo {volume} {124}},\
  \bibinfo {pages} {010504} (\bibinfo {year} {2020}{\natexlab{c}})}\BibitemShut
  {NoStop}%
\bibitem [{\citenamefont {Zhang}\ \emph {et~al.}(2020)\citenamefont {Zhang},
  \citenamefont {Zhang}, \citenamefont {Chen}, \citenamefont {Peng},
  \citenamefont {Xu}, \citenamefont {Yin}, \citenamefont {Yu}, \citenamefont
  {Ye}, \citenamefont {Han}, \citenamefont {Xu}, \citenamefont {Chen},
  \citenamefont {Li},\ and\ \citenamefont {Guo}}]{zhang2020experimental}%
  \BibitemOpen
  \bibfield  {author} {\bibinfo {author} {\bibfnamefont {W.-H.}\ \bibnamefont
  {Zhang}}, \bibinfo {author} {\bibfnamefont {C.}~\bibnamefont {Zhang}},
  \bibinfo {author} {\bibfnamefont {Z.}~\bibnamefont {Chen}}, \bibinfo {author}
  {\bibfnamefont {X.-X.}\ \bibnamefont {Peng}}, \bibinfo {author}
  {\bibfnamefont {X.-Y.}\ \bibnamefont {Xu}}, \bibinfo {author} {\bibfnamefont
  {P.}~\bibnamefont {Yin}}, \bibinfo {author} {\bibfnamefont {S.}~\bibnamefont
  {Yu}}, \bibinfo {author} {\bibfnamefont {X.-J.}\ \bibnamefont {Ye}}, \bibinfo
  {author} {\bibfnamefont {Y.-J.}\ \bibnamefont {Han}}, \bibinfo {author}
  {\bibfnamefont {J.-S.}\ \bibnamefont {Xu}}, \bibinfo {author} {\bibfnamefont
  {G.}~\bibnamefont {Chen}}, \bibinfo {author} {\bibfnamefont {C.-F.}\
  \bibnamefont {Li}},\ and\ \bibinfo {author} {\bibfnamefont {G.-C.}\
  \bibnamefont {Guo}},\ }\href {https://doi.org/10.1103/PhysRevLett.125.030506}
  {\bibfield  {journal} {\bibinfo  {journal} {Phys. Rev. Lett.}\ }\textbf
  {\bibinfo {volume} {125}},\ \bibinfo {pages} {030506} (\bibinfo {year}
  {2020})}\BibitemShut {NoStop}%
\bibitem [{\citenamefont {Tran}\ \emph {et~al.}(2015)\citenamefont {Tran},
  \citenamefont {Daki\ifmmode~\acute{c}\else \'{c}\fi{}}, \citenamefont
  {Arnault}, \citenamefont {Laskowski},\ and\ \citenamefont
  {Paterek}}]{Tran2015entanglement}%
  \BibitemOpen
  \bibfield  {author} {\bibinfo {author} {\bibfnamefont {M.~C.}\ \bibnamefont
  {Tran}}, \bibinfo {author} {\bibfnamefont {B.}~\bibnamefont
  {Daki\ifmmode~\acute{c}\else \'{c}\fi{}}}, \bibinfo {author} {\bibfnamefont
  {F.~m.~c.}\ \bibnamefont {Arnault}}, \bibinfo {author} {\bibfnamefont
  {W.}~\bibnamefont {Laskowski}},\ and\ \bibinfo {author} {\bibfnamefont
  {T.}~\bibnamefont {Paterek}},\ }\href
  {https://doi.org/10.1103/PhysRevA.92.050301} {\bibfield  {journal} {\bibinfo
  {journal} {Phys. Rev. A}\ }\textbf {\bibinfo {volume} {92}},\ \bibinfo
  {pages} {050301} (\bibinfo {year} {2015})}\BibitemShut {NoStop}%
\bibitem [{\citenamefont {Tran}\ \emph {et~al.}(2016)\citenamefont {Tran},
  \citenamefont {Daki\ifmmode~\acute{c}\else \'{c}\fi{}}, \citenamefont
  {Laskowski},\ and\ \citenamefont {Paterek}}]{Tran2016Correlations}%
  \BibitemOpen
  \bibfield  {author} {\bibinfo {author} {\bibfnamefont {M.~C.}\ \bibnamefont
  {Tran}}, \bibinfo {author} {\bibfnamefont {B.}~\bibnamefont
  {Daki\ifmmode~\acute{c}\else \'{c}\fi{}}}, \bibinfo {author} {\bibfnamefont
  {W.}~\bibnamefont {Laskowski}},\ and\ \bibinfo {author} {\bibfnamefont
  {T.}~\bibnamefont {Paterek}},\ }\href
  {https://doi.org/10.1103/PhysRevA.94.042302} {\bibfield  {journal} {\bibinfo
  {journal} {Phys. Rev. A}\ }\textbf {\bibinfo {volume} {94}},\ \bibinfo
  {pages} {042302} (\bibinfo {year} {2016})}\BibitemShut {NoStop}%
\bibitem [{\citenamefont {Ketterer}\ \emph {et~al.}(2019)\citenamefont
  {Ketterer}, \citenamefont {Wyderka},\ and\ \citenamefont
  {G\"uhne}}]{Ketterer2019Multipartite}%
  \BibitemOpen
  \bibfield  {author} {\bibinfo {author} {\bibfnamefont {A.}~\bibnamefont
  {Ketterer}}, \bibinfo {author} {\bibfnamefont {N.}~\bibnamefont {Wyderka}},\
  and\ \bibinfo {author} {\bibfnamefont {O.}~\bibnamefont {G\"uhne}},\ }\href
  {https://doi.org/10.1103/PhysRevLett.122.120505} {\bibfield  {journal}
  {\bibinfo  {journal} {Phys. Rev. Lett.}\ }\textbf {\bibinfo {volume} {122}},\
  \bibinfo {pages} {120505} (\bibinfo {year} {2019})}\BibitemShut {NoStop}%
\bibitem [{\citenamefont {Ketterer}\ \emph {et~al.}(2020)\citenamefont
  {Ketterer}, \citenamefont {Wyderka},\ and\ \citenamefont
  {G{\"{u}}hne}}]{Ketterer2020entanglement}%
  \BibitemOpen
  \bibfield  {author} {\bibinfo {author} {\bibfnamefont {A.}~\bibnamefont
  {Ketterer}}, \bibinfo {author} {\bibfnamefont {N.}~\bibnamefont {Wyderka}},\
  and\ \bibinfo {author} {\bibfnamefont {O.}~\bibnamefont {G{\"{u}}hne}},\
  }\href {https://doi.org/10.22331/q-2020-09-16-325} {\bibfield  {journal}
  {\bibinfo  {journal} {{Quantum}}\ }\textbf {\bibinfo {volume} {4}},\ \bibinfo
  {pages} {325} (\bibinfo {year} {2020})}\BibitemShut {NoStop}%
\bibitem [{\citenamefont {Knips}\ \emph {et~al.}(2020)\citenamefont {Knips},
  \citenamefont {Dziewior}, \citenamefont {Kobus}, \citenamefont {Laskowski},
  \citenamefont {Paterek}, \citenamefont {Shadbolt}, \citenamefont
  {Weinfurter},\ and\ \citenamefont {Meinecke}}]{Knips2020}%
  \BibitemOpen
  \bibfield  {author} {\bibinfo {author} {\bibfnamefont {L.}~\bibnamefont
  {Knips}}, \bibinfo {author} {\bibfnamefont {J.}~\bibnamefont {Dziewior}},
  \bibinfo {author} {\bibfnamefont {W.}~\bibnamefont {Kobus}}, \bibinfo
  {author} {\bibfnamefont {W.}~\bibnamefont {Laskowski}}, \bibinfo {author}
  {\bibfnamefont {T.}~\bibnamefont {Paterek}}, \bibinfo {author} {\bibfnamefont
  {P.~J.}\ \bibnamefont {Shadbolt}}, \bibinfo {author} {\bibfnamefont
  {H.}~\bibnamefont {Weinfurter}},\ and\ \bibinfo {author} {\bibfnamefont
  {J.~D.~A.}\ \bibnamefont {Meinecke}},\ }\href
  {https://doi.org/10.1038/s41534-020-0281-5} {\bibfield  {journal} {\bibinfo
  {journal} {npj Quantum Information}\ }\textbf {\bibinfo {volume} {6}},\
  \bibinfo {pages} {51} (\bibinfo {year} {2020})}\BibitemShut {NoStop}%
\bibitem [{\citenamefont {Gross}\ \emph {et~al.}(2007)\citenamefont {Gross},
  \citenamefont {Audenaert},\ and\ \citenamefont {Eisert}}]{Gross2007review}%
  \BibitemOpen
  \bibfield  {author} {\bibinfo {author} {\bibfnamefont {D.}~\bibnamefont
  {Gross}}, \bibinfo {author} {\bibfnamefont {K.}~\bibnamefont {Audenaert}},\
  and\ \bibinfo {author} {\bibfnamefont {J.}~\bibnamefont {Eisert}},\ }\href
  {https://doi.org/10.1063/1.2716992} {\bibfield  {journal} {\bibinfo
  {journal} {Journal of Mathematical Physics}\ }\textbf {\bibinfo {volume}
  {48}},\ \bibinfo {pages} {052104} (\bibinfo {year} {2007})},\ \Eprint
  {https://arxiv.org/abs/https://doi.org/10.1063/1.2716992}
  {https://doi.org/10.1063/1.2716992} \BibitemShut {NoStop}%
\bibitem [{\citenamefont {Webb}(2015)}]{Webb2015design}%
  \BibitemOpen
  \bibfield  {author} {\bibinfo {author} {\bibfnamefont {Z.}~\bibnamefont
  {Webb}},\ }\href {https://arxiv.org/abs/1510.02769} {\bibinfo {title} {The
  clifford group forms a unitary 3-design}} (\bibinfo {year} {2015}),\ \Eprint
  {https://arxiv.org/abs/1510.02769} {arXiv:1510.02769 [quant-ph]} \BibitemShut
  {NoStop}%
\bibitem [{\citenamefont {Zhu}\ \emph {et~al.}(2016)\citenamefont {Zhu},
  \citenamefont {Kueng}, \citenamefont {Grassl},\ and\ \citenamefont
  {Gross}}]{Zhu_2016}%
  \BibitemOpen
  \bibfield  {author} {\bibinfo {author} {\bibfnamefont {H.}~\bibnamefont
  {Zhu}}, \bibinfo {author} {\bibfnamefont {R.}~\bibnamefont {Kueng}}, \bibinfo
  {author} {\bibfnamefont {M.}~\bibnamefont {Grassl}},\ and\ \bibinfo {author}
  {\bibfnamefont {D.}~\bibnamefont {Gross}},\ }\href@noop {} {\bibinfo {title}
  {The clifford group fails gracefully to be a unitary 4-design}} (\bibinfo
  {year} {2016}),\ \Eprint {https://arxiv.org/abs/1609.08172} {arXiv:1609.08172
  [quant-ph]} \BibitemShut {NoStop}%
\bibitem [{\citenamefont {Kueng}\ and\ \citenamefont
  {Gross}(2015)}]{Kueng2015proj}%
  \BibitemOpen
  \bibfield  {author} {\bibinfo {author} {\bibfnamefont {R.}~\bibnamefont
  {Kueng}}\ and\ \bibinfo {author} {\bibfnamefont {D.}~\bibnamefont {Gross}},\
  }\href {https://arxiv.org/abs/1510.02767} {\bibinfo {title} {Qubit stabilizer
  states are complex projective 3-designs}} (\bibinfo {year} {2015}),\ \Eprint
  {https://arxiv.org/abs/1510.02767} {arXiv:1510.02767 [quant-ph]} \BibitemShut
  {NoStop}%
\bibitem [{\citenamefont {Brando}\ \emph {et~al.}(2016)\citenamefont {Brando},
  \citenamefont {Harrow},\ and\ \citenamefont {Horodecki}}]{Brando2016}%
  \BibitemOpen
  \bibfield  {author} {\bibinfo {author} {\bibfnamefont {F.~G. S.~L.}\
  \bibnamefont {Brando}}, \bibinfo {author} {\bibfnamefont {A.~W.}\
  \bibnamefont {Harrow}},\ and\ \bibinfo {author} {\bibfnamefont
  {M.}~\bibnamefont {Horodecki}},\ }\href
  {https://doi.org/10.1007/s00220-016-2706-8} {\bibfield  {journal} {\bibinfo
  {journal} {Communications in Mathematical Physics}\ }\textbf {\bibinfo
  {volume} {346}},\ \bibinfo {pages} {397} (\bibinfo {year}
  {2016})}\BibitemShut {NoStop}%
\bibitem [{\citenamefont {Haferkamp}\ \emph {et~al.}(2020)\citenamefont
  {Haferkamp}, \citenamefont {Montealegre-Mora}, \citenamefont {Heinrich},
  \citenamefont {Eisert}, \citenamefont {Gross},\ and\ \citenamefont
  {Roth}}]{Haferkamp2020homeopathy}%
  \BibitemOpen
  \bibfield  {author} {\bibinfo {author} {\bibfnamefont {J.}~\bibnamefont
  {Haferkamp}}, \bibinfo {author} {\bibfnamefont {F.}~\bibnamefont
  {Montealegre-Mora}}, \bibinfo {author} {\bibfnamefont {M.}~\bibnamefont
  {Heinrich}}, \bibinfo {author} {\bibfnamefont {J.}~\bibnamefont {Eisert}},
  \bibinfo {author} {\bibfnamefont {D.}~\bibnamefont {Gross}},\ and\ \bibinfo
  {author} {\bibfnamefont {I.}~\bibnamefont {Roth}},\ }\href@noop {} {\bibinfo
  {title} {Quantum homeopathy works: Efficient unitary designs with a
  system-size independent number of non-clifford gates}} (\bibinfo {year}
  {2020}),\ \Eprint {https://arxiv.org/abs/2002.09524} {arXiv:2002.09524
  [quant-ph]} \BibitemShut {NoStop}%
\bibitem [{\citenamefont {Hosur}\ \emph {et~al.}(2016)\citenamefont {Hosur},
  \citenamefont {Qi}, \citenamefont {Roberts},\ and\ \citenamefont
  {Yoshida}}]{Hosur_2016}%
  \BibitemOpen
  \bibfield  {author} {\bibinfo {author} {\bibfnamefont {P.}~\bibnamefont
  {Hosur}}, \bibinfo {author} {\bibfnamefont {X.-L.}\ \bibnamefont {Qi}},
  \bibinfo {author} {\bibfnamefont {D.~A.}\ \bibnamefont {Roberts}},\ and\
  \bibinfo {author} {\bibfnamefont {B.}~\bibnamefont {Yoshida}},\ }\bibfield
  {journal} {\bibinfo  {journal} {Journal of High Energy Physics}\ }\textbf
  {\bibinfo {volume} {2016}},\ \href {https://doi.org/10.1007/jhep02(2016)004}
  {10.1007/jhep02(2016)004} (\bibinfo {year} {2016})\BibitemShut {NoStop}%
\bibitem [{\citenamefont {Jaffe}\ \emph {et~al.}(2018)\citenamefont {Jaffe},
  \citenamefont {Liu},\ and\ \citenamefont {Wozniakowski}}]{Jaffe_2018}%
  \BibitemOpen
  \bibfield  {author} {\bibinfo {author} {\bibfnamefont {A.}~\bibnamefont
  {Jaffe}}, \bibinfo {author} {\bibfnamefont {Z.}~\bibnamefont {Liu}},\ and\
  \bibinfo {author} {\bibfnamefont {A.}~\bibnamefont {Wozniakowski}},\ }\href
  {https://doi.org/10.1007/s11425-017-9207-3} {\bibfield  {journal} {\bibinfo
  {journal} {Science China Mathematics}\ }\textbf {\bibinfo {volume} {61}},\
  \bibinfo {pages} {593–626} (\bibinfo {year} {2018})}\BibitemShut {NoStop}%
\bibitem [{\citenamefont {Elben}\ \emph
  {et~al.}(2019{\natexlab{b}})\citenamefont {Elben}, \citenamefont {Vermersch},
  \citenamefont {Roos},\ and\ \citenamefont {Zoller}}]{Elben_2019}%
  \BibitemOpen
  \bibfield  {author} {\bibinfo {author} {\bibfnamefont {A.}~\bibnamefont
  {Elben}}, \bibinfo {author} {\bibfnamefont {B.}~\bibnamefont {Vermersch}},
  \bibinfo {author} {\bibfnamefont {C.~F.}\ \bibnamefont {Roos}},\ and\
  \bibinfo {author} {\bibfnamefont {P.}~\bibnamefont {Zoller}},\ }\bibfield
  {journal} {\bibinfo  {journal} {Physical Review A}\ }\textbf {\bibinfo
  {volume} {99}},\ \href {https://doi.org/10.1103/physreva.99.052323}
  {10.1103/physreva.99.052323} (\bibinfo {year}
  {2019}{\natexlab{b}})\BibitemShut {NoStop}%
\bibitem [{\citenamefont {Collins}\ and\ \citenamefont
  {Nechita}(2010)}]{Collins_2010}%
  \BibitemOpen
  \bibfield  {author} {\bibinfo {author} {\bibfnamefont {B.}~\bibnamefont
  {Collins}}\ and\ \bibinfo {author} {\bibfnamefont {I.}~\bibnamefont
  {Nechita}},\ }\href {https://doi.org/10.1007/s00220-010-1012-0} {\bibfield
  {journal} {\bibinfo  {journal} {Communications in Mathematical Physics}\
  }\textbf {\bibinfo {volume} {297}},\ \bibinfo {pages} {345–370} (\bibinfo
  {year} {2010})}\BibitemShut {NoStop}%
\bibitem [{\citenamefont {Kotz}\ and\ \citenamefont
  {Johnson}(1992)}]{kotz1992breakthroughs}%
  \BibitemOpen
  \bibfield  {author} {\bibinfo {author} {\bibfnamefont {S.}~\bibnamefont
  {Kotz}}\ and\ \bibinfo {author} {\bibfnamefont {N.}~\bibnamefont {Johnson}},\
  }\href {https://books.google.com/books?id=iZoZAQAAIAAJ} {\emph {\bibinfo
  {title} {Breakthroughs in Statistics}}},\ \bibinfo {series} {Breakthroughs in
  Statistics}\ No.\ \bibinfo {number} {v. 3}\ (\bibinfo  {publisher} {U.S.
  Government Printing Office},\ \bibinfo {year} {1992})\BibitemShut {NoStop}%
\bibitem [{\citenamefont {Levine}\ \emph {et~al.}(2018)\citenamefont {Levine},
  \citenamefont {Keesling}, \citenamefont {Omran}, \citenamefont {Bernien},
  \citenamefont {Schwartz}, \citenamefont {Zibrov}, \citenamefont {Endres},
  \citenamefont {Greiner}, \citenamefont {Vuleti\ifmmode~\acute{c}\else
  \'{c}\fi{}},\ and\ \citenamefont {Lukin}}]{PhysRevLett.121.123603}%
  \BibitemOpen
  \bibfield  {author} {\bibinfo {author} {\bibfnamefont {H.}~\bibnamefont
  {Levine}}, \bibinfo {author} {\bibfnamefont {A.}~\bibnamefont {Keesling}},
  \bibinfo {author} {\bibfnamefont {A.}~\bibnamefont {Omran}}, \bibinfo
  {author} {\bibfnamefont {H.}~\bibnamefont {Bernien}}, \bibinfo {author}
  {\bibfnamefont {S.}~\bibnamefont {Schwartz}}, \bibinfo {author}
  {\bibfnamefont {A.~S.}\ \bibnamefont {Zibrov}}, \bibinfo {author}
  {\bibfnamefont {M.}~\bibnamefont {Endres}}, \bibinfo {author} {\bibfnamefont
  {M.}~\bibnamefont {Greiner}}, \bibinfo {author} {\bibfnamefont
  {V.}~\bibnamefont {Vuleti\ifmmode~\acute{c}\else \'{c}\fi{}}},\ and\ \bibinfo
  {author} {\bibfnamefont {M.~D.}\ \bibnamefont {Lukin}},\ }\href
  {https://doi.org/10.1103/PhysRevLett.121.123603} {\bibfield  {journal}
  {\bibinfo  {journal} {Phys. Rev. Lett.}\ }\textbf {\bibinfo {volume} {121}},\
  \bibinfo {pages} {123603} (\bibinfo {year} {2018})}\BibitemShut {NoStop}%
\bibitem [{\citenamefont {Yang}\ \emph {et~al.}(2020)\citenamefont {Yang},
  \citenamefont {Sun}, \citenamefont {Huang}, \citenamefont {Wang},
  \citenamefont {Deng}, \citenamefont {Dai}, \citenamefont {Yuan},\ and\
  \citenamefont {Pan}}]{Yang2020lattice}%
  \BibitemOpen
  \bibfield  {author} {\bibinfo {author} {\bibfnamefont {B.}~\bibnamefont
  {Yang}}, \bibinfo {author} {\bibfnamefont {H.}~\bibnamefont {Sun}}, \bibinfo
  {author} {\bibfnamefont {C.-J.}\ \bibnamefont {Huang}}, \bibinfo {author}
  {\bibfnamefont {H.-Y.}\ \bibnamefont {Wang}}, \bibinfo {author}
  {\bibfnamefont {Y.}~\bibnamefont {Deng}}, \bibinfo {author} {\bibfnamefont
  {H.-N.}\ \bibnamefont {Dai}}, \bibinfo {author} {\bibfnamefont {Z.-S.}\
  \bibnamefont {Yuan}},\ and\ \bibinfo {author} {\bibfnamefont {J.-W.}\
  \bibnamefont {Pan}},\ }\href {https://doi.org/10.1126/science.aaz6801}
  {\bibfield  {journal} {\bibinfo  {journal} {Science}\ }\textbf {\bibinfo
  {volume} {369}},\ \bibinfo {pages} {550} (\bibinfo {year} {2020})},\ \Eprint
  {https://arxiv.org/abs/https://science.sciencemag.org/content/369/6503/550.full.pdf}
  {https://science.sciencemag.org/content/369/6503/550.full.pdf} \BibitemShut
  {NoStop}%
\bibitem [{\citenamefont {Haah}\ \emph {et~al.}(2016)\citenamefont {Haah},
  \citenamefont {Harrow}, \citenamefont {Ji}, \citenamefont {Wu},\ and\
  \citenamefont {Yu}}]{Haah2016}%
  \BibitemOpen
  \bibfield  {author} {\bibinfo {author} {\bibfnamefont {J.}~\bibnamefont
  {Haah}}, \bibinfo {author} {\bibfnamefont {A.~W.}\ \bibnamefont {Harrow}},
  \bibinfo {author} {\bibfnamefont {Z.}~\bibnamefont {Ji}}, \bibinfo {author}
  {\bibfnamefont {X.}~\bibnamefont {Wu}},\ and\ \bibinfo {author}
  {\bibfnamefont {N.}~\bibnamefont {Yu}},\ }in\ \href
  {https://doi.org/10.1145/2897518.2897585} {\emph {\bibinfo {booktitle}
  {Proceedings of the Forty-Eighth Annual ACM Symposium on Theory of
  Computing}}},\ \bibinfo {series and number} {STOC '16}\ (\bibinfo
  {publisher} {Association for Computing Machinery},\ \bibinfo {address} {New
  York, NY, USA},\ \bibinfo {year} {2016})\ p.\ \bibinfo {pages}
  {913–925}\BibitemShut {NoStop}%
\bibitem [{\citenamefont {Yu}(2020)}]{Yu2020Pauli}%
  \BibitemOpen
  \bibfield  {author} {\bibinfo {author} {\bibfnamefont {N.}~\bibnamefont
  {Yu}},\ }\href@noop {} {\bibinfo {title} {Sample efficient tomography via
  pauli measurements}} (\bibinfo {year} {2020}),\ \Eprint
  {https://arxiv.org/abs/2009.04610} {arXiv:2009.04610 [quant-ph]} \BibitemShut
  {NoStop}%
\bibitem [{\citenamefont {Rozenbaum}\ \emph {et~al.}(2017)\citenamefont
  {Rozenbaum}, \citenamefont {Ganeshan},\ and\ \citenamefont
  {Galitski}}]{Rozenbaum_2017}%
  \BibitemOpen
  \bibfield  {author} {\bibinfo {author} {\bibfnamefont {E.~B.}\ \bibnamefont
  {Rozenbaum}}, \bibinfo {author} {\bibfnamefont {S.}~\bibnamefont
  {Ganeshan}},\ and\ \bibinfo {author} {\bibfnamefont {V.}~\bibnamefont
  {Galitski}},\ }\bibfield  {journal} {\bibinfo  {journal} {Physical Review
  Letters}\ }\textbf {\bibinfo {volume} {118}},\ \href
  {https://doi.org/10.1103/physrevlett.118.086801}
  {10.1103/physrevlett.118.086801} (\bibinfo {year} {2017})\BibitemShut
  {NoStop}%
\bibitem [{\citenamefont {Lin}\ and\ \citenamefont
  {Motrunich}(2018)}]{Lin_2018}%
  \BibitemOpen
  \bibfield  {author} {\bibinfo {author} {\bibfnamefont {C.-J.}\ \bibnamefont
  {Lin}}\ and\ \bibinfo {author} {\bibfnamefont {O.~I.}\ \bibnamefont
  {Motrunich}},\ }\bibfield  {journal} {\bibinfo  {journal} {Physical Review
  B}\ }\textbf {\bibinfo {volume} {97}},\ \href
  {https://doi.org/10.1103/physrevb.97.144304} {10.1103/physrevb.97.144304}
  (\bibinfo {year} {2018})\BibitemShut {NoStop}%
\bibitem [{\citenamefont {Xu}\ and\ \citenamefont {Swingle}(2019)}]{Xu_2019}%
  \BibitemOpen
  \bibfield  {author} {\bibinfo {author} {\bibfnamefont {S.}~\bibnamefont
  {Xu}}\ and\ \bibinfo {author} {\bibfnamefont {B.}~\bibnamefont {Swingle}},\
  }\href {https://doi.org/10.1038/s41567-019-0712-4} {\bibfield  {journal}
  {\bibinfo  {journal} {Nature Physics}\ }\textbf {\bibinfo {volume} {16}},\
  \bibinfo {pages} {199–204} (\bibinfo {year} {2019})}\BibitemShut {NoStop}%
\bibitem [{\citenamefont {Chen}\ \emph {et~al.}(2020)\citenamefont {Chen},
  \citenamefont {Yu}, \citenamefont {Zeng},\ and\ \citenamefont
  {Flammia}}]{Chen_2020}%
  \BibitemOpen
  \bibfield  {author} {\bibinfo {author} {\bibfnamefont {S.}~\bibnamefont
  {Chen}}, \bibinfo {author} {\bibfnamefont {W.}~\bibnamefont {Yu}}, \bibinfo
  {author} {\bibfnamefont {P.}~\bibnamefont {Zeng}},\ and\ \bibinfo {author}
  {\bibfnamefont {S.~T.}\ \bibnamefont {Flammia}},\ }\href@noop {} {\bibinfo
  {title} {Robust shadow estimation}} (\bibinfo {year} {2020}),\ \Eprint
  {https://arxiv.org/abs/2011.09636} {arXiv:2011.09636 [quant-ph]} \BibitemShut
  {NoStop}%
\bibitem [{\citenamefont {Koh}\ and\ \citenamefont {Grewal}(2020)}]{Koh_2020}%
  \BibitemOpen
  \bibfield  {author} {\bibinfo {author} {\bibfnamefont {D.~E.}\ \bibnamefont
  {Koh}}\ and\ \bibinfo {author} {\bibfnamefont {S.}~\bibnamefont {Grewal}},\
  }\href@noop {} {\bibinfo {title} {Classical shadows with noise}} (\bibinfo
  {year} {2020}),\ \Eprint {https://arxiv.org/abs/2011.11580} {arXiv:2011.11580
  [quant-ph]} \BibitemShut {NoStop}%
\bibitem [{\citenamefont {Swingle}\ and\ \citenamefont
  {Yunger~Halpern}(2018)}]{PhysRevA.97.062113}%
  \BibitemOpen
  \bibfield  {author} {\bibinfo {author} {\bibfnamefont {B.}~\bibnamefont
  {Swingle}}\ and\ \bibinfo {author} {\bibfnamefont {N.}~\bibnamefont
  {Yunger~Halpern}},\ }\href {https://doi.org/10.1103/PhysRevA.97.062113}
  {\bibfield  {journal} {\bibinfo  {journal} {Phys. Rev. A}\ }\textbf {\bibinfo
  {volume} {97}},\ \bibinfo {pages} {062113} (\bibinfo {year}
  {2018})}\BibitemShut {NoStop}%
\bibitem [{\citenamefont {Zhang}\ \emph {et~al.}(2019)\citenamefont {Zhang},
  \citenamefont {Huang},\ and\ \citenamefont {Chen}}]{Zhang_2019}%
  \BibitemOpen
  \bibfield  {author} {\bibinfo {author} {\bibfnamefont {Y.-L.}\ \bibnamefont
  {Zhang}}, \bibinfo {author} {\bibfnamefont {Y.}~\bibnamefont {Huang}},\ and\
  \bibinfo {author} {\bibfnamefont {X.}~\bibnamefont {Chen}},\ }\bibfield
  {journal} {\bibinfo  {journal} {Physical Review B}\ }\textbf {\bibinfo
  {volume} {99}},\ \href {https://doi.org/10.1103/physrevb.99.014303}
  {10.1103/physrevb.99.014303} (\bibinfo {year} {2019})\BibitemShut {NoStop}%
\bibitem [{\citenamefont {Kudler-Flam}\ \emph {et~al.}(2020)\citenamefont
  {Kudler-Flam}, \citenamefont {Nozaki}, \citenamefont {Ryu},\ and\
  \citenamefont {Tan}}]{Kudler_Flam_2020}%
  \BibitemOpen
  \bibfield  {author} {\bibinfo {author} {\bibfnamefont {J.}~\bibnamefont
  {Kudler-Flam}}, \bibinfo {author} {\bibfnamefont {M.}~\bibnamefont {Nozaki}},
  \bibinfo {author} {\bibfnamefont {S.}~\bibnamefont {Ryu}},\ and\ \bibinfo
  {author} {\bibfnamefont {M.~T.}\ \bibnamefont {Tan}},\ }\bibfield  {journal}
  {\bibinfo  {journal} {Journal of High Energy Physics}\ }\textbf {\bibinfo
  {volume} {2020}},\ \href {https://doi.org/10.1007/jhep01(2020)031}
  {10.1007/jhep01(2020)031} (\bibinfo {year} {2020})\BibitemShut {NoStop}%
\bibitem [{\citenamefont {Roberts}\ \emph {et~al.}(2018)\citenamefont
  {Roberts}, \citenamefont {Stanford},\ and\ \citenamefont
  {Streicher}}]{Roberts2018growth}%
  \BibitemOpen
  \bibfield  {author} {\bibinfo {author} {\bibfnamefont {D.~A.}\ \bibnamefont
  {Roberts}}, \bibinfo {author} {\bibfnamefont {D.}~\bibnamefont {Stanford}},\
  and\ \bibinfo {author} {\bibfnamefont {A.}~\bibnamefont {Streicher}},\ }\href
  {https://doi.org/10.1007/JHEP06(2018)122} {\bibfield  {journal} {\bibinfo
  {journal} {Journal of High Energy Physics}\ }\textbf {\bibinfo {volume}
  {2018}},\ \bibinfo {pages} {122} (\bibinfo {year} {2018})}\BibitemShut
  {NoStop}%
\bibitem [{\citenamefont {Qi}\ \emph {et~al.}(2019)\citenamefont {Qi},
  \citenamefont {Davis}, \citenamefont {Periwal},\ and\ \citenamefont
  {Schleier-Smith}}]{Qi_2019}%
  \BibitemOpen
  \bibfield  {author} {\bibinfo {author} {\bibfnamefont {X.-L.}\ \bibnamefont
  {Qi}}, \bibinfo {author} {\bibfnamefont {E.~J.}\ \bibnamefont {Davis}},
  \bibinfo {author} {\bibfnamefont {A.}~\bibnamefont {Periwal}},\ and\ \bibinfo
  {author} {\bibfnamefont {M.}~\bibnamefont {Schleier-Smith}},\ }\href@noop {}
  {\bibinfo {title} {Measuring operator size growth in quantum quench
  experiments}} (\bibinfo {year} {2019}),\ \Eprint
  {https://arxiv.org/abs/1906.00524} {arXiv:1906.00524 [quant-ph]} \BibitemShut
  {NoStop}%
\bibitem [{\citenamefont {Luitz}\ and\ \citenamefont
  {Bar~Lev}(2017)}]{PhysRevB.96.020406}%
  \BibitemOpen
  \bibfield  {author} {\bibinfo {author} {\bibfnamefont {D.~J.}\ \bibnamefont
  {Luitz}}\ and\ \bibinfo {author} {\bibfnamefont {Y.}~\bibnamefont
  {Bar~Lev}},\ }\href {https://doi.org/10.1103/PhysRevB.96.020406} {\bibfield
  {journal} {\bibinfo  {journal} {Phys. Rev. B}\ }\textbf {\bibinfo {volume}
  {96}},\ \bibinfo {pages} {020406} (\bibinfo {year} {2017})}\BibitemShut
  {NoStop}%
\bibitem [{\citenamefont {Da\ifmmode~\breve{g}\else \u{g}\fi{}}\ \emph
  {et~al.}(2020)\citenamefont {Da\ifmmode~\breve{g}\else \u{g}\fi{}},
  \citenamefont {Duan},\ and\ \citenamefont {Sun}}]{PhysRevB.101.104415}%
  \BibitemOpen
  \bibfield  {author} {\bibinfo {author} {\bibfnamefont {C.~B.}\ \bibnamefont
  {Da\ifmmode~\breve{g}\else \u{g}\fi{}}}, \bibinfo {author} {\bibfnamefont
  {L.-M.}\ \bibnamefont {Duan}},\ and\ \bibinfo {author} {\bibfnamefont
  {K.}~\bibnamefont {Sun}},\ }\href
  {https://doi.org/10.1103/PhysRevB.101.104415} {\bibfield  {journal} {\bibinfo
   {journal} {Phys. Rev. B}\ }\textbf {\bibinfo {volume} {101}},\ \bibinfo
  {pages} {104415} (\bibinfo {year} {2020})}\BibitemShut {NoStop}%
\bibitem [{\citenamefont {Gorin}\ \emph {et~al.}(2006)\citenamefont {Gorin},
  \citenamefont {Prosen}, \citenamefont {Seligman},\ and\ \citenamefont
  {{\v{Z}}nidari{\v{c}}}}]{Gorin2006}%
  \BibitemOpen
  \bibfield  {author} {\bibinfo {author} {\bibfnamefont {T.}~\bibnamefont
  {Gorin}}, \bibinfo {author} {\bibfnamefont {T.}~\bibnamefont {Prosen}},
  \bibinfo {author} {\bibfnamefont {T.~H.}\ \bibnamefont {Seligman}},\ and\
  \bibinfo {author} {\bibfnamefont {M.}~\bibnamefont {{\v{Z}}nidari{\v{c}}}},\
  }\href {http://www.sciencedirect.com/science/article/pii/S0370157306003310}
  {\bibfield  {journal} {\bibinfo  {journal} {Physics Reports}\ }\textbf
  {\bibinfo {volume} {435}},\ \bibinfo {pages} {33} (\bibinfo {year}
  {2006})}\BibitemShut {NoStop}%
\bibitem [{\citenamefont {Yan}\ \emph {et~al.}(2020)\citenamefont {Yan},
  \citenamefont {Cincio},\ and\ \citenamefont
  {Zurek}}]{PhysRevLett.124.160603}%
  \BibitemOpen
  \bibfield  {author} {\bibinfo {author} {\bibfnamefont {B.}~\bibnamefont
  {Yan}}, \bibinfo {author} {\bibfnamefont {L.}~\bibnamefont {Cincio}},\ and\
  \bibinfo {author} {\bibfnamefont {W.~H.}\ \bibnamefont {Zurek}},\ }\href
  {https://doi.org/10.1103/PhysRevLett.124.160603} {\bibfield  {journal}
  {\bibinfo  {journal} {Phys. Rev. Lett.}\ }\textbf {\bibinfo {volume} {124}},\
  \bibinfo {pages} {160603} (\bibinfo {year} {2020})}\BibitemShut {NoStop}%
\bibitem [{\citenamefont {Liu}\ \emph {et~al.}(2017)\citenamefont {Liu},
  \citenamefont {Wozniakowski},\ and\ \citenamefont {Jaffe}}]{Liu_2017}%
  \BibitemOpen
  \bibfield  {author} {\bibinfo {author} {\bibfnamefont {Z.}~\bibnamefont
  {Liu}}, \bibinfo {author} {\bibfnamefont {A.}~\bibnamefont {Wozniakowski}},\
  and\ \bibinfo {author} {\bibfnamefont {A.~M.}\ \bibnamefont {Jaffe}},\ }\href
  {https://doi.org/10.1073/pnas.1621345114} {\bibfield  {journal} {\bibinfo
  {journal} {Proceedings of the National Academy of Sciences}\ }\textbf
  {\bibinfo {volume} {114}},\ \bibinfo {pages} {2497–2502} (\bibinfo {year}
  {2017})}\BibitemShut {NoStop}%
\bibitem [{\citenamefont {Collins}\ and\ \citenamefont
  {Śniady}(2006)}]{Collins_2006}%
  \BibitemOpen
  \bibfield  {author} {\bibinfo {author} {\bibfnamefont {B.}~\bibnamefont
  {Collins}}\ and\ \bibinfo {author} {\bibfnamefont {P.}~\bibnamefont
  {Śniady}},\ }\href {https://doi.org/10.1007/s00220-006-1554-3} {\bibfield
  {journal} {\bibinfo  {journal} {Communications in Mathematical Physics}\
  }\textbf {\bibinfo {volume} {264}},\ \bibinfo {pages} {773–795} (\bibinfo
  {year} {2006})}\BibitemShut {NoStop}%
\end{thebibliography}%

\onecolumngrid
\newpage
\begin{appendix}

\section{Interpretation of higher-point OTOCs}\label{sec:interpret}
We propose a physical interpretation of higher-point OTOCs. For simplicity, we approximate the expectation value over the maximally mixed state as one over a Haar random state $\ket{\psi_h}$ \cite{PhysRevB.96.020406,PhysRevB.101.104415}. The approximate four-point OTOC is
\begin{equation}
    \tilde{C}_{4}(t)=\bra{\psi_h}W^\dagger (t)V^\dagger W(t)V\ket{\psi_h}=\bra{\psi_h}U_H^\dagger W^\dagger U_H V^\dagger U^\dagger_H W U_H V\ket{\psi_h}.
\end{equation}

\begin{figure}[t]
  \centering
  \includegraphics[scale=.2]{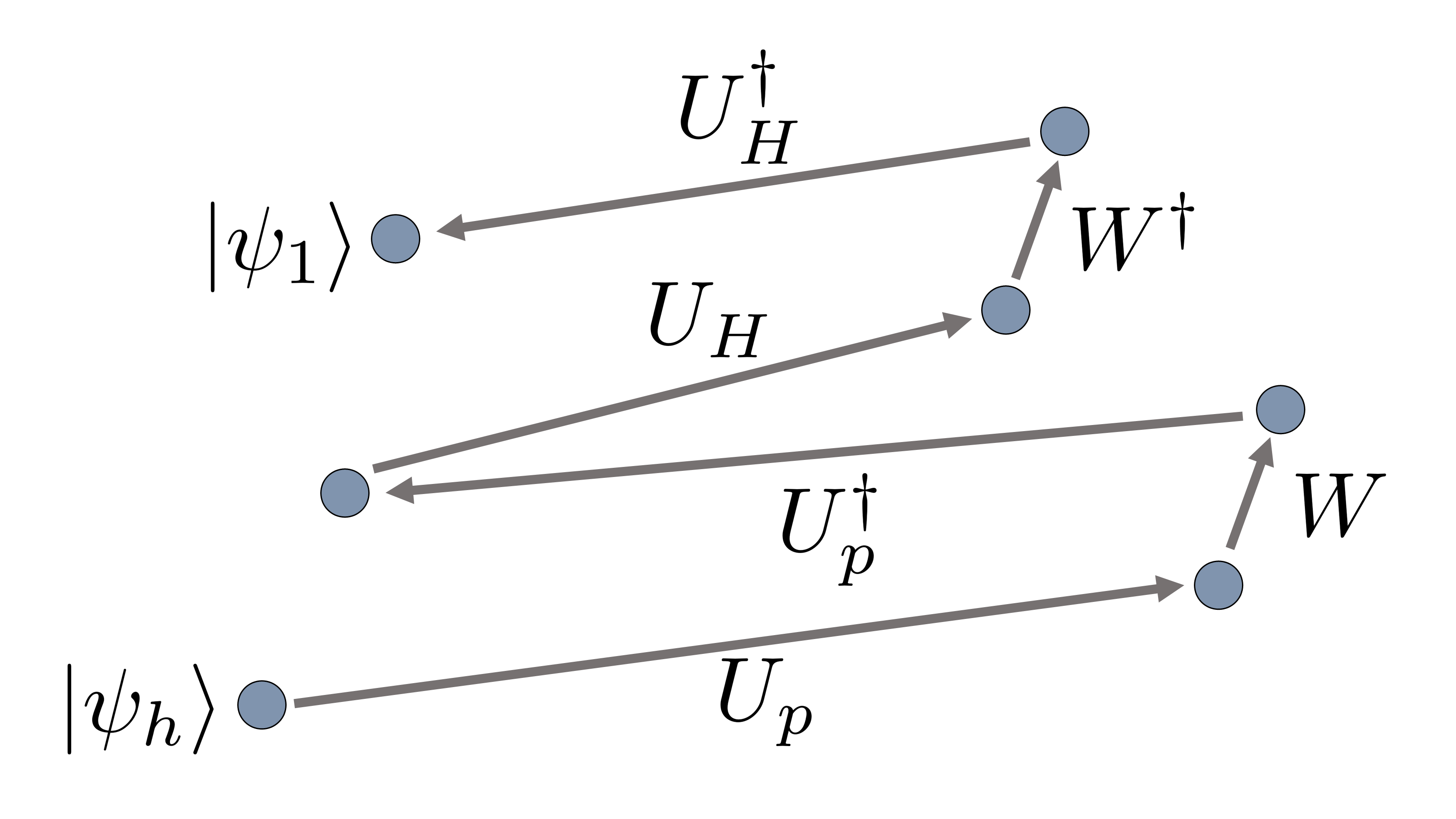}
  \vspace*{-3mm}
  \caption{Schematic of a single perturbed evolution procedure evolving $\ket{\psi_h}$ to $\ket{\psi_1}$.}
  \label{fig:Los}
\end{figure}

In analogy to the Loschmidt echo \cite{Gorin2006,PhysRevLett.124.160603}, define the perturbed evolution unitary as $U_p(t)=U_H(t) V$. The state $U_p(t)\ket{\psi_h}$ represents a small `kick', $V$, perturbing $\ket{\psi_h}$ before evolution by $U_H(t)$. The approximate OTOC in terms of $U_p(t)$ is
\begin{equation}
    \tilde{C}_4(t) =\bra{\psi_h}U_H^\dagger W^\dagger U_H U^\dagger_p W U_p \ket{\psi_h}.
\end{equation}
The state $\ket{\psi_1}=U_H^\dagger W^\dagger U_H U^\dagger_p W U_p \ket{\psi_h}$ is interpreted through following perturbation procedure (see Fig.~\ref{fig:Los}). Evolve $\ket{\psi_h}$ by $U_p(t)$, then apply $W$. Evolve the state using $U_p^\dagger(t)$, which corresponds to backwards time evolution by $U^\dagger_H(t)=U_H(-t)$ followed by $V^\dagger$. Evolve with $U_H(t)$, apply $W^\dagger$, and finally evolve backwards in time using $U^\dagger_H(t)$. The correlator is the overlap between the initial state $\ket{\psi_h}$ and the evolved state $\ket{\psi_1}$. The role of $W$ is to perturb the state and ensure $U_p(t)$ is not undone by $U_p^\dagger(t)$. $V$ probes how chaotic $U_H(t)$ is. In the case where $V$ is the identity, $U_p(t)=U_H(t)$ and the evolution procedure evolves $\ket{\psi_h}$ away from and back to itself. No chaos is detected. For a nontrivial $V$, the perturbation procedure evolves the state away from $\ket{\psi_h}$. $\tilde{C}_4(t)$ indicates the sensitivity of the evolution of $U_H(t)$ to perturbations.

Consider the approximate $4k$-point OTOC,
\begin{equation}
   \tilde{C}_{4k}(t) =\bra{\psi_h}(U_H^\dagger W^\dagger U_H U^\dagger_p W U_p)^k \ket{\psi_h}.
\end{equation}
This correlator gives the overlap of $\ket{\psi_h}$ with the state resulting from $k$ applications of the perturbation procedure to $\ket{\psi_h}$. Each application evolves the state successively further from $\ket{\psi_h}$. The higher-point correlators correspond to a larger number of perturbations to the initial state, which probe the chaos induced by $U_H(t)$ in finer detail.

We return to the discussion of exact OTOCs. A higher-point OTOC $C_{4k}(t)$ reveals the finer scrambling dynamics due to $k$ perturbations by $V$. After each perturbation, evolution by $U_H(t)$ further delocalizes the quantum information of the initially local operator $W$. Higher-point correlators therefore scramble information more quickly, resulting in a faster OTOC decay. In a different context, an interpretation of higher-point correlators is given in terms of multiple shock waves \cite{Shenker_2014}.

\section{Graphical calculus}\label{sec:diagrams}
We present a graphical calculus for matrices adapted from \cite{Collins_2010,Elben_2019} and mention a related picture-language for quantum information \cite{Jaffe_2018,Liu_2017}. An operator $A_1$ is drawn as a box with input and output legs
\begin{equation}
\begin{tikzpicture}
    \node[] (v0) at (.5,0) {$A_1=$};
    \node[rectangle, fill=egg, rounded corners, minimum width=1em, minimum size =2em, draw] (v1) at (2,0) {$A_1$};
    
    \draw [thick,color=dullblue]    (1,0)--(v1)--(3,0);
\end{tikzpicture}.
\end{equation}
The purpose of the coloring is solely to distinguish these diagrams from quantum circuits. By expanding $A_1$ in a basis, $A_{1}=\sum_{i,j}a^1_{i,j}\ket{i}\bra{j}$, we let the left leg correspond to $\ket{i}$ and the right leg correspond to $\bra{j}$. Unless otherwise stated, each leg index corresponds to a state in Hilbert space $\mathcal{H}_d$. The transpose of $A_1$ is represented by
\begin{equation}\label{eq:transpose}
\begin{tikzpicture}
    \node[] (v0) at (0,0) {$A_1^T=$};
    \node[rectangle, fill=egg, rounded corners, minimum width=1em, minimum size =2em, draw] (v1) at (2,0) {$A_1$};
    
    \draw [thick,color=dullblue]    
    (v1)--(2.5,0)--(2.5,.5)--(1.2,.5)--(1.2,0)--(.7,0)
    (v1)--(1.5,0)--(1.5,-.5)--(2.7,-.5)--(2.7,0)--(3.2,0);
	
\end{tikzpicture}.
\end{equation}
A product of two operators is drawn by connecting the output leg of one with the input leg of the other
\begin{equation}
\begin{tikzpicture}
    \node[] (v0) at (-2,0) {$A_1 A_2=$};     
	 \node[rectangle, fill=egg, rounded corners, minimum width=1em, minimum size =2em, draw] (v1) at (0,0) {$A_1$};       
    \node[rectangle, fill=egg, rounded corners, minimum width=1em, minimum size =2em, draw] (v2) at (1,0) {$A_2$};      
    \draw [thick,color=dullblue] 
    (-1,0) -- (v1)    
    (v1) -- (v2)
    (v2) -- (2,0);
\end{tikzpicture}.
\end{equation}
More formally, a connection between legs is an index contraction and hence represents matrix multiplication.

The trace $\tr\{A_1\}=\sum_i a^1_{i,i}\bra{i}i\rangle$ is drawn by connecting the legs of $A_1$
\begin{equation}
\begin{tikzpicture}
    \node[] (v0) at (0,0) {$\tr\{A_1\}=$};     
	 \node[rectangle, fill=egg, rounded corners, minimum width=1em, minimum size =2em, draw] (v1) at (2,0) {$A_1$};         
    \draw [thick,color=dullblue] 
    (1,0) -- (v1) -- (3,0)
    (1,0)--(1,0.5)--(3,0.5)--(3,0);
	
\end{tikzpicture}.
\end{equation}
A tensor product of operators is represented as a `stacking' of boxes
\begin{equation}
\begin{tikzpicture}
    \node[] (v0) at (-2,-.5) {$A_1 \otimes A_2=$};     
	 \node[rectangle, fill=egg, rounded corners, minimum width=1em, minimum size =2em, draw] (v1) at (0,0) {$A_1$};       
    \node[rectangle, fill=egg, rounded corners, minimum width=1em, minimum size =2em, draw] (v2) at (0,-1) {$A_2$};      
    \draw [thick,color=dullblue] 
    (-1,0) -- (v1) -- (1,0)    
    (-1,-1) -- (v2) -- (1,-1);
\end{tikzpicture}.
\end{equation}
The trace of this tensor product is drawn by connecting the input and output legs at each level. For a permutation $\pi$, we may define the permutation operator $T_{\pi}$ through
\begin{equation}
    T_{\pi}\ket{a_1}\otimes \ket{a_2}\otimes \cdots \otimes \ket{a_k}=\ket{a_{\pi(1)}}\otimes\ket{a_{\pi(2)}}\otimes\cdots \otimes \ket{a_{\pi(k)}}.
\end{equation}
For concreteness, consider the permutation $\pi=(1)(2,3)$ and a tensor product of three states $\ket{a_1}\otimes\ket{a_2}\otimes\ket{a_3}$. The corresponding permutation operator acts as follows
\begin{equation}
    T_{(1)(2,3)}\ket{a_1}\otimes\ket{a_2}\otimes\ket{a_3}=\ket{a_1}\otimes\ket{a_3}\otimes\ket{a_2}.
\end{equation}
Diagrammatically, this permutation operator is
\begin{equation}
\begin{tikzpicture}
	\node[] (v0) at (0,.5) {$T_{(1)(2,3)}=$}; 
	\draw [thick,color=dullblue] 
	(1,1.5)--(3,1.5)
    (1,0.5) -- (1.5,0.5)  -- (2.5,-.5) -- (3,-.5)
    (1,-.5) -- (1.5,-.5) -- (2.5,.5) -- (3, .5);
\end{tikzpicture}.
\end{equation}
Diagrams for other permutation operators can be constructed in a similar manner. As an example, we compute the following trace
\begin{equation}
\begin{tikzpicture}
	\node[] (v0) at (-1.7,0) {$\tr\{T_{(1)(2,3)}A_1\otimes A_2\otimes A_3\}=$}; 
	\node [rectangle, fill=egg, rounded corners, minimum width=1em, minimum size=2em, draw] (v1) at (3.5,1) {$A_1$};
	\node [rectangle, fill=egg, rounded corners, minimum width=1em, minimum size=2em, draw] (v2) at (3.5,0) {$A_2$};
	\node [rectangle, fill=egg, rounded corners, minimum width=1em, minimum size=2em, draw] (v3) at (3.5,-1) {$A_3$};
	\draw [thick,color=dullblue] 
	(1,1)--(v1)--(4.3,1)
    (1,0) -- (1.5,0)  -- (2.5,-1) -- (v3) -- (4.3,-1)
    (1,-1) -- (1.5,-1) -- (2.5,0) -- (v2)--(4.3,0);
    \draw [thick,color=dullblue] 
    (1,1)--(1,1.5)--(4.3,1.5)--(4.3,1)
    (1,0)--(1,0.5)--(4.3,0.5)--(4.3,0)
    (1,-1)--(1,-1.5)--(4.3,-1.5)--(4.3,-1);
    \node[] (v4) at (4.7,0) {$=$}; 
    \node [rectangle, fill=egg, rounded corners, minimum width=1em, minimum size=2em, draw] (v5) at (6.3,0.5) {$A_1$};
	\node [rectangle, fill=egg, rounded corners, minimum width=1em, minimum size=2em, draw] (v6) at (5.8,-0.5) {$A_2$};
	\node [rectangle, fill=egg, rounded corners, minimum width=1em, minimum size=2em, draw] (v7) at (6.8,-0.5) {$A_3$};
	\draw [thick,color=dullblue] 
	(5.1,0.5) -- (v5) -- (7.5,0.5)
    (5.1,-0.5) -- (v6)-- (v7) -- (7.5,-0.5)
    (5.1,-0.5)--(5.1,-1)--(7.5,-1)--(7.5,-0.5)
    (5.1,0.5)--(5.1,1)--(7.5,1)--(7.5,0.5);
    \node[] (v5) at (9.2,0) {$=\tr\{A_1\} \tr\{A_2A_3\}$}; 
\end{tikzpicture}.
\end{equation}

The diagram for the Bell state $\ket{\Phi}=\frac{1}{d^{1/2}}\sum_{i=1}^d \ket{i}\otimes \ket{i}$ is defined as
\begin{equation}
\begin{tikzpicture}
	\node[] (v0) at (0,0) {$\ket{\Phi}=\frac{1}{d^{1/2}}\cdot$};
	\draw [thick,color=dullblue] 
    (1,.5) -- (1.5,.5) -- (1.5,-.5) -- (1,-.5);
\end{tikzpicture}.
\end{equation}
As an outer product, the state is
\begin{equation}
\begin{tikzpicture}
	\node[] (v0) at (0,0) {$\ket{\Phi}\bra{\Phi}=\frac{1}{d}\cdot$};
	\draw [thick,color=dullblue] 
    (1,.5) -- (1.5,.5) -- (1.5,-.5) -- (1,-.5)
    (2.25,.5)--(1.75,.5)--(1.75,-.5)--(2.25,-.5);
\end{tikzpicture}.
\end{equation}
The diagram for $\ket{\Phi}$ can be used to construct the useful identity
\begin{equation}
\begin{tikzpicture}
	\node[rectangle, fill=egg, rounded corners, minimum width=1em, minimum size=2em, draw] (v0) at (.75,.5) {$A_1$};
	\draw [thick,color=dullblue] 
    (0,.5)--(v0) -- (1.5,.5) -- (1.5,-.5) -- (0,-.5);
    \node[] (v1) at (2,0) {$=$};
    \node[rectangle, fill=egg, rounded corners, minimum width=1em, minimum size=2em, draw] (v2) at (3.25,-.5) {$A_1^T$};
	\draw [thick,color=dullblue] 
    (2.5,.5) --(4,.5) -- (4,-.5) -- (v2)--(2.5,-.5);
\end{tikzpicture}.
\end{equation}
This can be directly verified using the diagrammatic representation of $A_1^T$ from Eq.~\eqref{eq:transpose}.

\section{Background on random matrix theory}
This section relates permutation operators to random matrices, which allow for the computation of OTOCs through random measurements. For a more complete discussion, see \cite{Roberts_2017, Collins_2006}. Let $U$ be a unitary on Hilbert space $\mathcal{H}_d$. Define the operator
\begin{equation}
    A=\bigotimes_{i=1}^{k}A_i,
\end{equation}
where $A_i$ is an operator on $\mathcal{H}_d$. By the Schur-Weyl duality, if $[A,U^{\otimes k}]=0$ for all unitaries $U$, then $A$ can be written as a linear combination of permutations operators $T_\pi$,
\begin{equation}
    A=\sum_{\pi\in S_{k}} c_\pi T_{\pi}.
\end{equation}
The sum is carried out over $S_k$, the set of all permutations of the set $\{1,2,\ldots,k\}$. Define the $k$-fold twirling channel of $A$ as
\begin{equation}
    \Phi_{\Hr}^{(k)}(A)=\int_{\Hr}(U^{\dagger\otimes k}) A U^{\otimes k} dU.
\end{equation}
The integral is over the Haar measure $dU$ on the unitary group. As the Haar measure is invariant under the action of the group, we write the twirling channel as a superposition of permutation operators
\begin{equation}\label{eq:ChannelWein}
    \Phi_{\Hr}^{(k)}(A)=\sum_{\pi,\sigma\in S_k}C_{\pi,\sigma}T_{\pi}\tr\{T_\sigma A\},
\end{equation}
where the coefficients $C_{\pi,\sigma}$ are known as the Weingarten matrix. They satisfy
\begin{equation}
    (C^{-1})_{\pi,\sigma}=d^{f(\pi\circ \sigma)}.
\end{equation}
where $f(\pi\circ \sigma)$ is the number of cycles of the composition of permutations $\pi$ and $\sigma$.

\vspace{3mm}
\noindent
\begin{fact}\label{Fact:cor} Let $\{A_i\}_{i=1}^k$ be a set of traceless operators $A_i$ on Hilbert space $\mathcal{H}_d$ of dimension $d$. Let $T_{\sigma}$ be a permutation operator on $\mathcal{H}_d^{\otimes k}$ and let $U$ be a Haar random unitary on $\mathcal{H}_d$. Then
\begin{equation}
\sum_{\sigma\in D_k}\tr\{T_\sigma A_1\otimes\cdots\otimes A_k\}=\frac{(d-1+k)!}{(d-1)!}\overline{\langle U^\dagger A_1 U \rangle_{\rho_0}\cdots\langle U^\dagger A_k U \rangle_{\rho_0}}.
\end{equation}
The sum is carried out over $D_k$, the set of derangements (permutations with no fixed points) in $S_k$. The notation 
$\langle \boldsymbol{\cdot} \rangle_{\rho_0}=\tr\{ \rho_0 \ \boldsymbol{\cdot}\}$ denotes the expectation value over a pure state $\rho_0\in\mathcal{H}_d$. The notation $\overline{(\cdots)}$ denotes an integral over the Haar measure on the unitary group: $\overline{(\cdots)}=\int_{\Hr} (\cdots) dU $.
\end{fact}
\noindent
\begin{proof} 
Consider the k-fold twirling channel on $\otimes_{i=1}^k A_i$,
\begin{align}
\Phi^{(k)}_{\Hr}(A_1\otimes\cdots\otimes A_k)=\int_{\Hr}dU (U^{\dagger\otimes k}) (A_1\otimes\cdots \otimes A_k)(U^{\otimes k}).
\end{align}
Introducing pure state $\rho_0$, we compute the following trace
\begin{align}
\tr\{\rho_0^{\otimes k}\Phi^{(k)}_{\Hr}(A_1\otimes\cdots\otimes A_k)\}=\overline{\langle U^\dagger A_1 U\rangle_{\rho_0}\cdots \langle U^\dagger A_k U \rangle_{\rho_0}}.
\end{align}
Using Eq.~\eqref{eq:ChannelWein},
\begin{equation}
\Phi^{(k)}_{\Hr}(A_1\otimes\cdots\otimes A_k)=\sum_{\pi,\sigma\in S_k}C_{\pi,\sigma}T_{\pi}\tr\{T_{\sigma} A_1\otimes\cdots\otimes A_k\},
\end{equation}
so
\begin{align}
&\tr\{\rho_0^{\otimes k}\Phi^{(k)}_{\Hr}(A_1\otimes\cdots\otimes A_k)\}=\sum_{\pi,\sigma\in S_k}C_{\pi,\sigma}\tr\{\rho_0^{\otimes k}T_{\pi}\}\tr\{T_{\sigma} A_1\otimes\cdots\otimes A_k\}.
\end{align}
For any $\pi$, we can write
\begin{align}
\tr\{\rho_0^{\otimes k}T_\pi\}=\tr\{\rho_0^k\}^{n_k}\tr\{\rho_0^{k-1}\}^{n_{k-1}}\cdots \tr\{\rho_0\}^{n_1},
\end{align}
where $n_1,..., n_k$ are some integers. Since $\rho_0$ is a pure state, $\tr\{\rho_0^m\}=\tr\{\rho_0\}=1$ for any positive integer $m$. This yields $\tr\{\rho_0^{\otimes k}T_\pi\}=1$.
Our expression for the trace then becomes
\begin{align}
&\tr\{\rho_0^{\otimes k}\Phi^{(k)}_{\Hr}(A_1\otimes\cdots\otimes A_k)\}=\sum_{\sigma\in S_k}\tr\{T_{\sigma} A_1\otimes\cdots\otimes A_k\}\sum_{\pi\in S_k}C_{\pi,\sigma}.
\end{align}
It can be shown that the Weingarten matrix satisfies
\begin{equation}
\sum_{\pi\in S_k}C_{\pi,\sigma}=\frac{(d-1)!}{(d-1+k)!}.
\end{equation}
The trace then becomes
\begin{equation}
\tr\{\rho_0^{\otimes k}\Phi^{(k)}_{\Hr}(A_1\otimes\cdots\otimes A_k)\}=\frac{(d-1)!}{(d-1+k)!}\sum_{\sigma\in S_k}\tr\{T_{\sigma} A_1\otimes\cdots\otimes A_k\}.
\end{equation}
We now have two expressions for $\tr\{\rho_0^{\otimes k}\Phi^{(k)}_{\Hr}(A_1\otimes\cdots\otimes A_k)\}$. Together, they yield
\begin{equation}
\sum_{\sigma \in S_k}\tr\{T_{\sigma} A_1\otimes\cdots\otimes A_k\}=\frac{(d-1+k)!}{(d-1)!}\overline{\langle U^\dagger A_1 U\rangle_{\rho_0}\cdots \langle U^\dagger A_k U \rangle_{\rho_0}}.
\end{equation}
We have not yet used the traceless property of $A_i$, so the above equation is valid even for an $A_i$ with a non-vanishing trace.
We can split the sum over $S_k$ into two sums
\begin{equation}
\sum_{\sigma\in S_k}\tr\{T_{\sigma} A_1\otimes\cdots\otimes A_k\}=\sum_{\sigma\in D_k}\tr\{T_{\sigma} A_1\otimes\cdots\otimes A_k\}+\sum_{\sigma\in F_k}\tr\{T_{\sigma} A_1\otimes\cdots\otimes A_k\}.
\end{equation}
$D_k$ is the set of derangements and $F_k$ is the set of permutations in $S_k$ containing at least one fixed point. For $\sigma\in F_k$, the trace $\tr\{T_{\sigma} A_1\otimes\cdots\otimes A_k\}$ introduces at least one trace of the form $\tr\{A_i\}$, which vanishes since all $A_i$ are traceless. The sum over $F_k$ then vanishes and we may write
\begin{equation}
\sum_{\sigma\in S_k}\tr\{T_{\sigma} A_1\otimes\cdots\otimes A_k\}=\sum_{\sigma\in D_k}\tr\{T_{\sigma} A_1\otimes\cdots\otimes A_k\}.
\end{equation}
This yields
\begin{equation}
\sum_{\sigma \in D_k}\tr\{T_{\sigma} A_1\otimes\cdots\otimes A_k\}=\frac{(d-1+k)!}{(d-1)!}\overline{ \langle U^\dagger A_1 U\rangle_{\rho_0}\cdots \langle U^\dagger A_k U \rangle_{\rho_0}},
\end{equation}
proving the fact.
\end{proof}

\section{Relating correlators to global random unitaries}\label{sec:HaarProof}
The goal of this section is to write the correlators $\langle A_1 A_2A_1A_2\rangle$ and $\langle A_1 A_2 \rangle$ (where all $A_i$ are traceless, unitary, and Hermitian) in terms of simpler correlators that can be measured experimentally. This is done by introducing random unitaries.

The correlator $\langle A_1 A_2 \rangle$ may be rewritten in terms of a permutation operator
\begin{equation}
    \langle A_1 A_2 \rangle=\frac{1}{d}\tr\{A_1 A_2\}=\frac{1}{d}\tr\{T_{(1,2)}A_1\otimes A_2\}.
\end{equation}
This equation can be proven diagrammatically:
\begin{equation}
\begin{tikzpicture}
	 \node[rectangle, fill=egg, rounded corners, minimum width=1em, minimum size =2em, draw] (v0) at (0,0) {$A_1$};       
    \node[rectangle, fill=egg, rounded corners, minimum width=1em, minimum size =2em, draw] (v1) at (1,0) {$A_2$};      
    \node[] (v2) at (2.5,0) {$=$};   
    \draw [thick,color=dullblue] 
    (-1,0) -- (v0) -- (v1) -- (2,0) -- (2,.5)--(-1,0.5)--(-1,0);
	\node[rectangle, fill=egg, rounded corners, minimum width=1em, minimum size =2em, draw] (v0) at (6,0.5) {$A_1$};       
    \node[rectangle, fill=egg, rounded corners, minimum width=1em, minimum size =2em, draw] (v1) at (6,-.5) {$A_2$};         
    \draw [thick,color=dullblue] 
    (3.5,0.5) -- (4,0.5)  -- (5,-.5) -- (v1) -- (7,-.5) 
    (3.5,-.5) -- (4,-.5) -- (5,.5) -- (v0) -- (7,.5)
    (3.5,0.5)--(3.5,1)--(7,1)--(7,.5)
    (3.5,-0.5)--(3.5,-1)--(7,-1)--(7,-.5);
\end{tikzpicture}.
\end{equation}
Using Fact \ref{Fact:cor} with $k=2$ yields
\begin{equation}
\tr\{T_{(1,2)}A_1 \otimes A_2\}=d(d+1)\overline{\langle U^\dagger A_1 U\rangle_{\rho_0}\langle U^\dagger A_2 U\rangle_{\rho_0}}.
\end{equation}
Note that permutation $(1,2)$ is the only derangement in $D_2$. The correlator then becomes
\begin{equation}
    \langle A_1 A_2 \rangle=(d+1)\overline{\langle U^\dagger A_1 U\rangle_{\rho_0}\langle U^\dagger A_2 U\rangle_{\rho_0}}.
\end{equation}
The correlators on the right may be measured using the protocol discussed in Sec.~\ref{sec:global}.

We now express $\langle A_1 A_2 A_1 A_2 \rangle$ in terms of similar correlators. We begin by introducing a permutation operator
\begin{equation}
\langle A_1 A_2 A_1 A_2 \rangle =\frac{1}{d}\tr\{T_{(1,2,3,4)}A_1\otimes A_2\otimes A_1\otimes A_2\}.
\end{equation}
The application of Fact \ref{Fact:cor} with $k=4$ simplifies to
\begin{equation}\label{eq:der}
\sum_{\sigma \in D_4}\tr\{T_{\sigma} A_1\otimes A_2\otimes A_1\otimes A_2\}=d(d+1)(d+2)(d+3)\overline{\langle U^\dagger A_1 U\rangle_{\rho_0}^2\langle U^\dagger A_2 U \rangle_{\rho_0}^2}.
\end{equation}
The corresponding set of derangements is
\begin{equation}
\begin{split}
    D_4=\{&(1,2,3,4), (1,2,4,3), (1,3,2,4), (1,3,4,2), \\
    &(1,4,2,3), (1,4,3,2), (1,2)(3,4), (1,3)(2,4), (1,4)(2,3)\}.
\end{split}
\end{equation}
The sum over $D_4$ simplifies to
\begin{equation}
\begin{split}
\sum_{\sigma \in D_4}\tr\{T_{\sigma} A_1\otimes A_2\otimes A_1\otimes A_2\}=2\tr\{&A_1A_2A_1A_2\}+2\tr\{A_1 A_2\}^2\\
+&\tr\{A_1^2\}\tr\{A_2^2\}+4\tr\{A_1^2A_2^2\}.
\end{split}
\end{equation}
Since each $A_i$ is Hermitian and unitary, $A_i^2=I$. This yields $\tr\{A_1^2\}\tr\{A_2^2\}=d^2$ and $\tr\{A_1^2A_2^2\}=d$. Note that $\tr\{A_1A_2\}=d\langle A_1 A_2\rangle$. The sum then simplifies to
\begin{equation}
\sum_{\sigma \in D_4}\tr\{T_{\sigma} A_1\otimes A_2\otimes A_1\otimes A_2\}=2\tr\{A_1A_2A_1A_2\}
+2d^2\langle A_1 A_2 \rangle^2
+d(d+4).
\end{equation}
Equating this to Eq.~\eqref{eq:der},
\begin{equation}
2\tr\{A_1A_2A_1A_2\}
+2d^2\langle A_1 A_2\rangle^2
+d(d+4)=d(d+1)(d+2)(d+3)\overline{\langle U^\dagger A_1 U\rangle_{\rho_0}^2\langle U^\dagger A_2 U \rangle_{\rho_0}^2}.
\end{equation}
Rewriting,
\begin{equation}
\tr\{A_1A_2A_1A_2\}=\frac{1}{2}d(d+1)(d+2)(d+3)\overline{\langle U^\dagger A_1 U\rangle_{\rho_0}^2\langle U^\dagger A_2 U \rangle_{\rho_0}^2}
-d^2\langle A_1 A_2\rangle^2
-\frac{1}{2}d(d+4).
\end{equation}
The correlator then becomes
\begin{equation}
\langle A_1A_2A_1A_2\rangle=\frac{1}{2}(d+1)(d+2)(d+3)\overline{\langle U^\dagger A_1 U\rangle_{\rho_0}^2\langle U^\dagger A_2 U \rangle_{\rho_0}^2}
-d\langle A_1 A_2\rangle^2
-\frac{1}{2}(d+4).
\end{equation}
All correlators on the right can be measured using the protocol in Sec.~\ref{sec:global}.

\section{OTOCs at late times}
We calculate the average OTOC values where the evolution unitary is drawn randomly from the Haar measure on the unitary group. Since the Haar random unitary ensemble can be used in place of large-time chaotic evolution, these values serve as a benchmark for the evolution due to physical, chaotic
Hamiltonians. Consider the $4k$-point OTOC 
\begin{equation}
C_{4k}(t)=\frac1{d}\tr\{(W^\dagger(t) V^\dagger W(t)V)^k \},
\end{equation}
and take $W$ and $V$ as Pauli operators. For instance, $W=Z_1X_2$ and $V=Z_{N-1}Z_N$. Due to the random feature of the Haar measure, the specific choice of Pauli operators does not affect the result. Based on the Hermitian property of $W$ and $V$, the average OTOC can be written as  
\begin{equation}\label{avHaar}
\begin{aligned}
\overline{C_{4k}}&=\int_{\mathrm{Haar}}\frac1{d}\tr\{[U^{\dag}WUV]^{2k} \}dU\\
&=\frac1{d}\tr\Big\{\int_{\mathrm{Haar}}[U^{\dag}WU]^{\otimes 2k}dU\ V^{\otimes 2k} T_{\sigma_0}\Big\}\\
&=\frac1{d}\tr\left\{\Phi^{(2k)}_{\mathrm{Haar}}(W^{\otimes 2k})V^{\otimes 2k} T_{\sigma_0}\right\}\\
&=\frac1{d}\sum_{\pi,\sigma\in S_{2k}}C_{\pi,\sigma}\tr\left\{T_\pi W^{\otimes 2k}\right\} \tr\left\{V^{\otimes 2k}T_{\sigma'}\right\},\\
\end{aligned}
\end{equation}
where in the second line we write the equation over $2k$ copies of the original Hilbert space $\mathcal{H}_d$ and move the integral inside the trace. In the final line we apply the Weingarten formula. We also use the permutations $\sigma_0=(1,2,\ldots,2k)$ and $\sigma'=\sigma_0\circ\sigma$. Since any Pauli operator $W$ satisfies $W^2=I^{\otimes N}$ and $\tr\left\{W\right\}=0$, the 
term $\tr\left\{T_\pi W^{\otimes 2k}\right\}$ simplifies to
\begin{equation}
\begin{aligned}
\tr\left\{T_\pi W^{\otimes 2k}\right\}&=&0\ \ \ \mathrm{, \ if \ } \pi  \ \mathrm{has \  an \ odd \ cycle}\\
&=&d^{f(\pi)} \ \mathrm{, \ if \ } \pi  \ \mathrm{has \  no \ odd \ cycles}
\end{aligned}
\end{equation}
where $f(\pi)$ is the number of cycles of the permutation $\pi$. For example, $(12)(34)$ and $(123)(4)$ both contain two cycles, but both cycles are odd in the latter case.
This relation also holds for $\tr\left\{V^{\otimes 2k}T_{\sigma'}\right\}$. As a result, Eq.~\eqref{avHaar} simplifies to
\begin{equation}\label{avHaar1}
\begin{aligned}
\overline{C_{4k}}=\frac1{d}\sum_{\substack{\pi,\sigma\in S_{2k}, \\\pi, \sigma'=\sigma_0\circ\sigma \mathrm{\ even \  cycles}}}C_{\pi,\sigma}d^{f(\pi)+f(\sigma')}.
\end{aligned}
\end{equation}
A more general formula is derived in Ref.~\cite{Roberts_2017}.
\subsection{Four-point OTOC}
We first compute $\overline{C}_{4}$ where $k=1$. Noting $S_2=\{(1)(2),(1,2)\}$, the Weingarten matrix in this case is
\begin{equation}
C_{\pi,\sigma}=\frac{1}{d^2-1}\left(
\begin{array}{cc}
1 & -\frac1{d} \\
-\frac1{d} & 1
\end{array}
\right).
\end{equation}
We take $\sigma_0=(1,2)$. The sum in Eq.~\eqref{avHaar1}, only allows $\pi=(1,2)$ and $\sigma=(1)(2)$. As a result, the four-point OTOC in the large-time limit satisfies
\begin{equation}\label{}
\begin{aligned}
\overline{C_{4}}=\frac1{d}\frac1{d^2-1}\Big(-\frac1{d}\Big)d^2=-\frac1{d^2-1}
\end{aligned}.
\end{equation}

\subsection{Eight-point OTOC and (non)commutator types}\label{sec:type}
Reference \cite{Roberts_2017} has found that there are two general definitions of eight-point OTOCs that display distinct large-time behavior. For operators $A,B,C,D$, one can define the commutator type correlator as $C_{ct}=\langle \tilde{A}B\tilde{C}D\tilde{A}D\tilde{C}B\rangle$ and the non-commutator type correlator as $C_{nc}=\langle \tilde{A}B\tilde{C}D\tilde{A}B\tilde{C}D\rangle$. Define $\tilde{A}=U^{\dag}AU$, where $U$ is a Haar random unitary. The non-commutator type correlator attains a large-time value which scales as $\sim \frac1{d^2}$. This occurs even if $U$ is sampled from the Clifford ensemble. The four-point OTOC also scales as $\sim \frac1{d^2}$, indicating that non-commutator type correlators saturate the same floor value as the four-point correlator and therefore cannot reveal any new scrambling information in the large-time limit. The correlator $C_8(t)=\langle W(t)VW(t)VW(t)VW(t)V\rangle$ defined in our work is a non-commutator type correlator. Thus, it can only be used to study early-time scrambling dynamics. 

The commutator type correlator scales as $\sim\frac{1}{d^4}$ at long times, allowing us to probe systems which display higher-point scrambling in the large-time limit. By taking $A=B$ and $C=D$, the commutator type correlator can be written as
\begin{equation}\label{newC8}
\begin{aligned}
C_{ct}=\langle \tilde{A}A\tilde{C}C\tilde{A}C\tilde{C}A\rangle
\end{aligned}.
\end{equation}
We can adapt our measurement protocols to estimate this correlator.

\subsection{Estimating commutator type correlators}\label{sec:NonComEst}

Using the technique from Sec.~\ref{sec:2qubit}, we introduce a single Bell state to compute an estimator for the following commutator type correlator
\begin{align}\label{AP:C8comm}
C_{ct}(t)&=\langle W(t)WV(t)VW(t)VV(t)W\rangle,\\
&=\frac{1}{d}\tr\{ U_HWU_H^\dagger VU_HVU_H^\dagger W U_H VU_H^\dagger VU_HWU_H^\dagger  W\}.
\end{align}
The cyclic property of the trace is used in the second line. Eq.~\eqref{AP:C8comm} looks similar to the original eight-point OTOC defined in Eq.~\eqref{C4k}, but displays a different ordering of $W$ and $V$. Taking $W=Z_1$ and $V=X_N$, the correlator becomes

\begin{equation}
\scalebox{.75}{
\begin{tikzpicture}
     \draw [thick,color=dullblue] 
    (-1,1)--(16,1)
    
    (-1,0)--(16,0)
    
    (-1,-1)--(16,-1);

    \draw[thick,color=dullblue]
    (-.7,-.1)--(-.6,.1)
    (-.7+1.65,-.1)--(-.6+1.65,.1)
    (-.7+3.65,-.1)--(-.6+3.65,.1)
    (-.7+5.65,-.1)--(-.6+5.65,.1)
    (-.7+7.65,-.1)--(-.6+7.65,.1)
    (-.7+9.65,-.1)--(-.6+9.65,.1)
    (-.7+11.65,-.1)--(-.6+11.65,.1)
    (-.7+13.65,-.1)--(-.6+13.65,.1)
    (-.7+15.65,-.1)--(-.6+15.65,.1);
    
    \node[] (vtext) at (-2.2,0) {\Large $C_{ct}(t)=\frac{1}{d}\cdot$};
	 \node[rectangle, fill=egg, rounded corners, minimum width=2em, minimum height =7.2em, draw] (v0) at (0,0) {$U_H$};
	 \node[rectangle, fill=egg, rounded corners, minimum width=2em, minimum size =2em, draw] (v1) at (1,1) {$Z$};      
    \node[rectangle, fill=egg, rounded corners, minimum width=2em, minimum height =7.2em, draw] (v2) at (2,0) {$U_H^\dagger$};
    \node[rectangle, fill=egg, rounded corners, minimum width=2em, minimum height =2em, draw] (v3) at (3,-1) {$X$};
    \node[rectangle, fill=egg, rounded corners, minimum width=2em, minimum height =7.2em, draw] (v4) at (4,0) {$U_H$};
	 \node[rectangle, fill=egg, rounded corners, minimum width=2em, minimum size =2em, draw] (v5) at (5,-1) {$X$};      
    \node[rectangle, fill=egg, rounded corners, minimum width=2em, minimum height =7.2em, draw] (v6) at (6,0) {$U_H^\dagger$};
    \node[rectangle, fill=egg, rounded corners, minimum width=2em, minimum height =2em, draw] (v7) at (7,1) {$Z$};
    
    \node[rectangle, fill=egg, rounded corners, minimum width=2em, minimum height =7.2em, draw] (v8) at (8,0) {$U_H$};
	 \node[rectangle, fill=egg, rounded corners, minimum width=2em, minimum size =2em, draw] (v9) at (9,-1) {$X$};      
    \node[rectangle, fill=egg, rounded corners, minimum width=2em, minimum height =7.2em, draw] (v10) at (10,0) {$U_H^\dagger$};
    \node[rectangle, fill=egg, rounded corners, minimum width=2em, minimum height =2em, draw] (v11) at (11,-1) {$X$};
    \node[rectangle, fill=egg, rounded corners, minimum width=2em, minimum height =7.2em, draw] (v12) at (12,0) {$U_H$};
	 \node[rectangle, fill=egg, rounded corners, minimum width=2em, minimum size =2em, draw] (v13) at (13,1) {$Z$};      
    \node[rectangle, fill=egg, rounded corners, minimum width=2em, minimum height =7.2em, draw] (v14) at (14,0) {$U_H^\dagger$};
    \node[rectangle, fill=egg, rounded corners, minimum width=2em, minimum height =2em, draw] (v16) at (15,1) {$Z$};
\end{tikzpicture},
}
\end{equation}

\begin{equation}\label{eq:NonComm}
\scalebox{.75}{
\begin{tikzpicture}
     \draw [thick,color=dullblue] 
    (-1,1)--(.75,1) (1.25,1)--(12.75,1) (13.25,1)--(16,1)
    
    (-1,0)--(0.6,0) (0.9,0)--(1.1,0) (1.4,0)--(12.6,0) (12.9,0)--(13.1,0) (13.4,0)--(16,0)
    
    (-1,-1)--(0.6,-1) (0.9,-1)--(1.1,-1) (1.4,-1)--(4.75,-1) (5.25,-1)--(8.75,-1) (9.25,-1)--(12.6,-1) (12.9,-1)--(13.1,-1) (13.4,-1)--(16,-1)

    (-1,-2)--(0.75,-2) (1.25,-2)--(4.6,-2) (4.9,-2)--(5.1,-2) (5.4,-2)--(8.6,-2)
    (8.9,-2)--(9.1,-2) (9.4,-2)--(12.6,-2) (12.9,-2)--(13.1,-2) (13.4,-2)--(16,-2)
    
    (-1,-3)--(4.75,-3) (5.25,-3)--(8.6,-3)
    (8.9,-3)--(9.1,-3) (9.4,-3)--(12.6,-3) (12.9,-3)--(13.1,-3) (13.4,-3)--(16,-3)
    
    (-1,-4)--(8.75,-4)
    (9.25,-4)--(12.6,-4) (12.9,-4)--(13.1,-4) (13.4,-4)--(16,-4)
    
    (-1,-5)--(12.75,-5) (13.25,-5)--(16,-5);
    
    \draw [thick,color=dullblue] 
    (0.75,1)--(0.75,-2) (1.25,1)--(1.25,-2) 
    (4.75,-1)--(4.75,-3) (5.25,-1)--(5.25,-3)
    (8.75,-1)--(8.75,-4) (9.25,-1)--(9.25,-4) 
    (12.75,1)--(12.75,-5) (13.25,1)--(13.25,-5);

    \draw[thick,color=dullblue]
    (-.7,-.1)--(-.6,.1)
    (-.7+1.65,-.1)--(-.6+1.65,.1)
    (-.7+3.65,-.1)--(-.6+3.65,.1)
    (-.7+5.65,-.1)--(-.6+5.65,.1)
    (-.7+7.65,-.1)--(-.6+7.65,.1)
    (-.7+9.65,-.1)--(-.6+9.65,.1)
    (-.7+11.65,-.1)--(-.6+11.65,.1)
    (-.7+13.65,-.1)--(-.6+13.65,.1)
    (-.7+15.65,-.1)--(-.6+15.65,.1);
    
    \node[] (vtext) at (-2.2,-2) {\Large $C_{ct}(t)=\frac{1}{d}\cdot$};
	 \node[rectangle, fill=egg, rounded corners, minimum width=2em, minimum height =7.2em, draw] (v0) at (0,0) {$U_H$};
	 \node[rectangle, fill=egg, rounded corners, minimum width=2em, minimum size =2em, draw] (v1) at (3,-2) {$Z^T$};      
    \node[rectangle, fill=egg, rounded corners, minimum width=2em, minimum height =7.2em, draw] (v2) at (2,0) {$U_H^\dagger$};
    \node[rectangle, fill=egg, rounded corners, minimum width=2em, minimum height =2em, draw] (v3) at (3,-1) {$X$};
    \node[rectangle, fill=egg, rounded corners, minimum width=2em, minimum height =7.2em, draw] (v4) at (4,0) {$U_H$};
	 \node[rectangle, fill=egg, rounded corners, minimum width=2em, minimum size =2em, draw] (v5) at (7,-3) {$X^T$};      
    \node[rectangle, fill=egg, rounded corners, minimum width=2em, minimum height =7.2em, draw] (v6) at (6,0) {$U_H^\dagger$};
    \node[rectangle, fill=egg, rounded corners, minimum width=2em, minimum height =2em, draw] (v7) at (7,1) {$Z$};
    
    \node[rectangle, fill=egg, rounded corners, minimum width=2em, minimum height =7.2em, draw] (v8) at (8,0) {$U_H$};
	 \node[rectangle, fill=egg, rounded corners, minimum width=2em, minimum size =2em, draw] (v9) at (11,-4) {$X^T$};      
    \node[rectangle, fill=egg, rounded corners, minimum width=2em, minimum height =7.2em, draw] (v10) at (10,0) {$U_H^\dagger$};
    \node[rectangle, fill=egg, rounded corners, minimum width=2em, minimum height =2em, draw] (v11) at (11,-1) {$X$};
    \node[rectangle, fill=egg, rounded corners, minimum width=2em, minimum height =7.2em, draw] (v12) at (12,0) {$U_H$};
	 \node[rectangle, fill=egg, rounded corners, minimum width=2em, minimum size =2em, draw] (v13) at (15,-5) {$Z^T$};      
    \node[rectangle, fill=egg, rounded corners, minimum width=2em, minimum height =7.2em, draw] (v14) at (14,0) {$U_H^\dagger$};
    \node[rectangle, fill=egg, rounded corners, minimum width=2em, minimum height =2em, draw] (v16) at (15,1) {$Z$};
\end{tikzpicture}.
}
\end{equation}
Retrieve the definition of $\rho_{H,N,a_1}$ from Sec.~\ref{sec:2qubit}. Define $\rho_{1,a_1}$ as the state in which system qubit 1 forms a Bell state with one ancillary qubit, while the remaining qubits are in the maximally mixed state. Define the time-evolved state
\begin{equation}
\rho_{H,1,a_1}=(U_H\otimes I_{a_1})\rho_{1,a_1}(U_H^\dagger\otimes I_{a_1}).  
\end{equation}
The correlator from Eq.~\eqref{eq:NonComm} can be written as
\begin{equation}
    C_{ct}(t)=d^{3}\tr\{(\rho_{H,1,a_1}\otimes\rho_{H,N,a_1}\otimes\rho_{H,N,a_1}\otimes\rho_{H,1,a_1})O_{ct}\},
\end{equation}
where
\begin{equation}
    O_{ct}=(X_NZ^T_{a_1}\otimes Z_1X^T_{a_1} \otimes X_NX^T_{a_1}\otimes Z_1Z^T_{a_1})\prod_{l=1}^NT_{(l,N+1+l,2(N+1)+l,3(N+1)+l)}.
\end{equation}
Prepare a shadow of size $K\geq 2$ for $\rho_{H,1,a_1}$ and $\rho_{H,N,a_1}$. These two shadows can be used to compute the estimator
\begin{equation}
    \hat{C}_{ct}(t)=\frac{d^3}{4}\binom{K}{2}^{-2}\sum_{\substack{i_1\neq i_4\\i_2\neq i_3}}^K\tr\Big\{(\hat{\rho}_{H,1,a_1}^{(i_1)}\otimes\hat{\rho}_{H,N,a_1}^{(i_2)}\otimes\hat{\rho}_{H,N,a_1}^{(i_3)}\otimes\hat{\rho}_{H,1,a_1}^{(i_4)}) O_{ct}\Big\}.
\end{equation}

The commutator protocol to estimate $C_{ct}(t)$ is:
\begin{enumerate}
\begin{samepage}
    \item Prepare $N$ system qubits and $1$ ancillary qubit. Create a Bell state between system qubit $N$ and the ancillary qubit. Prepare the remaining system qubits in the maximally mixed state. 
    \item Evolve the system qubits with $U_H(t)$.
    \item Construct a shadow of size $K\geq 2$ for the state.
    \item Prepare $N$ system qubits and $1$ ancillary qubit. Create a Bell state between system qubit $1$ and the ancillary qubit. Prepare the remaining system qubits in the maximally mixed state. 
    \item Evolve the system qubits with $U_H(t)$.
    \item Construct a shadow of size $K\geq 2$ for the state.
    \item Use both shadows to compute $\hat{C}_{ct}(t)$.
\end{samepage}
\end{enumerate}
This requires the preparation of two different states. However, by introducing two ancillary qubits and generating two Bell states, this protocol can be implemented through the preparation of just one state.

\section{Proofs}
\subsection{Proof of Lemma \ref{Lemma:2copy}}\label{ApLem2copy}
\begin{proof}
We begin the proof by first taking $O_2=T_{(1,2)}$, and later show that it can be used to bound the original case $O_2=T_{(1,2)}W^{\otimes 2}$. The swap operator $T_{(1,2)}$ on $\mathcal{H}_d^{\otimes 2}$ can be decomposed in the Pauli basis
\begin{equation}
T_{(1,2)}=\frac1{d}\sum_i P_i\otimes P_i,
\end{equation}
where $P_i$ is a Pauli operator and the sum is carried out over the $N$-qubit Pauli group. Following the variance analysis for the shadow norm (see Eq.~(S53) in Ref.~\cite{Huang_2020}), the variance can be upper bounded by 
\begin{equation}\label{VarBetter}
\begin{aligned}
&\max_{\sigma=\rho^{\otimes 2}} \mbb{E}_{U\sim \mr{Cl}(2)^{\otimes {2N}}}\sum_{b\in\{0,1\}^{2N}}\bra{b}U\sigma U^{\dag}\ket{b} \bra{b}U(\mc{D}^{-1}_{\frac1{3}})^{\otimes 2N}(T_{(1,2)}) U^{\dag}\ket{b}^2\\
=&\max_{\sigma=\rho^{\otimes 2}} \frac1{d^2}\sum_{i,j}\mbb{E}_{U\sim \mr{Cl}(2)^{\otimes {2N}}}\sum_{b\in\{0,1\}^{2N}} \bra{b}U\sigma U^{\dag}\ket{b} \bra{b}U(\mc{D}^{-1}_{\frac1{3}})^{\otimes 2N}(P_i^{\otimes 2}) U^{\dag}\ket{b}
\bra{b}U(\mc{D}^{-1}_{\frac1{3}})^{\otimes 2N}(P_j^{\otimes 2}) U^{\dag}\ket{b}
\\
=&\max_{\sigma=\rho^{\otimes 2}} \frac1{d^2}\sum_{i,j} f(i,j)^2 \tr\{\sigma P_i^{\otimes 2}P_j^{\otimes 2}\}\\
=&\max_{\rho} \frac1{d^2}\sum_{i,j} f(i,j)^2 \tr\{\rho P_iP_j\}^2,
\end{aligned}
\end{equation}
where in the second line we insert the decomposition of the swap operator, and $f(i,j)$ is a function for two Pauli operators defined in Lemma 4 in Ref.~\cite{Huang_2020}. $f(i,j)=0$ if there exists a qubit index $k$ such that the k-th qubit operators $P_i^k\neq P_j^k$ and $P_i^k,P_j^k\neq I$. Otherwise, $f(i,j) = 3^s$, where s is the number of non-identity Pauli indices that match.
We remark that, different than the shadow norm, our maximization is on the subset of the states in $\mathcal{H}_d^{\otimes 2}$ with a 2-copy structure, i.e. states of the form $\sigma=\rho^{\otimes 2}$. This tensor structure leads to the tighter bound for the variance. 

We introduce three functions denoted by $a, b, c$ for the two $N$-qubit Pauli operators $P_i, P_j$ indexed by $i,j$. $a(i,j)$ denotes the qubit positions for which $P_i, P_j$ share the same single-qubit Pauli operator; $b(i,j)$ denotes the positions where $P_i$ has a single-qubit Pauli operator and $P_j$ has $I$ (or vice versa); $c(i,j)$ denotes the positions where $P_i, P_j$ both have $I$.
For example, for a 6-qubit system with $P_i=X_1Y_2X_3I_4I_5I_6$ and $ P_j=X_1Y_2I_3Z_4I_5I_6$, the functions are: $a(i,j)=\{1,2\}$, $b(i,j)=\{3,4\}$ and $c(i,j)=\{5,6\}$. We use $|a(i,j)|$ to denote the number of elements in $a(i,j)$ and omit the $i,j$ indices when there is no ambiguity. $P^a_i$ is the $|a|$-qubit operator restricted on subsystem $a$ from $P_i$.

One needs not to consider the $P_i, P_j$ pair with different Pauli operators acting on the same qubit, since they return $f(i,j)=0$. As a result, the summation in Eq.~\eqref{VarBetter} can be transformed to the summation on all possible $a,b,c$ with $|a|+|b|+|c|=n$. We use $(i, j)\vdash(a,b,c) $ to denote that $P_i, P_j$ satisfy the operator constraints on the given $a,b,c$ subsystems. We write

\begin{equation}\label{Varabc}
\begin{aligned}
\sum_{i,j} f(i,j)^2 \tr\{\rho P_iP_j\}^2
=&\sum_{a,b,c} 3^{2|a|} \sum_{(i,j)\vdash(a,b,c)}\tr\{\rho I_{a}\otimes P_i^b P_j^b\otimes I_{c}\}^2\\
=&\sum_{a,b,c} 3^{2|a|}\sum_{(i,j)\vdash(a,b,c)} \tr\{\rho_b P_i^b P_j^b\}^2.
\end{aligned}
\end{equation}
Here $3^{2|a|}$ accounts for the function $f(i,j)^2$, and we use the fact that the qubit Pauli operators of $P_i, P_j$ are the same in $a$ and they both have $I$ on $c$. 

There are $3^{|a|}$ possible choices of Paulis $\{X,Y,Z\}^{|a|}$ on subsystem $a$. $P^i_b P^j_b$ can take all Pauli operators $\{X,Y,Z\}^{|b|}$ on $b$ by the property of the $b$ function. For each choice of operators on $a$, we can sum the result of all of these possible Paulis on $b$. One thus has 
\begin{equation}\label{Paulib}
\begin{aligned}
\sum_{(i,j)\vdash(a,b,c)} \tr\{\rho_b P^i_b P^j_b\}^2
&=3^{|a|} 2^{|b|}\sum_{\tilde{P_b}\in \{X,Y,Z\}^{|b|}} \tr\{\rho_b \tilde{P_b}\}^2\\
&\leq 3^{|a|} 2^{|b|}\sum_{\tilde{P_b}\in \{\id,X,Y,Z\}^{|b|}} \tr\{\rho_b \tilde{P_b}\}^2\\
&=3^{|a|}2^{|b|} 2^{|b|} \tr\{\rho_b ^2\}\\
&\leq 3^{|a|}2^{2|b|}.
\end{aligned}
\end{equation}
The $2^{|b|}$ in the first line is due to the two possibilities for $P_i$ or $P_j$ taking a Pauli or the identity on the qubit in $b$. The last inequality is due to the purity $\tr\{\rho_b ^2\}\leq 1$.
Inserting Eq.~\eqref{Paulib} into Eq.~\eqref{Varabc}, we get the upper bound for the variance
\begin{equation}
\begin{aligned}
&\frac1{d^2}\sum_{i,j} f(i,j)^2 \tr\{\rho P_iP_j\}^2\\
=&\frac1{d^2}\sum_{a,b,c} 3^{2|a|} \sum_{(i,j)\vdash(a,b,c)}\tr\{\rho_b P^i_b P^j_b\}^2\\
\leq &\frac1{d^2}\sum_{|a|+|b|+|c|=N} \binom{N}{|a|}\binom{N-|a|}{|b|}\binom{N-|a|-|b|}{|c|}   3^{3|a|} 2^{2|b|}=\frac{(3^3+2^2+1)^N}{d^2}=d^3.\\
\end{aligned}
\end{equation}

For the case $O_2=T_{(1,2)}W^{\otimes 2}$ with $W$ a Pauli operator, as in Eq.~\eqref{Varabc} one has that the variance is upper bounded by
\begin{equation}
\begin{aligned}
\mathrm{Var}\left[\tr\left\{ O_2 \hat{\rho}\otimes\hat{\rho}'\right\}\right]\leq&\max_{\sigma=\rho^{\otimes 2}} \frac1{d^2}\sum_{i,j} f(i,j)^2 \tr\{\sigma P_i^{\otimes 2}W^{\otimes 2}P_j^{\otimes 2}W^{\otimes 2}\}\\
=&\max_{\rho} \frac1{d^2}\sum_{i,j} f(i,j)^2 \tr\{\rho P_iWP_jW\}^2\\
=&\max_{\rho} \frac1{d^2}\sum_{i,j} f(i,j)^2 \tr\{\rho P_iP_j\}^2\leq d^3,
\end{aligned}
\end{equation}
where we use the fact $\tr\{\rho P_iWP_jW\}^2=\tr\{\rho P_iP_j\}^2$ for any Pauli operator.
\comments{
\begin{equation}
\begin{aligned}
T_{(1,2)}W^{\otimes 2}=&\frac1{d}\left[\sum_{\{i:[P_i,W]=0\}} (P_iW)\otimes (P_iW)+\sum_{\{j:{P_j,W}=0\}} (P_jW)\otimes (P_jW)\right]\\
=&\frac1{d}\left[\sum_{\{i:[P_i,W]=0\}} (P_{i'})\otimes (P_{i'})-\sum_{\{j:{P_j,W}=0\}} P_{j'}\otimes P_{j'}\right]
\end{aligned}
\end{equation}
where $P_{i'}=P_iW, P_{j'}=iP_jW$, and there are half of Pauli operator which (anti)commute with $W$. }
\end{proof}

\subsection{Proof of Proposition \ref{prop:D8}}\label{Ap:D8Var} 
\begin{proof}
We rewrite the variance of $\hat{L}_{8}$ from main text here.
\begin{equation*}
\begin{aligned}
\mathrm{Var}(\hat{L}_{8})=&\left[
d^3\binom{K}{4}^{-1}\right]^2\mathrm{Var}\left(\sum_{\vec{i}  }\hat{D}_{8}\left(\vec{i}\right)\right)  \\ 
=&
d^6\binom{K}{4}^{-2}\sum_{\vec{i},\vec{j}}\left\{\mathbb{E}\left[\hat{D}_{8}\left(\vec{i}\right)\hat{D}_{8}\left(\vec{j}\right)\right]-D_8^2 \right\}\\
=&
d^6\binom{K}{4}^{-2}\sum_{\vec{i},\vec{j}}\left[V_8\left(\vec{i},\vec{j}\right)-D_{8}^2\right],
\end{aligned}
\end{equation*}
where the summation over $\vec{i}$ labels the summation over $i_1<i_2<i_3<i_4$, and similarly for $\vec{j}$. 

Similar to the analysis of $\hat{C}_{4}$ in Sec.~\ref{Sec:Var4}, the variance $\mathrm{Var}(\hat{L}_{8})$ depends on the coincidences of the indices. For simplicity of notation, we use $T_t$ to denote the shift operator $T_{(1,2\cdots,t)}$ in the following discussion.
\begin{itemize}
      \item No coincidence: $V_8=D_8^2$, since the snapshots are independent and one can evaluate the expectation value separately. 
      \item One coincidence: there are a total of $\binom{M}{1}\binom{M-1}{6}\binom{6}{3}$ such terms. We have
\begin{equation}
\begin{aligned}
V_8&=\mathbb{E}\left[\tr\left\{\Theta_4(T_4) W^{\otimes 4} \rho_{V}^{\otimes3}\otimes\hat{\rho}_{V}\right\}^2\right]\\
&=\mathbb{E}\left[\tr\left\{T_4 W^{\otimes 4} \rho_{V}^{\otimes3}\otimes\hat{\rho}_{V}\right\}^2\right]\\
&=\mathbb{E}\left[\tr\left\{ W\rho_{V}W\rho_{V}W\rho_{V}W\  \hat{\rho}_{V}\right\}^2\right].
\end{aligned}
\end{equation}
The second line is due to the six different permutation orders returning the same result. We can take $O_1=W\rho_{V}W\rho_{V}W\rho_{V}W$, and $\mathbb{E}\left[\tr\left\{ O_1 \hat{\rho}_{V}\right\}\right]=D_8$. In this way, $V_8-D_8^2=\mathrm{Var}[\tr\left\{ O_1 \hat{\rho}_{V}\right\}]$, and one can analyze the variance based on Fact \ref{Fact:Var} with $\tr\{O_1^2\}=\frac{4}{d^2}D_8$. 
\item Two coincidences: there are a total of $\binom{M}{2}\binom{M-2}{4}\binom{4}{2}$ such terms. Due to the redundancy of the twirling channel $\Theta_4$, we only need to consider the average on the following three orders
\begin{equation}
\begin{aligned}
V_8=&\mathbb{E}\Big[\Big(\frac1{3}\tr\left\{T_4 W^{\otimes 4} \rho_{V}^{\otimes2}\otimes\hat{\rho}^{(1)}_{V}\otimes\hat{\rho}^{(2)}_{V}\right\}+\frac1{3}\tr\left\{T_4 W^{\otimes 4} \rho_{V}^{\otimes2}\otimes\hat{\rho}^{(2)}_{V}\otimes\hat{\rho}^{(1)}_{V}\right\}+
\frac1{3}\tr\left\{T_4 W^{\otimes 4}
\rho_{V}\otimes\hat{\rho}^{(1)}_{V}\otimes \rho_{V}\otimes\hat{\rho}^{(2)}_{V}\right\}\Big)^2\Big]\\
=&\mathbb{E}\Big[\Big(\frac1{3}\tr\left\{T_2 O'\otimes W\ \hat{\rho}^{(1)}_{V}\otimes\hat{\rho}^{(2)}_{V}\right\}+\frac1{3}\tr\left\{T_2 W\otimes O'\ \hat{\rho}^{(1)}_{V}\otimes\hat{\rho}^{(2)}_{V}\right\}+
\frac1{3}\tr\left\{T_2 [W\rho_{V}W]^{\otimes 2}\ \hat{\rho}^{(1)}_{V}\otimes\hat{\rho}^{(2)}_{V}\right\}\Big)^2\Big],
\end{aligned}
\end{equation}
where $O'=W\rho_{V}W\rho_{V}W$. We take the final observable 
$O_2=\frac1{3} \left(T_2 O'\otimes W+T_2 W\otimes O' +T_2 [W\rho_{V}W]^{\otimes 2}\right)$ to analyze the variance based on Fact \ref{Fact:Var} with 
\begin{equation}
\begin{aligned}
\tr\{O_2^2\}&=\frac{1}{9}\tr\left\{\left( O'\otimes W+W\otimes O' +[W\rho_{V}W]^{\otimes 2}\right)^2\right\}\\
&=\frac{1}{9}\left(2d\tr\{O'^2\}+2\tr\{O'W\}^2+4\tr\{O'W\rho_{V}W\}\tr\{\rho_{V}W\}
+\tr\{(W\rho_{V}W)^2\}^2\right)\\
&=\frac{1}{9}\left(\frac{8}{d}D_4+2D_4^2+\frac{16}{d^2}D_2^2+\frac{4}{d^2}\right)
\end{aligned}
\end{equation}
with $D_2=\tr\{W\rho_{V}\}$ and $D_4=\tr\{W\rho_{V}W\rho_{V}\}=(C_4+1)/d$.
\item Three coincidences: there are a total of $\binom{M}{3}\binom{M-3}{2}\binom{2}{1}$ such terms. Similar to the two previous coincidence cases, one can get
\begin{equation}
\begin{aligned}
V_8&=\mathbb{E}\left[\tr\left\{O_3  \hat{\rho}^{(1)}_{V}\otimes\hat{\rho}^{(2)}_{V}\otimes\hat{\rho}^{(3)}_{V}\right\}^2\right],
\end{aligned}
\end{equation}
where $O_3=\frac1{2}(T_3+T_3^{\dag})\Theta_3(O''\otimes W^{\otimes 2})$, with $O''=W\rho_{V}W$. The variance can be bounded by Fact \ref{Fact:Var} with
\begin{equation}
\begin{aligned}
\tr\{O_3^2\}&=\tr\left\{\frac1{4}(2\id+T^2_3+T_3^{\dag 2})\Theta_3(O''\otimes \id^{\otimes 2})^2\right\}\\
&=\frac1{2}\tr\left\{(\id+T_3)\Theta_3(O''\otimes \id^{\otimes 2})^2\right\}\\
&=\frac1{6}\left[\tr\{O''^2\}d^2+2\tr\{O''W\}^2d\right]+\frac1{2}\tr\{O''^2\}\\
&=\frac{d(1+D_2^2)}{3}+\frac{1}{d},
\end{aligned}
\end{equation}
where we use $\tr\{O''^2\}=\frac{2}{d}$.
\item Four coincidences: this corresponds to $\vec{i}=\vec{j}$, with $\binom{M}{4}$ such terms
\begin{equation}
\begin{aligned}
V_8&=\mathbb{E}\left[\tr\left\{O_4 \hat{\rho}^{(1)}_{V}\otimes\hat{\rho}^{(2)}_{V}\otimes\hat{\rho}^{(3)}_{V}\otimes\hat{\rho}^{(4)}_{V}\right\}^2\right],\\
\end{aligned}
\end{equation}
where $O_4=\Theta_4(T_4) W^{\otimes 4}$ for the 4-copy state, and 
\begin{equation}
\begin{aligned}
\tr\{O_4^2\}&=\tr\left\{\Theta_4\left(T_4\right)\Theta_4\left(T_4\right)\right\}\\
&=\tr\left\{\Theta_4\left(T_4\right)T_4\right\}\\
&=\frac{d^4+5d^2}{6}.
\end{aligned}
\end{equation}
\end{itemize}

For simplicity, denote the number of terms in each case as $N_c=\binom{K}{c}\binom{K-c}{8-2c}\binom{8-2c}{4-2c}=\binom{K}{4}\binom{4}{c}\binom{K-4}{4-c}$ for $c=1,2,3,4$, and denote the variance of $O_c$ on $c$ snapshots as $\mathrm{Var}(\hat{O}_c)$. 
Combining the above cases, one has
\begin{equation}
\begin{aligned}
\mathrm{Var}(\hat{L}_{8})=& d^6\binom{K}{4}^{-2}\sum_{\vec{i},\vec{j}}\left[V_8\left(\vec{i},\vec{j}\right)-D_{8}^2\right]\\
=&d^6\binom{K}{4}^{-2} \sum_{c=1}^4 N_c \mathrm{Var}(\hat{O}_c)\\
\leq& d^6\binom{K}{4}^{-1}\sum_{c=1}^4\binom{4}{c}\binom{K-4}{4-c}\tr(O_c^2)d^c\\
=&\frac{24d^6}{K(K-1)(K-2)(K-3)}\Big\{ 4\cdot\frac1{6}(K-4)(K-5)(K-6)\frac{4}{d^2}D_8d+\\
&6/2(K-4)(K-5)\frac{1}{9}\left(\frac{8}{d}D_4+2D_4^2+\frac{16}{d^2}D_2^2+\frac{4}{d^2}\right)d^2+ \\
&4(K-4)\left[\frac{d(1+D_2^2)}{3}+\frac{1}{d}\right]d^3 +\frac{d^4+5d^2}{6}d^4\Big\}\\
\leq &\frac{64d^5D_8}{K}+ \frac{16(4dD_4+d^2D_4^2+8D_2^2+2)}{K^2}+\frac{32[d^{10}(1+D_2^2)+3d^8]}{K^3}+\frac{4(d^{14}+5d^6)}{K^4}.
\end{aligned}
\end{equation}
Focusing on the early-time behaviour where $d^3D_8\sim 8,dD_4 \sim 2$ and also using $|D_2|\leq 1$, the variance satisfies
\begin{equation}
\begin{aligned}
\mathrm{Var}(\hat{L}_{8})\leq \frac{512d^2}{K}+\frac{352}{K^2}+\frac{32(2d^{10}+3d^8)}{K^3}+\frac{4(d^{14}+5d^6)}{K^4}.
\end{aligned}
\end{equation}

It is clear that in the large $d$ limit, the final term is dominating and the variance scales as $O(d^{14}/K^4)$. To suppress the error to $\epsilon$, one needs $K=O(d^{3.5}/\sqrt{\epsilon})$. If one naively uses quantum state tomography to reconstruct $\rho_V$ and calculate $L_8=d^3D_8=d^3\tr\{(W\rho_V)^4\}$ within an error $\epsilon$, one needs the precision of tomography to be about  $\epsilon'=\epsilon/d^3$. Even with the optimistic estimation obtained by making the trace distance comparable to the infidelity, the necessary number of measurements in quantum tomography scales as $K=\Omega(d^2/\epsilon')=\Omega(d^5/\epsilon)$. Furthermore, if one is restricted to independent measurements on a single-copy of $\rho_V$ (like in the classical shadow protocol here), the scaling gets worse.
\end{proof}

\end{appendix}

\end{document}